\documentclass[12pt]{iopart} 
\usepackage{graphics} 
\usepackage{graphicx}
\usepackage{caption}  
\usepackage{iopams} 
\expandafter\let\csname equation*\endcsname\relax
 \expandafter\let\csname endequation*\endcsname\relax
\usepackage{amsmath}
\newcommand{\istate}{$|I\rangle$}

\newcommand{\gstate}{$|G\rangle$}  

\begin{document}

\title{Resonant Elastic Soft  X-Ray Scattering}

\author{J. Fink$^1$, E. Schierle$^2$,  E. Weschke$^2$, and J. Geck$^1$}
\address{$^1$ Leibniz-Institute for Solid State and Materials Research Dresden, P.O.Box
270116, D-01171 Dresden, Germany\\ $^2$Helmholtz-Zentrum Berlin f\"ur Materialien und
Energie, Albert-Einstein-Str. 15, D-12489 Berlin, Germany\\} 
\ead{J.Fink@ifw-dresden.de;enrico.schierle@helmholtz-berlin.de;\\eugen.weschke@helmholtz-berlin.de;J.Geck@ifw-dresden.de}

\begin{abstract} 
Resonant (elastic) soft  x-ray scattering (RSXS) offers a unique element, site, and valence specific probe to study spatial modulations of charge, spin, and orbital degrees of freedom  in solids on the nanoscopic length scale. It cannot only be used to investigate single crystalline materials. This method also enables to examine electronic ordering phenomena in thin films and to zoom into electronic properties emerging at buried interfaces in artificial heterostructures. During the last 20 years, this  technique, which combines x-ray scattering with x-ray absorption spectroscopy, has  developed into a powerful probe to study electronic ordering phenomena in complex materials and furthermore delivers  important information on the electronic structure  of condensed matter. This review provides an introduction to the technique, covers the progress in experimental equipment, and gives a survey on recent RSXS studies of ordering in correlated electron systems and at interfaces.

(Some figures in this article are in colour only in the electronic version)
\end{abstract}


\vspace{2pc} 
\submitto{\RPP}
\vspace*{20mm}
Dated: \the\day. \the\month. \the\year
\newpage
\tableofcontents
\newpage

\maketitle

\section{Introduction}
Scattering experiments are amongst the most powerful tools to obtain information on the microscopic structure of matter. In solid state physics, conventional elastic scattering of photons, electrons, and neutrons are the standard methods to study the precise spacing and the location of atoms in solids.  One obtains this information by measuring the angular distribution of the diffracted particles which depends on the microscopic lattice structure of the studied solid.  In conventional x-ray and electron diffraction experiments, the incident particles interact with all the electrons in the sample and, since the majority of the electrons is located close to the nucleus, these experiments yield the averaged position of the atoms. The situation is very similar for neutron diffraction, for which scattering occurs from the nuclei directly. In addition to the sensitivity to the lattice structure,  neutron  and non-resonant  x-rays also  interact with the magnetic moments and therefore information on the magnetic structures in solids can be derived.

However, in many of the currently most intensively studied systems not only the behaviour of lattice and spins needs to be studied. Also the charge and orbital degrees of freedom often play an essential role. In some materials a collective electronic ordering of spins, charges, and orbitals occurs which typically only affects a small fraction of the valence electrons. These phenomena can lead to novel exciting ground states. For instance,  complex electronic phenomena involving the cooperative ordering of various electronic degrees of freedom are discussed intensively  in relation with such outstanding phenomena like  high-temperature superconductivity  in cuprates\,\cite{Bednorz1986,Hackl2011} or the colossal magneto-resistance in manganites\,\cite{Searle1970,Tokura2006}. 

 As we will describe later on, charge and orbital orders are very difficult to observe with the aforementioned traditional scattering techniques. A new experimental probe, which enables to observe complex electronic order was therefore urgently needed.
 Resonant (elastic) soft x-ray scattering (RSXS) provides exactly this: a highly sensitive probe for spacial modulations  of spins, charges, and orbitals in complex  materials. This unique sensitivity is achieved by merging diffraction and x-ray absorption spectroscopy (XAS) into a single experiment, where the scattering part provides the information of spatial modulations and the XAS part provides the sensitivity to the electronic structure. 
More precisely, RSXS close to an absorption edge involves virtual transitions from core levels into unoccupied states close to the Fermi level and these virtual transitions depend strongly on the  spin, charge and orbital configuration of the resonant scattering centers.
The ordering of spins, charges, and orbitals typically results  in electronic superlattices with periodicities of several nanometers, which matches very well with the wavelengths of soft x-rays lying between  between $\approx$  6 to 0.6 nm corresponding to  photon energies  between $\approx$ 200 eV to 2000 eV.

Moreover, since the virtual excitations in RSXS are related to specific core level excitations, the excitation energy of which changes from element to element, the method is element specific like XAS. Thus RSXS, different from magnetic neutron scattering can probe magnetic structures related to specific elements.

The price paid for the unique sensitivity  offered by RSXS  is a very limited Ewald sphere and a rather short photon penetration depth limiting studies to only the topmost 100 atomic layers or, depending on the resonance used, even less. But these limitations are very often compensated by the gain in sensitivity, which is extremely important
when turning to magnetic and electronic properties of samples characterized by a very small amount of contributing material, i.e., for studying thin films, nanostructures as well as surfaces and interfaces. As a consequence of the broken translational symmetry at the surface, nanosystems can show ordering phenomena which differ locally or macroscopically from those of the respective bulk systems,  and even completely new phenomena can arise. For such systems, RSXS has been established as a very powerful and unique tool to study complex electronic ordering phenomena on a microscopic scale. The need for such studies has strongly increased within the last decade since the fabrication of high quality samples with macroscopic properties tunable by composition, strain, size and dimensionality has become possible, raising the hope for future multifunctional heterostructures characterized by so far unexpected and unexplored novel material properties and functionality.   

Although  RSXS offers many important ingredients to study condensed matter, it has only recently developed its full power. The reason for this is that intense  soft x-ray sources with tunable energy only became available with the advent of 2nd and 3rd generation synchrotron radiation  facilities. Furthermore those soft x-rays are absorbed in air and therefore the diffraction stations have to operate in vacuum, which poses another complication and made the development of RSXS challenging.  Finally, as mentioned above, the penetration depth into the solids is of the order of nanometers which means that surface effects may become important. Thus in many surface sensitive systems the surfaces should be prepared under ultra-high vacuum (UHV) conditions or at least should be kept under UHV during the measurements.

This review covers the recent progress  in RSXS and gives a survey of the application of the technique. After the introduction, the principles of RSXS are presented, followed by a description of the experimental development. The main part is then devoted to the application of RSXS in the study of magnetic structures, charge order, and orbital order in thin layers, artificial structures, interfaces, and correlated systems. The review concludes with a summary and an outlook. 

\section{Principles of resonant x-ray scattering}

\begin{figure}[t!]
\centering
\includegraphics[width=7.5cm]{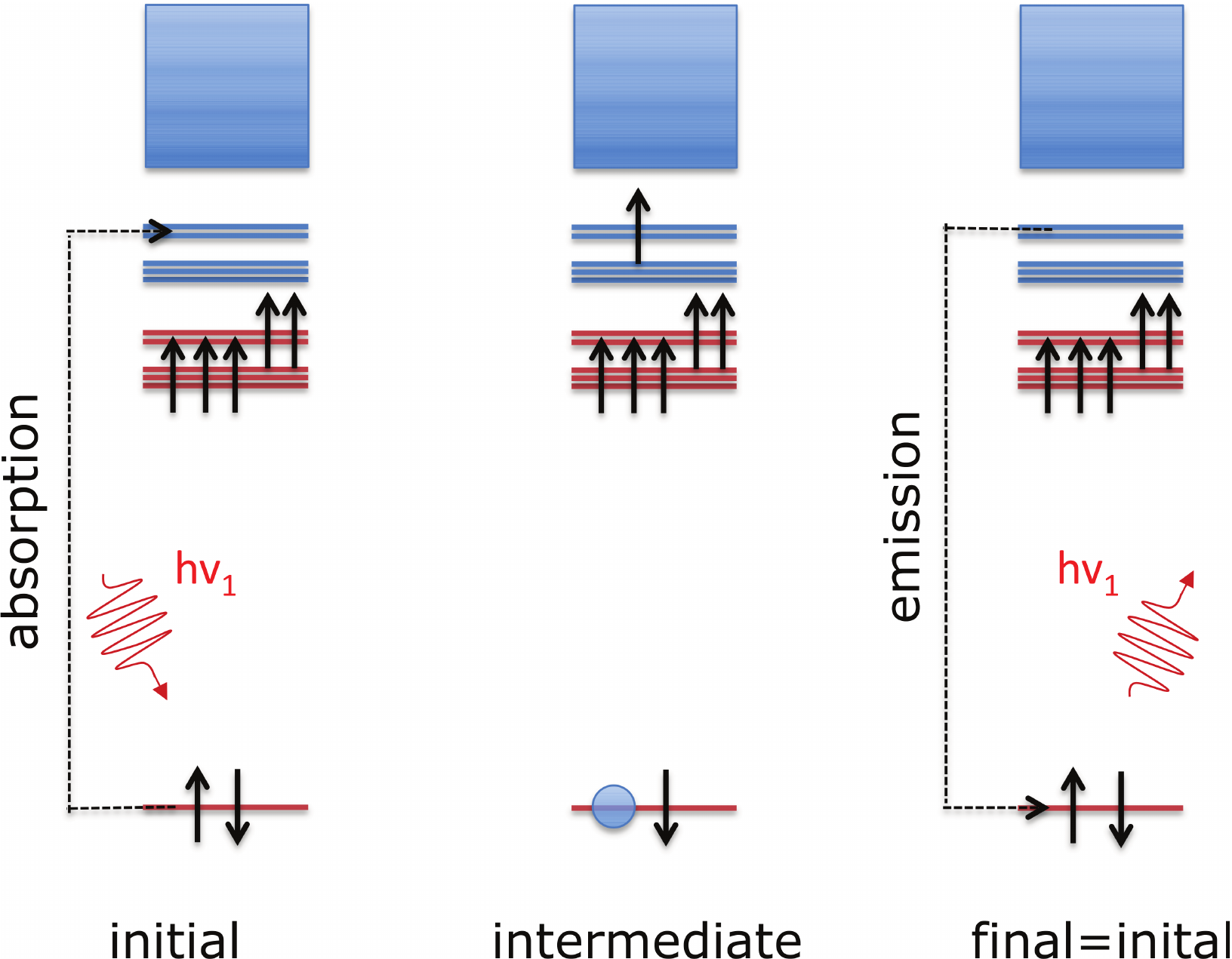}
\caption{Schematic illustration of the elastic resonant scattering process. In the first step an incoming photon is absorbed by the scatterer (left) and a core-electron is promoted into an unoccupied state close to the Fermi level. This results in the intermediate state of the scattering process (middle). The intermediate state then decays via recombination of the excited electron with the core-hole and a photon is emitted. In the resulting final state the electronic configuration of the scatterer (right) is identical to that of the initial state.}\label{fig:rsxs_scheme}
\end{figure}

Elastic resonant x-ray scattering  combines x-ray spectroscopy and x-ray diffraction in one single experiment. Roughly speaking, x-rax diffraction  provides the information about the spatial order, while the spectroscopic part provides the sensitivity to the electronic states involved in the ordering. A first qualitative understanding of this strongly enhanced sensitivity to electronic order can be gleaned from the schematic illustration of the resonant scattering process shown in figure\,\ref{fig:rsxs_scheme}: the incoming photon virtually excites a core electron into the unoccupied states close to the Fermi level, thereby creating the so-called intermediate state $|I\rangle$ of the resonant scattering process. This virtual transition depends very strongly on the properties of the valence shell and, therefore, results in the  tremendously enhanced sensitivity of resonant x-ray scattering to electronic ordering.  The state $|I\rangle$ then decays back into the ground state $|G\rangle$ and a photon with the same energy as the incoming one is re-emitted. This combination of spectroscopy and diffraction will be described in the following.

\subsection{Diffraction from a crystal}\label{sec:scattering}

In this section, \,\ aspects of the diffraction of x-rays by a crystal will be summarized. For more detailed and extensive descriptions of this topic the reader is referred to the literature\,\cite{Lovesey1996,Als-Nielson2001,James1982,Warren1990}.

\subsubsection{Diffracted intensity in the kinematic approximation}\hfill\\
\\
In the following we will consider a crystal that is formed by a perfectly periodic arrangement of lattice sites, which act as scattering centers for the incident x-ray field. The scattering from site $n$ in the crystal is described in terms of a scattering length $f_n$, which is also called the form factor, and can be represented as\,\cite{Hill1996}
\begin{equation}\label{eq:f_general}
 f_n=f_n^T+f_n^M+\Delta f_n'+i \Delta f_n''.
\end{equation}
The first two terms $f_n^T$ and $f_n^M$ represent the non-resonant charge and magnetic scattering, respectively, where the scattering described by $f_n^T$, which is proportional to the total number of electrons of the scatterer,  is called Thomson scattering. $\Delta f_n= \Delta f_n'+i \Delta f_n''$ denotes the so-called dispersion correction, which is not only a function of the photon energy $\hbar \omega$,  but also of the polarisation of the incoming (${\bf e}$) and scattered beam (${\bf e}'$). This correction becomes very important close to absorption edges and describes the resonant scattering processes illustrated in figure\,\ref{fig:rsxs_scheme}.

\begin{figure}[t!]
\centering
\includegraphics[width=7.5cm]{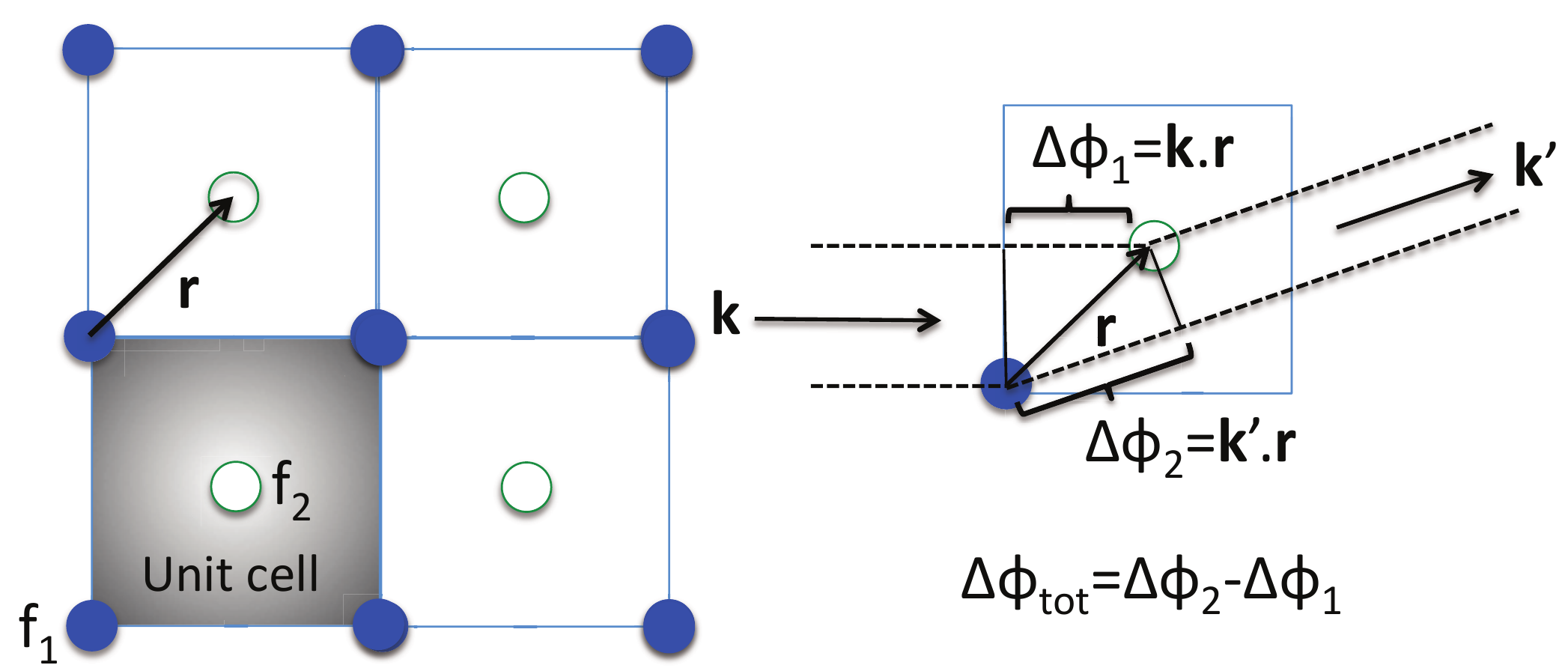}
\caption{Scattering from a crystal. The periodic structure of a crystal can be described using a unit cell and a lattice basis (left panel). The unit cell is indicated by the shaded area. In this example the lattice basis contains two sites  
with scattering length $f_1$ and $f_2$, respectively. The whole crystal can be described by discrete translations of the unit cell with its basis. The relative phase of x-rays scattered by the two sites of the basis  is $\mathbf{k}'.\mathbf{r}-\mathbf{k}.\mathbf{r}=\mathbf{Q}.\mathbf{r}$ (right panel).}\label{fig:scatter_scheme}
\end{figure}

Physically the scattering length $f_n$ describes the change in amplitude and phase suffered by the incident wave during the scattering process: a scatterer with scattering length $f_n$ exposed to an incident plane wave $\propto e^{i\mathbf{k}.\mathbf{r}}$ causes a scattered radial wave $\propto f_n e^{i k r}/r$, which, far away from the scattering center, can be approximated by a plane wave. 

In order to calculate the intensity of the total scattered wave field with wave vector $\mathbf{k}'$, all the radial waves have to be summed with the correct relative phases. As illustrated in figure\,\ref{fig:scatter_scheme}, the geometric relative phase of two scatterers is given by $\mathbf{Q}.\mathbf{r}$, where $\mathbf{Q}=\mathbf{k}'-\mathbf{k}$ is the scattering vector and $\mathbf{r}$ is the difference in position. The total intensity detected in a distant detector is therefore

\begin{eqnarray}\label{eq:kinematic}
I({\bf Q}) & \propto & | \sum_{n,m} f_n(\hbar \omega, {\bf e}, {\bf e}') e^{i{\bf Q}.({\bf R}_m+{\bf r}_n)}|^2  \nonumber\\
        & = & |\underbrace{ \sum_n f_n(\hbar \omega, {\bf e}, {\bf e}') e^{i {\bf Q}.{\bf r}_n}}_{\doteq F(\hbar \omega, {\bf e}, {\bf e}')}|^2 \times |\underbrace{\sum_m e^{i {\bf Q}.{\bf R}_m}}_{\doteq L({\bf Q})}|^2 \, ,
\end{eqnarray}
where ${\bf R}_m$ is the vector pointing to the origin of unit cell $m$ and ${\bf r}_n$ is the position of the scatterer $f_n$ measured from that origin. $F(\hbar \omega, {\bf e}, {\bf e}')$ is called the unit cell structure factor and describes the interference of the waves scattered from the different sites within a unit cell. The lattice sum $L({\bf Q})$ is due to the interference of the scattering from the different unit cells at ${\bf R}_m$. Its ${\bf Q}$-dependence therefore provides information about the number of sites scattering coherently, i.e., it is related to the correlation length of the studied order. For an infinite number of coherently scattering unit cells,  $L({\bf Q}) \rightarrow \sum_{\bf G\in \mathcal{B}^*}\delta(\bf{G}-\bf{Q})$,  where $\mathcal{B}^*$ represents the reciprocal lattice. This corresponds to the well-known Laue condition, which states that a reflection can only be observed if  $\bf{Q}$ is a reciprocal lattice vector.

The treatment described above is known as the kinematic approximation. This approximation is valid as long as the intensity of the scattered wave is much weaker than that of the incidence wave, which means that the interaction between the incoming and outgoing waves or multiple scattering events can be neglected. Since in resonant x-ray diffraction experiments one usually deals with weak superlattice reflections, this approximation is valid in most cases. If, however, the scattered wave becomes too strong, then one has to resort to the so-called dynamical theory of diffraction. This description will not be presented  here and the interested reader is referred to the literature\,\cite{Als-Nielson2001,Authier2001}.

\begin{figure}[t!]
\centering
\includegraphics[width=7.5cm]{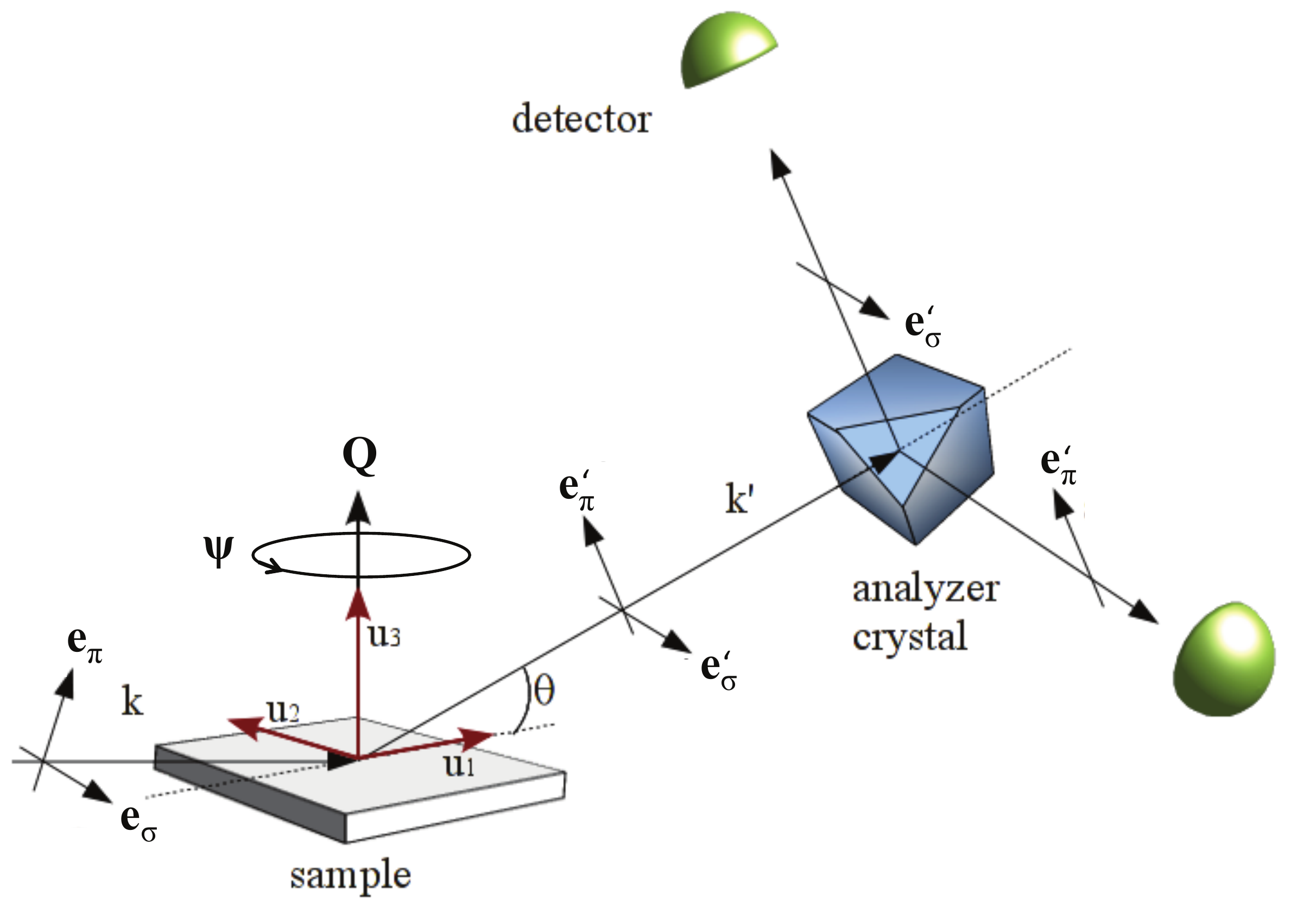}
\caption{Schematic geometry of an RSXS experiment and conventions used (courtesy of J.E. Hamann-Borrero). The scattering plane, which is defined by the incoming and outgoing wave vectors $\mathbf{k}$ and $\mathbf{k}'$, is vertical in this example. Incoming and outgoing polarisations are denoted as $\mathbf{e}_{\sigma,\pi}$ and  $\mathbf{e}_{\sigma,\pi}'$, respectively, where $\sigma$ indicates a polarisation perpendicular to the scattering plane and $\pi$ refers to a polarisation parallel to the scattering plane. The polarisation of the scattered radiation can be determined using the polarisation analyzer. $\mathbf{u}_{1,2,3}$ define the reference frame in which the polarisations, wave vectors and spin-directions will be expressed.}\label{fig:exp_scheme}
\end{figure}

\subsubsection{Scattering geometry and conventions}\label{sec:geometry}\hfill\\
\\
A typical geometry of a resonant scattering experiment is shown in figure\,\ref{fig:exp_scheme}. An incoming photon beam with defined wave vector $\mathbf{k}$ impinges on the sample and is scattered elastically into the direction defined by $\mathbf{k}'$, corresponding to a scattering vector $\mathbf{Q}=\mathbf{k}'-\mathbf{k}$. In the so-called specular geometry the incoming and the outgoing beam are at an angle $\Theta$ with the sample surface. In this case, the scattering angle between $\mathbf{k}$ and $\mathbf{k}$ is $\angle(\mathbf{k},\mathbf{k}')=2\Theta$.

The wave vectors of the incident and scattered beam define the scattering plane, which is vertical in the present example. Polarisation directions parallel or perpendicular to this scattering plane are referred to as $\pi$- and $\sigma$-polarisation, respectively. Correspondingly, the incoming and outgoing polarisations in figure\,\ref{fig:exp_scheme} are denoted as $\mathbf{e}_{\sigma,\pi}$ and  $\mathbf{e}_{\sigma,\pi}'$.

An important aspect of resonant x-ray scattering is given by the polarisation dependence of  $\Delta f_n$ at resonance. This can cause the polarisation of the outgoing beam to be different from that of the incoming  beam. A typical example, which is frequently encountered in resonant scattering experiments, is a change of the incoming polarisation from $\mathbf{e}_{\sigma}$ to an outgoing polarisation $\mathbf{e}_{\pi}'$; so-called $\sigma\pi$-scattering. Such changes in the polarisation contain important information about the symmetry of the studied order, as will be discussed in more detail later on. 

Controlling both the incoming and the outgoing polarisation can therefore be a big advantage. The incoming polarisation can be routinely  controlled by modern insertion devices like  elliptical undulators. The determination of the outgoing polarisation in RSXS  is currently less common, but can be achieved  using a polarisation analyzer as sketched in figure\,\ref{fig:exp_scheme}. By switching between the two configurations shown in the figure, for example, it is possible to observe either the $\mathbf{e}_{\pi}'$- or the $\mathbf{e}_{\sigma}'$-component of the scattered beam.

In the following discussion of the various case studies, a number of different scan-types will be described, which we will introduce briefly here:
\begin{itemize}
\item Radial scan or $\Theta2\Theta$ scan. This is a scan along the direction defined by the scattering vector $\mathbf{Q}$, which therefore provides information about the correlations along $\mathbf{Q}$. In practice this scan is done by rotating the sample and the detector in steps of $\delta\Theta$ and $2\delta\Theta$ about $\mathbf{u}_2$, respectively, which is equivalent to changing $\Theta$ of the incoming and outgoing beam by the same amount. 
\item Transverse scan. This is a scan perpendicular to the direction defined by  $\mathbf{Q}$. It hence contains information about correlation in the plane perpendicular to $\mathbf{Q}$, but is also often affected by the sample mosaic. A transverse scan can be done by rotating the sample about $\mathbf{u}_1$ or $\mathbf{u}_2$, while keeping the detector position fixed.
\item Azimuth scan. This is another way of investigating the polarisation dependence of the resonant scattering. In this type of scan, the resonant scattered intensity is recorded while rotating the  sample by an angle $\psi$ about the scattering vector $\mathbf{Q}$ (cf. figure \,\ref{fig:exp_scheme}). Azimuthal scans will be described in more detail in Sec.\,\ref{sec:pol_dep}.
\item Energy scan. Here the photon energy dependence of a given reflection at $\mathbf{Q}$ is measured. In many cases this measurement is performed by scanning the photon energy, while  recording the scattered intensity at $\mathbf{Q}=\mathbf{k}'-\mathbf{k}$. Since both $\mathbf{k}$ and $\mathbf{k}'$ change with the photon energy, this means that for each photon energy the scattering geometry has to be adjusted such as to keep $\mathbf{Q}$ constant.
\end{itemize}

\subsection{Scattering of light by an atom}\label{sec:res_scatter}

In the following sections\,\ref{sec:lightmatter}-\ref{sec:RSXSandXAS} the resonant scattering length and its relation to x-ray absorption will be described in a rather formal way. Readers who are mainly interested in the application of the method, can skip this theoretical part and continue with Sec.\,\ref{sec:Hannon}. Readers who would like to learn more about the theoretical description of the interaction between light and matter can find a very nice treatment with more details in Ref.\,\cite{Sakurai1967}.

\subsubsection{Interaction of light and matter}\label{sec:lightmatter}\hfill\\
\\
In general, the scattering of photons caused by matter is due to the interactions between the electromagnetic wave and the particles in a solid. In the present case only the interactions between the electromagnetic wave and the electrons in a solid are important. The corresponding coupling term  $\mathcal H_{int}$ is obtained by replacing the momentum operator ${\bf p}_l$ in the free electron Hamiltonian $\mathcal H_0=\sum_l p_l^2/2m$ by $({\bf p}_l-e {\bf \hat A}/c)$, which gives
\begin{equation}\label{Hint}
\mathcal H_{int}=\sum_l \left( -\frac{e}{m c} {\bf p}_l.{\bf \hat
  A}({\bf r}_l,t)+\frac{e^2}{2 m c^2} {\bf \hat A}({\bf r}_l,t).{\bf
  \hat A}({\bf r}_l,t) \right).
\end{equation}  
Here, ${\bf p}_l$ and ${\bf r}_l$ correspond  to the {\it l}\,th electron of the atom. Furthermore, ${\bf p}_l.{\bf \hat
  A}={\bf \hat A}.{\bf p}_l$ for a transversal radiation field ${\bf  \hat A}$ has been used. Note also, that the spin dependent part of $\mathcal H_{int}$, which is  proportional to \mbox{${\bf s}_l.(\nabla \times {\bf \hat A})$} and  \mbox{${\bf s}_l.({\bf \hat E} \times {\bf \hat A})$}, has been neglected, because these relativistic terms scale as $\hbar\omega/mc^2$ and hence are very small compared to the terms in Eq.\,\ref{Hint}.  
The quantized electromagnetic field that couples to the electrons can be expressed as
\begin{equation}\label{Aquant}
{\bf \hat A}({\bf r},t)=\sqrt{\frac{2\pi \hbar c}{V}} \sum_{i,{\bf k}}
   \frac{1}{\sqrt{k}} {\bf e}_i\, \left( a_{{\bf k},i}(t)\,  e^{i{\bf k}.{\bf r}} + a_{{\bf k},i}^+(t)\, e^{-i{\bf k}.{\bf r}} \right)
\end{equation}
where  $a_{{\bf k},i}(t)= a_{{\bf k},i}(0)\exp{[-i\omega_{{\bf k}}t]}$ and  $a^+_{{\bf k},i}(t)= a_{{\bf k},i}^+(0)\exp{[i\omega_{{\bf k}}t]}$. The operators $a_{{\bf k},i}^+$ and $a_{{\bf k},i}$, respectively, create and annihilate a photon with polarisation ${\bf e}_i$ and wave vector ${\bf k}$. 

Since the  first term in Eq.\,\ref{Hint} is proportional to ${\bf p}.{\bf \hat A}$ it is linear in $a_{{\bf k},i}^+$ and $a_{{\bf k},i}$ and therefore describes processes involving the emission and absorption of one photon. In other words, the ${\bf p}.{\bf \hat A}$ term changes the number of photons by $\pm1$. The second term in Eq.\,\ref{Hint} is proportional to ${\bf \hat A}.{\bf  \hat A}$ and is therefore quadratic in the $a_{{\bf k},i}^+$ and $a_{{\bf k},i}$. As a result, this term changes the number of photons by 0 or $\pm2$.

\subsubsection{Kramers-Heisenberg formula}\hfill\\
\\
As introduced above, the scattering of a photon from a specific site $n$ in the crystal is described by the scattering length $f_n$, which is determined by the interactions given by Eq.\,\ref{Hint}. The most important terms entering $f_n$ can be calculated  by means of time dependent perturbation theory up to second order in  $\mathcal H_{int}$. In the following we will briefly sketch the main steps of this calculation and will present the final result. A detailed derivation of the expression for $f_n$ can be found, for example, in Ref.\,\cite{Sakurai1967}. Since in the following we will consider the scattering from a single site, we will drop the site index $n$.

The following scattering event will be considered: a single photon with wave vector ${\bf k}$ and polarisation ${\bf e}$ impinges on
a lattice site in the initial state $|G\rangle$ and is scattered into a state with wave vector ${\bf k}'$ and polarisation ${\bf e}'$. Since we are dealing with elastic scattering, the lattice site will remain in state $|G\rangle$; i.e.
\begin{equation}\label{scatter}
|G,{\bf k},{\bf e}\rangle\doteq|G\rangle\cdot |{\bf k}, {\bf  e}\rangle  \rightarrow |G\rangle \cdot |{\bf k}',{\bf e}'\rangle  \doteq
|G,{\bf k}',{\bf e}'\rangle.
\end{equation}
Before and after the scattering event only one photon exists and, therefore, the only non-vanishing contributions of $\mathcal H_{int}$ (Eq.\,\ref{Hint}) to the scattering process must contain one creation  and one annihilation  operator corresponding to  $ |{\bf k}', {\bf  e}'\rangle$ and $ |{\bf k}, {\bf  e}\rangle$, respectively.

The terms first order in $\mathcal H_{int}$ are given by the matrix element   $M_1= \langle G,{\bf k}',{\bf e}'|\mathcal H_{int} |G,{\bf k},{\bf e}\rangle$. Since the interaction term ${\bf p}.{\bf \hat A}$ is linear in the $a_{{\bf k},i}$ and $a^+_{{\bf k},i}$, it changes the net number of photons and therefore does  not contribute to $M_1$. Only the term proportional to ${\bf \hat A}.{\bf \hat A}$  contains products of $a^+_{{\bf k}',i'}$ and $a_{{\bf k},i}$, which do not change the number of photons and, hence, only this term gives non-vanishing contributions to $M_1$. 

While the ${\bf p}.{\bf \hat A}$ term does not contribute to the first order matrix element $M_1$, there are contributions of this term in second order perturbation theory, which are of the same order of magnitude as $M_1$.  These second order contributions have to be considered as well. The second order matrix element $M_2$ involves two successive  ${\bf  p}.{\bf \hat A}$ interactions. The dominant contributions  to $M_2$ are due to terms, where the incoming photon is annihilated first and then the scattered photon created: $M_2\sim  \langle G,{\bf k}',{\bf e}'|({\bf  p}.{\bf \hat A})^+ |I\rangle \langle I| {\bf  p}.{\bf \hat A}|G,{\bf k},{\bf e}\rangle$. Here, $|I\rangle$ denotes an intermediate state of the system without a photon.  

Taken together, the  terms given by $M_1$ and $M_2$ yield the following expression for the differential cross-section $d\sigma/d\Omega$, i.e., the probability that a photon is scattered into the solid angle $d\Omega$:
\begin{center}
\begin{eqnarray}\label{KHeqn}
\frac{d\sigma}{d\Omega}&=& r_0^2 
\Bigg| {\bf e}.{\bf e}'\,
 \langle G|\rho({\bf Q})|G\rangle \nonumber \\
 &-& m \sum_{I} \frac{\langle G|{\bf e}'.{\bf J}({\bf k}')|\,I\,\rangle \langle \,I\,|{\bf e}.{\bf J}(-{\bf k})|G\rangle}{E_I-E_{G}-\hbar \omega- i \Gamma_I/2} \Bigg|^2
\end{eqnarray}
\end{center}
In this equation, $r_0=e^2/(m\,c^2)\simeq 2.82\times10^{-13}\,\rm cm$ is the classical electron radius, $\rho(\bf Q)$ corresponds to the Fourier amplitude at ${\bf Q}={\bf k}'-{\bf k}$ of the charge density, $|G \rangle$ ($|I \rangle$) is the ground state (intermediate state) of the scatterer (cf. figure\,\ref{fig:rsxs_scheme}), $E_G$ ($E_I$) is the energy of  $|G \rangle$ ($|I \rangle$), $\Gamma_I$ is the life time of $|I\rangle$ and ${\bf J}({\bf k})=1/m\sum_l \mathbf{p}_l\,e^{i \mathbf{k}.\mathbf{r}_l}$ is the current operator that describes the virtual transitions between  $|G \rangle$  and $|I \rangle$.

Equation\,\ref{KHeqn} is the famous Kramers-Heisenberg formula applied to elastic scattering of photons. By definition the differential cross-section is related to the scattering length by  $d\sigma/d\Omega=|f|^2$, i.e., the above equation also provides the expression for $f_n$. The first term proportional to $ \langle G|\rho({\bf Q})|G\rangle$ describes the non-resonant Thomson scattering of photons from the total charge density $\rho$.  The scattering due to the Thomson term is given by the first order matrix element $M_1$, which describes the direct scattering of a photon caused by the ${\bf \hat A}.{\bf \hat A}$ interaction. This non-resonant scattering scales with the total number of electrons and, hence, usually  does not provide high sensitivity to electronic ordering phenomena, which typically affect only a very small fraction of $\rho$. 

The second order matrix element $M_2$ results in the second term in Eq.\,\ref{KHeqn}. This term describes the resonant scattering processes of the kind illustrated in  figure\,\ref{fig:rsxs_scheme}. The matrix elements in the nominators describe the virtual \gstate $\rightarrow$ \istate $\rightarrow$ \gstate \/ processes, which correspond to the two successive ${\bf  p}.{\bf \hat A}$ interactions. Close to an absorption edge, the photon energy $\hbar\omega$ is close to some of the $E_I-E_G$, in which case the corresponding contribution to the sum becomes large and can completely dominate the scattering.

According to the above discussion, the scattering length $f$ can be written as
\begin{equation}\label{eq:scattKH}
f(\omega,{\bf e}, {\bf e'})=f^T({\bf Q},{\bf e}, {\bf e}')+\Delta f(\hbar \omega,{\bf e}, {\bf e}',{\bf k},{\bf k}') ,
\end{equation}
where $f_T$ represents the non-resonant Thomson scattering and $\Delta f$ is the anomalous dispersion correction. Equation\,\ref{eq:scattKH} corresponds exactly to Eq.\,\ref{eq:f_general} given above with the magnetic scattering neglected.

\subsubsection{Dipole approximation}\label{sec:dipoleapprox}\hfill\\
\\
In many cases, the current operator ${\bf J}$ is treated in the dipole approximation, which means that for $\langle\bf{k}.\bf{r}_l\rangle\simeq 0$ one can use $e^{i {\bf k}.{\bf r}_l}\simeq1$ and therefore ${\bf J}({\bf k})\rightarrow  1/m \sum_l {\bf p}_l$. In the dipole approximation the dependence of $\Delta f$ on ${\bf k}$ and ${\bf k}'$ is therefore neglected. Furthermore, since $2m [\mathcal{H}_0,{\bf r}]=-2i\hbar {\bf p}$, one can replace the momentum operators ${\bf p}_l$ appearing in  ${\bf J}$  by the commutator  $[\mathcal{H}_0,{\bf r}_l]$. This together with the resonant term in Eq.\,\ref{KHeqn} then yields the following expression for the resonant scattering length:
\begin{center}
\begin{eqnarray}\label{eqn:fdipole}
\Delta f & =&
 k^2
\sum_{I}   \frac{\langle G|({\bf e}'.{\bf D})^+|\,I\,\rangle \langle \,I\,|{\bf e}.{\bf D}|G\rangle}{E_I-E_{G}-\hbar \omega- i 
\Gamma_I/2}
\end{eqnarray}
\end{center}
Here, ${\bf D}=e \sum_l {\bf r}_l$ is the dipole operator. The polarisation dependence in the above equation can also be expressed using the tensor $\Delta \hat f$, which is defined by $\mathbf{e}'^+.\Delta \hat f(\hbar\omega).\mathbf{e}=\Delta f(\hbar\omega,\mathbf{e},\mathbf{e}')$. This tensor formalism is often found in the literature and will also be used in the following.

\subsubsection{Resonant scattering length, x-ray absorption and index of refraction}\label{sec:RSXSandXAS}\hfill\\
\\
As can be already realized in figure\,\ref{fig:rsxs_scheme}, the excitation from the ground state into the intermediate state corresponds to an x-ray absorption process. Therefore, there should be a relation between the resonant scattering length $\Delta f$ and the x-ray absorption cross section. That such a direct relation indeed exists can be seen in the following way: firstly, $\Delta f$ can also be written more compactly as
\begin{equation}\label{eqn:dfgreen}
\Delta f =r_0\,m\,\langle G|\,[{\bf e}'.{\bf J}({\bf k}')]^+ \, \mathcal{G}(\hbar \omega)\, [{\bf e}.{\bf J}(-{\bf k})]\,|G\rangle
\end{equation}
using the Greens function
\begin{equation}\label{eqn:green}
\mathcal{G}(\hbar \omega)=\left(\mathcal{H}_I-E_G-\hbar \omega-i\Gamma/2\right)^{-1}.
\end{equation}
Note that $\mathcal{G}$ is an operator, which is related to the hamiltonian $\mathcal{H}_I$ describing the intermediate state, i.e., the state with the core-hole.

Secondly, the transition probability per unit time $w_{abs}$ for the x-ray absorption is given by Fermi's golden rule:
\begin{equation}\label{eqn:fermi}
w_{abs}=\frac{2 \pi}{\hbar} \, \sum_F \left|\, \langle F | \mathcal{H}_{int} |G \rangle\,\right|^2 \, \delta (E_F-E_G-\hbar \omega)
\end{equation} 
As explained in Sec.\,\ref{sec:lightmatter}, only the  ${\bf p}.{\bf \hat A}$ is linear in the photon annihilators and, hence, it  is the only term active in the x-ray absorption process. The effective vector field  ${\bf A}^{abs}$, which couples $|G\rangle$ to $|F\rangle$ can be expressed as\,\cite{Sakurai1967}.
\begin{equation}\label{Aabsorp}
{\bf A}^{abs}=c \,  \sqrt{\frac{2 \pi n_{{\bf k},i}\, \hbar}{\omega_{{\bf k}}\,
    V}} {\bf e} \ e^{i({\bf k.r}-\omega_{{\bf k}}\,t)}.
\end{equation}
The action of the operator $a_{{\bf k},i}$ on the multiphoton state yields the factor $\sqrt{n_{\mathbf{k},i}}$, where $n_{\mathbf{k},i}$ is the number of photons with wave vector $\mathbf{k}$ and polarisation $\mathbf{e}$. Furthermore, $w_{abs}$ together with the photon flux $\Phi=c \, n_{{\bf k},i}/V$ gives the the absorption cross-section $d\sigma_{abs}/d\Omega=w_{abs}/\Phi$. This relation together with Eqs.\,\ref{eqn:fermi} and \ref{Aabsorp} gives the following result:
 \begin{eqnarray}\label{eqn:abs}
 \frac{d\sigma_{abs}}{d\Omega}&=&\frac{4 \pi^2}{k}\, r_0\,m \sum_F|\, \langle F|\, {\bf e}.{\bf J}({\bf k})\,|G\rangle|^2\,\, \delta (E_F-E_G-\hbar \omega)\nonumber\\
 &=& \frac{4 \pi}{k}\, r_0\,m \,\mathrm{Im}\left[\, \langle G|\,[{\bf e}.{\bf J}({\bf k})]^+ \, \mathcal{G}(\hbar \omega)\, [{\bf e}.{\bf J}(-{\bf k})]\,|G\rangle\,\right]
\end{eqnarray}
Since the Greens function entering $d\sigma_{abs}/d\Omega$ is also given by Eq.\,\ref{eqn:green}, the comparison of Eqs.\,\ref{eqn:dfgreen} and \ref{eqn:abs} shows that
\begin{center}
\begin{equation}\label{eqn:opticaltheorem}
\frac{d\sigma_{abs}}{d\Omega}= \frac{4 \pi}{k}\,\mathrm{Im}\left[\Delta f\right]
\end{equation}
\end{center}
for ${\bf k}={\bf k}'$ and ${\bf e}={\bf e}'$. This important relation is known as the optical theorem. It enables one  to determine the imaginary part of the resonant scattering length from x-ray absorption. Once $\mathrm{Im}\left[\Delta f\right]$ is known, the corresponding real part can be calculated by means of the Kramers-Kronig relations, which read
\begin{eqnarray}\label{eqn:KKR}
{\rm Re}[\Delta f(\epsilon)]&=&\frac{2}{\pi} \mathcal P \int_0^{\infty}
\frac{\epsilon'{\rm Im}[\Delta f(\epsilon')]}{\epsilon'^2-\epsilon^2} d\epsilon'\quad {\rm and}\nonumber\\
{\rm Im}[\Delta f(\epsilon)]&=& -\frac{2\epsilon}{\pi} \mathcal P \int_0^{\infty}
\frac{{\rm Re}[\Delta f(\epsilon')]}{\epsilon'^2-\epsilon^2} d\epsilon'\, ,
\end{eqnarray}
with $\mathcal P \int_0^{\infty}$ the principal part $\lim_{\delta\rightarrow 0} \int_0^{\omega-\delta}+ \int_{\omega+\delta}^{\infty}$ of the integral and $\epsilon=\hbar \omega$ the photon energy.

\begin{figure}[t!]
\centering
\includegraphics[width=8.5cm]{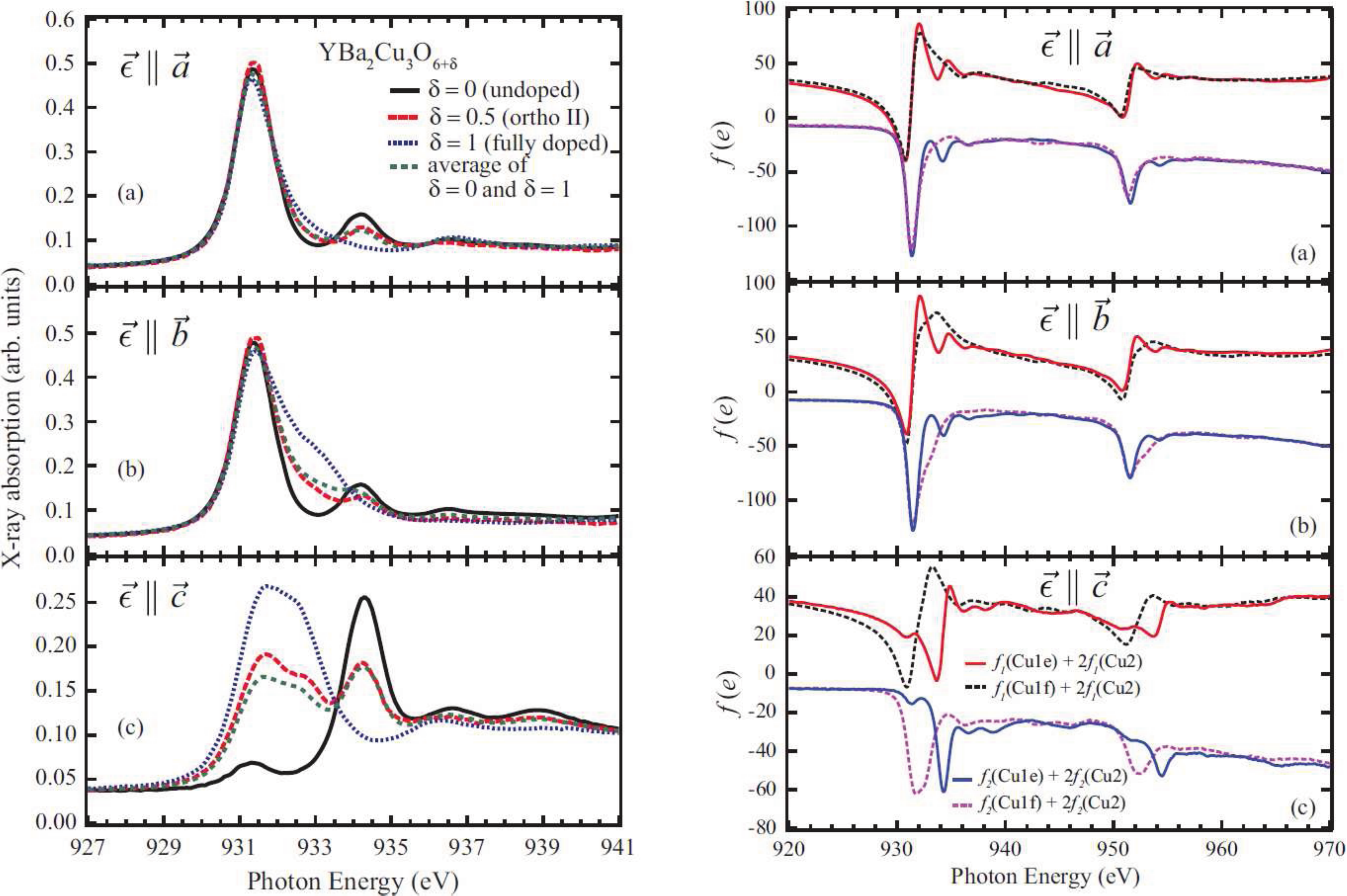}
\caption{Relation of x-ray absorption spectroscopy (XAS) and RSXS\,\cite{Hawthorn2011}. The left panel shows the XAS measured at the Cu $L_3$ edge for  YBa$_2$Cu$_3$O$_{6+\delta}$. The XAS spectra exhibit a characteristic polarisation dependence, which is caused by the local symmetries of the different Cu sites in the lattice. The right panel shows the resonant structure factors as a function of energy and polarisation, which were deduced from a Kramers-Kronig analysis. It can be seen that the energy dependent lineshape of $\Delta f$ depends strongly on the polarisation. (Reprinted from \cite{Hawthorn2011}. Copyright © 2011, American Physical Society.)}\label{fig:KKanalysis}
\end{figure}

Using these relations the energy dependence of $\Delta f(\hbar\omega,\mathbf{e},\mathbf{e})=\mathbf{e}^+.\Delta \hat f(\hbar\omega).\mathbf{e}$ can be calculated, where $\mathbf{e}$ is the polarisation of the absorbed photon. This is demonstrated in figure\,\ref{fig:KKanalysis}, were such an analysis is shown for the high-temperature cuprate superconductor  YBa$_2$Cu$_3$O$_{6+\delta}$. It is important to realize however, that the above analysis allows one to obtain only matrix elements of the form ${\mathbf e}^+.\Delta \hat f. {\mathbf e}$. The matrix elements ${\mathbf e}'^+.\Delta \hat f. {\mathbf e}$ with ${\mathbf e}\neq{\mathbf e}'$ can therefore not directly be determined by x-ray absorption. In principle it is possible, however, to measure the x-ray absorption with 3 different polarisations with respect to the crystal axes and 6 other linear combinations of those. This enables to retrieve all 9 matrix elements of $\Delta \hat f$, by solving a set of linear equations.

The scattering length can also be related to the index of refraction $n$. We will only briefly discuss the origin of this relation. Readers interested in the detailed derivation are referred to the literature\,\cite{Als-Nielson2001,Attwood2007}. The relation between $n$ and $f$ is essentially given by the fact that the transmission of a beam through a plate of some material can be described in two equivalent ways: (i) behind the plate there will be an additional phase shift that was suffered by the wave during the passage through the material of the plate. At normal incidence, this phase shift is simply given by $(1-n)\,k\,W$, where $k$ and $W$ are the the wave vector of the light and the thickness of the plate, respectively. (ii) The plate can also be considered as a set of scatterers, which cause an additional scattered wave behind the plate. The total wave field behind the plate is therefore a superposition of the incident wave and the scattered wave, where the latter is related to the scattering length $f$ of the scatterers within the plate. Only identical scattering centers are assumed. Since both descriptions must give the same result for the wave field behind the plate one can deduce that
\begin{equation}\label{eq:nfrelation}
n=1-\frac{2\pi\,\rho_a\,r_0}{k^2}\times(f^T+\Delta f),
\end{equation}
where $\rho_a$ is the number density of the material. Since $\Delta f$ is a complex number, the above Eq.\,\ref{eq:nfrelation} relates the real and imaginary part of $f$ to the dispersion terms $\delta$ and $\beta$ of  $n=1-\delta+i\beta$.

\subsubsection{Free magnetic atom}\label{sec:Hannon}\hfill\\
\\
The general expressions derived in the previous sections are useful to understand the general process of resonant diffraction. However, for the analysis of experimental data it is necessary to evaluate $\Delta f$ for a specific situation. The simplest case for which explicit expressions can be derived, is the one of a free atom with a magnetic moment. In this situation only the magnetic moment breaks the spherical symmetry of the free atom. For this special case the resonant scattering length in the dipole approximation can be expressed as\,\cite{Hannon1988,Hill1996}
 \begin{equation}\label{eqn:Hannon}
\Delta f = R^{(0)}\,{\bf e}'^*.{\bf e}-i \,R^{(1)}\,({\bf e}'^*\times{\bf e}).{\bf \hat s}+R^{(2)}\,({\bf e}'^*.{\bf \hat s})\,({\bf e}.{\bf \hat s}),
\end{equation}

where  ${\bf \hat s}$ is the unit vector defining the local spin moment and the $ R^{(j)}= R^{j)}(\hbar \omega)$ ($j=0,1,2$) are photon energy dependent resonance factors. These factors depend on the products $\langle G|({\bf e}'.{\bf D})^+|\,I\,\rangle \langle \,I\,|{\bf e}.{\bf D}|G\rangle$ appearing in Eq.\,\ref{eqn:fdipole} and define the strength of the resonant scattering.

There are 3 different terms in Eq.\,\ref{eqn:Hannon}: the first term does not depend on the spin direction and corresponds to non-magnetic resonant scattering.  This term simply adds to the Thomson scattering in Eqs.\,\ref{eq:f_general} and \ref{KHeqn}, which has the same polarisation dependence. 
The second term is proportional to $({\bf e}^*\times{\bf e}).{\bf \hat s}$ and  is linear in ${\bf \hat s}$ and therefore describes magnetic resonant scattering. By virtue of the optical theorem this part of $\Delta f$ can be related to the x-ray magnetic circular dichroism in absorption\,\cite{Lovesey1996}. 
The third term is proportional to $({\bf e}'^*.{\bf \hat s})\,({\bf e}.{\bf \hat s})$, i.e., this term is quadratic in ${\bf \hat s}$ and hence also corresponds to resonant magnetic x-ray scattering. This scattering is related to the linear dichroism in x-ray absorption\,\cite{Lovesey1996}.

Equation\,\ref{eqn:Hannon} is very often used for the analysis of resonant scattering from magnetically ordered materials\,\cite{Lovesey1996,Lovesey1987}. However, it is important to remember that this expression has been derived for a free atom, where only the magnetic moment breaks the otherwise spherical symmetry. 

\subsubsection{Magnetic atoms in a crystal}\label{sec:Haverkort}\hfill\\
\\
Whenever an atom is embedded in a crystal, the chemical environment will  break the spherical symmetry that  was used to derive Eq.\,\ref{eqn:Hannon}.  This equation  has therefore to be considered as an approximation for describing magnetic scattering in a crystal lattice.
For instance, the x-ray absorption of a non-magnetic and non-cubic material will change with the polarisation ${\bf e}$ pointing along the inequivalent lattice directions. The optical theorem therefore immediately implies that $\Delta f(\hbar\omega, \mathbf{e},\mathbf{e})$ also varies with $\mathbf{e}$. This polarisation dependence is not captured by Eq.\,\ref{eqn:Hannon}, because in the non-magnetic case, this formula implies $\Delta f(\hbar\omega, \mathbf{e},\mathbf{e})=R^{(0)}$, which is isotropic and the same for all ${\mathbf e}$ .

In the following we will provide a simplified description of the effects described  by Carra and Thole\,\cite{Carra1994}, and by Haverkort\,\cite{Haverkort2010}. In particular, we will restrict ourselves to the dipole approximation and to magnetic scattering terms that are linear in the spin direction  $\mathbf{ \hat s}=(x,y,z)$. This is sufficient to illustrate the effects of the local environment, but in general the higher order terms can be relevant as well. For the complete expressions including higher order terms in $\mathbf{ \hat s}$ the reader is referred to Ref.\,\cite{Haverkort2010}.

In the most general case of a magnetic site in  a low symmetry environment, the tensor  $\Delta \hat f(\hbar \omega)=\left\{\Delta f(\hbar \omega,\mathbf{e},\mathbf{e}')\right\}_{i,j}$ (cf. \ref{sec:dipoleapprox}), describing the polarisation and energy dependence of the resonant scattering has the form:
\begin{displaymath}
\Delta \hat f =
\left(
\begin{array}{ccc}
  R_{xx}   & R_{xy} & R_{xz}  \\
  R_{yx}   & R_{yy} & R_{yz} \\
  R_{zx}   & R_{zy} & R_{zz}   
\end{array}
\right)
\end{displaymath}
with complex energy dependent matrix elements $R_{ij}=R_{ij}(\hbar \omega)$. 
However, depending on the local point group symmetry of the scatterer,  this tensor can be simplified and especially in high symmetries the simplifications are significant: in the extreme case of a free non-magnetic atom with spherical symmetry one has $\Delta \hat f=R^{(0)}\times I$ with the identity matrix $I$. If the spherical symmetry is broken only by the spin moment of the scatterer, the first two terms of Eq.\,\ref{eqn:Hannon} translate to
\begin{equation}\label{eqn:fsph}
\Delta \hat f =
 R^{(0)}\,I+ R^{(1)}\,\left(
\begin{array}{ccc}
0 & - z & y\\
 z & 0 & -x \\
-y &x & 0
\end{array}
\right)
\end{equation}

It is this expression that is very often used for the analysis of resonant x-ray scattering data. It is important to know in which situations this approximate expression can be used and in which situation it yields wrong results. We therefore discuss two other cases to clarify this point. First, we consider a local environment of $O_h$-symmetry. In this situation  $\Delta \hat f$ can be expressed as
\begin{equation}\label{eqn:fOh}
\Delta \hat f =
R^{(0)}_{a1g}\,I+R^{(1)}_{t1u}\,
\left(
\begin{array}{ccc}
0 & - z & y\\
 z & 0 & -x \\
-y &x & 0
\end{array}
\right)
\end{equation}

Comparing Eqs.\,\ref{eqn:fsph} and \ref{eqn:fOh} it can be seen, that the spherical approximation may still be used, as long as the magnetic term linear in  $\mathbf{ \hat s}$ dominates the resonant scattering signal. We mention that already for $O_h$-symmetry the terms quadratic in $\mathbf{ \hat s}$ differ from Eq.\,\ref{eqn:Hannon}. As a second example we consider a local $D_{4h}$ symmetry. Depending on the deviation from $O_h$, the resonant scattering length can now  deviate significantly from the spherical approximation:

\begin{eqnarray}\label{eqn:fD4h}
\Delta \hat f &=&
\left(
\begin{array}{ccc}
R^{(0)}_{a1g, B} & 0 & 0\\
 0 & R^{(0)}_{a1g,B} & 0 \\
0 &0 & R^{(0)}_{a1g,A}
\end{array}
\right) \nonumber \\
 &+&
\left(
\begin{array}{ccc}
0 & - z R^{(1)}_{a2u} & y R^{(1)}_{eu}\\
 z R^{(1)}_{a2u}& 0 & -x R^{(1)}_{eu}\\
-y R^{(1)}_{eu}&x R^{(1)}_{eu}& 0
\end{array}
\right),
\end{eqnarray}

As can be seen in the above equation, the energy dependence of the magnetic resonant scattering now depends on the spin direction, i.e., depending on the direction of  $\mathbf{ \hat s}$,  $R^{(1)}_{a2u}$ and $R^{(1)}_{eu}$ contribute with different weights. This effect is therefore given by the difference between $R^{(1)}_{a2u}$ and  $R^{(1)}_{eu}$.  It is also important to note that even the non-magnetic scattering in Eq.\,\ref{eqn:fD4h} is no longer isotropic. This means that for local symmetries lower than cubic the non-magnetic resonant scattering also  becomes polarisation dependent, in agreement with the qualitative arguments given in the beginning of this section.

\subsection{Polarisation effects}\label{sec:pol_dep}

We already mentioned that the polarisation dependence of resonant x-ray scattering provides important information about the symmetry of the studied ordering. In this section we will provide a simple  example for this, namely the resonant magnetic scattering of a bcc antiferromagnet. In this case the polarisation dependence of the resonant scattering allows one to determine  the directions of the ordered moment.

\subsubsection{General description of polarisation effects}\label{sec:general_pol}\hfill\\
\\
Before we discuss our example, it is useful to introduce an efficient way to describe polarisation dependent resonant scattering.
The resonant scattering length as a function of the incoming and outgoing polarisation can be expressed, using a $2\times2$ scattering matrix $\mathcal{S}$\,\cite{Bergevin1981}:

\begin{equation}
\mathcal{S}_n=\left(
\begin{array}{cc}
\epsilon_{\sigma}^t.(\Delta \hat f_n).\epsilon_{\sigma}' & \epsilon_{\sigma}^t.(\Delta \hat f_n).\epsilon_{\pi}'\\
\epsilon_{\pi}^t.(\Delta \hat f_n).\epsilon_{\sigma}' & \epsilon_{\pi}^t.(\Delta \hat f_n).\epsilon_{\pi}'
\end{array}
\right)
\end{equation}

The matrix elements of $\mathcal{S}_n$ therefore correspond to $\sigma\sigma$-, $\sigma\pi$-, $\pi\sigma$- and $\pi\pi$-scattering of lattice site $n$. Using these  $\mathcal{S}_n$,  the corresponding structure factor matrix $\mathcal{F}$ can be calculated according to Eq.\,\ref{eq:kinematic}: $\mathcal{F}(\mathbf{Q})=\sum_n \mathcal{S}_n e^{i \mathbf{Q}.\mathbf{r}_n}$.

For the special case described by Eq.\,\ref{eqn:Hannon}, the scattering matrix as a function of the local spin-direction reads\,\cite{Hill1996}:

\begin{eqnarray}\label{eqn:SOh}
\mathcal{S}&=& R^{(0)}
\left( 
\begin{array}{cc}
1&0\\
0&\cos(2\Theta)
\end{array}
\right)\\
&- &
i R^{(1)}
\left( 
\begin{array}{cc}
0 &x \cos(\Theta) + z \sin(\Theta) \\
 z \sin(\Theta)-x \cos(\Theta) &-y \sin(2\Theta)
\end{array}
\right) \nonumber \\
&+&
R^{(2)}
\left( 
\begin{array}{cc}
y^2 & -y(x \sin(\Theta) - z \cos(\Theta)) \\
 y( x \sin(\Theta)+z \cos(\Theta)) &-\cos^2(\Theta)(x^2 \tan^2(\Theta)+z^2)
\end{array}
\right),\nonumber 
\end{eqnarray}

where the spin-direction $\mathbf{\hat s}=(x,y,z)$ is expressed with respect to the $\mathbf{u}_{1,2,3}$ and $\Theta$ denotes the scattering angle (cf. figure\,\ref{fig:exp_scheme}).

\begin{figure}[t!]
\centering
\includegraphics[width=8.5cm]{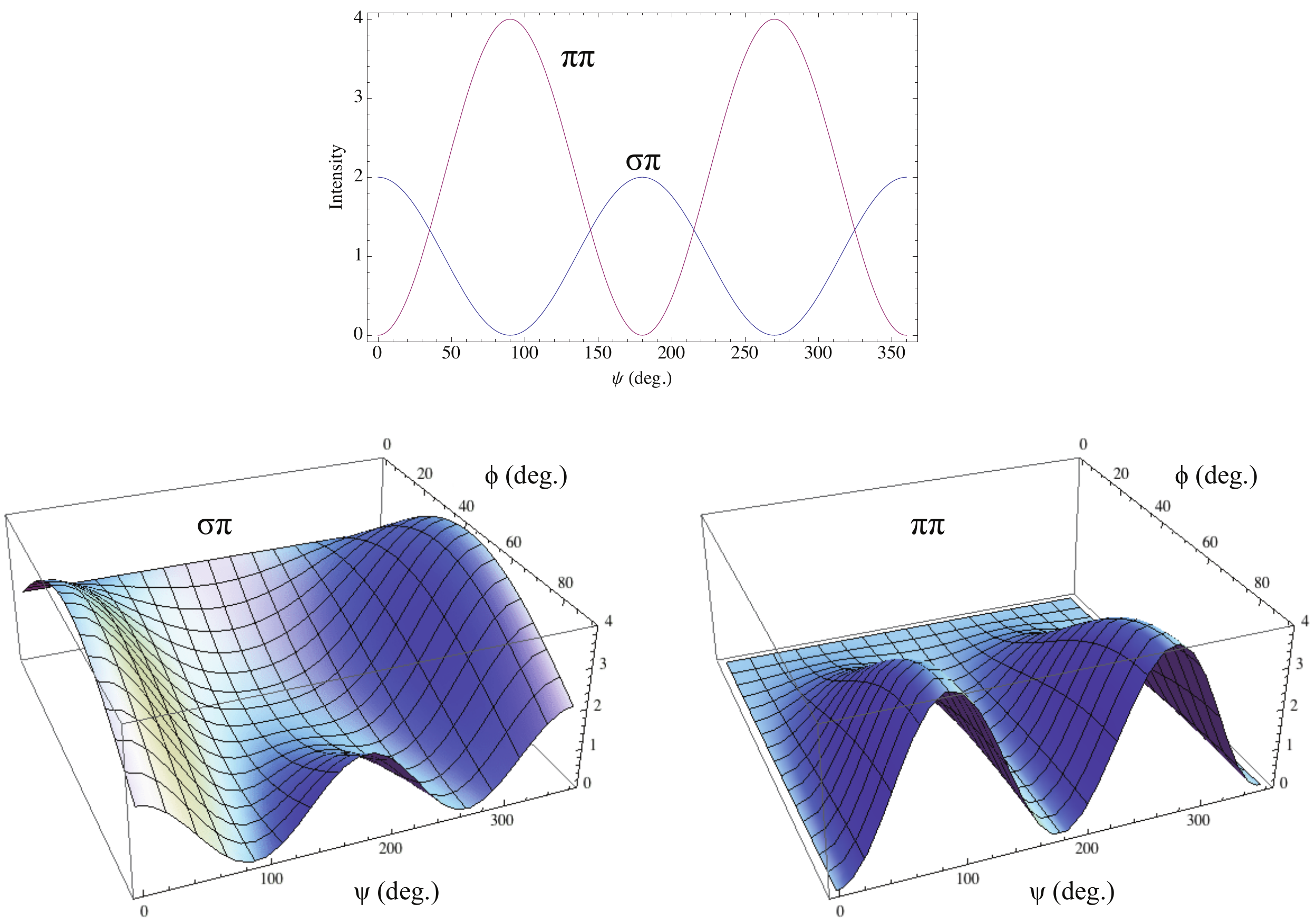}
\caption{Polarisation and azimuthal dependence of resonant magnetic scattering as obtained for the bcc antiferromagnet described in the text. The top panel shows the azimuthal dependence of the $\sigma\pi$- and $\pi\pi$-scattering, where the spin directions are $\pm(1,0,0)$ and $\psi$ is the rotation angle about the (00L)-scattering vector. The lower two panels display the azimuthal dependences of the $\sigma\pi$- (left) and $\pi\pi$-scattering (right) for the spins pointing along $(\sin{\Phi},0,\cos{\Phi)}$. In this example, the polarisation dependence of the resonant scattering enables  the spin-directions to be determined. $\Theta=45^{\circ}$ in all cases.}\label{fig:azi}.
\end{figure}

\subsubsection{Example of a bcc antiferromagnet}\hfill\\
\\
In order to illustrate the polarisation dependence of resonant magnetic x-ray scattering, we consider the example of a bcc antiferromagnet. The unit cell contains two inequivalent lattice sites: one at $\mathbf{r}_1=(0,0,0)$ with $\mathbf{\hat s}_1=(1,0,0)$ and one at $\mathbf{r}_2=(0.5,0.5,0.5)$ with $\mathbf{\hat s}_2=-(1,0,0)$. For the sake of simplicity we assume that the two lattice sites differ only in the local spin moment and are identical otherwise. Note that the above coordinates refer to the so-called crystal frame, which is defined by the crystal axes.

We will consider the (001)-reflection for which $\mathcal{F}(001)=\mathcal{S}_1-\mathcal{S}_2$. Since the system is cubic and only the terms linear in $\mathbf{\hat s}$  contribute to resonant $\mathcal{F}(001)$, we can use the expression given in Eq.\,\ref{eqn:fOh} which is proportional to $({\bf e}'\times{\bf e}).{\bf \hat s}$, i.e., the  $\mathcal{S}_{1,2}$ are simply given by the term proportional to $R^{(1)}$ in Eq.\,\ref{eqn:SOh}. This then yields:

\begin{equation}
\mathcal{F}(001)=-2 R^{(1)} i \left(
\begin{array}{cc}
 0 & \cos{\psi}\cos{\Theta}\\
-\cos{\psi}\cos{\Theta} & \sin{\psi}\sin{2\Theta}
\end{array}
\right)
\end{equation}

In this equation $\psi$ is the azimuthal angle and describes a rotation of the sample about $\mathbf{u}_{3}$ as indicated in figure\,\ref{fig:exp_scheme}.  When the sample is rotated about $\mathbf{u}_{3}$, the crystal frame and with it the spin directions rotate within the $\mathbf{u}_{1,2}$-plane. In the current setting the spin directions are along  $\mathbf{u}_2$ for  $\psi=0$.
The square modulus of the matrix elements of $\mathcal{F}$ is proportional to the intensities in the different scattering channels. For instance, the intensities in the $\sigma\pi$- and the $\pi\pi$-channels are $I_{\sigma\pi}\propto ( \cos{\psi}\cos{\Theta})^2$ and $I_{\pi\pi}\propto(\sin{\psi}\sin{2\Theta})^2$, respectively. The calculated azimuth-dependences for $I_{\sigma\pi}$ and $I_{\pi\pi}$ are shown in the top panel figure\,\ref{fig:azi}. The azimuthal dependence of $I_{\sigma\pi}$ is due to the fact that $({\bf e}_{\sigma}\times{\bf e}_\pi').{\bf \hat s}\propto \mathbf{k}'.{\bf \hat s}$, i.e., this intensity is maximal if the spin directions are parallel to the scattering plane. Similarly, it is easy to see that $({\bf e}_{\pi}\times{\bf e}_\pi').{\bf \hat s}\propto (\mathbf{k}\times\mathbf{k}').{\bf \hat s}$, which means that $I_{\pi\pi}$ maximal for spins aligned perpendicular to the scattering plane. Figure\,\ref{fig:azi} also shows the azimuthal dependences of $I_{\sigma\pi}$ and $I_{\pi\pi}$ for spins along $\pm(\sin{\Phi},0,\cos{\Phi)}$. It can be seen that the azimuthal dependence changes drastically with the inclination angle $\Phi$ between $\mathbf{u}_3$ and ${\bf \hat s}$. 

In this simple example of a cubic antiferromagnetic system, the polarisation dependence of the resonant scattering provides a powerful means to identify magnetic scattering and to determine spin-directions: (i) in a cubic system, the observation of $\sigma\pi$-scattering directly implies magnetic scattering. (ii) Analyzing the azimuthal dependences enables to determine the direction of the spins.

However, great care must be taken when dealing with non-cubic systems. In this case, $\sigma\pi$-scattering no longer unambiguously identifies magnetic scattering. A very prominent example is the orbitally ordered and magnetically disordered phase of LaMnO$_3$, where the $\sigma\pi$-scattering is caused by orbital ordering and not by spin order\,\cite{Murakami1998}. The $\sigma\pi$-scattering in this case is related to the polarisation dependence of the first term in Eq.\,\ref{eqn:fD4h}. In materials with low local symmetries, the analysis of the azimuthal dependences is also more involved and the cubic approximation might lead to wrong results\,\cite{Haverkort2010}. 

\subsection{Spectroscopic effects}

One of the most outstanding properties of resonant x-ray diffraction is its greatly enhanced sensitivity to electronic ordering phenomena. This sensitivity is  a spectroscopic effect, which is related to the resonance factors $R^{(i)}(\hbar\omega)$ of the resonant scattering length. It should be noted, however, that resonant scattering involves transitions into unoccupied states, i.e., these experiments probe modulations of the unoccupied states close to the Fermi level. Besides the high sensitivity, the inherent element selectivity is a second advantage of resonant x-ray scattering: by tuning the photon energy to a specific absorption edge, the scattering is usually dominated solely by the scatterers at resonance. 

In the following we very briefly summarize some of the main spectroscopic effects at the O $K$, the TM $L_{2,3}$ and the RE $M_{4,5}$ edges, which will be relevant for the results presented in section\,\ref{sec:CaseStudies}. A much more complete account of spectroscopic effects in core-level spectroscopies, including the core-levels described below, can be found in \cite{DeGroot2008}. 

\subsubsection{Oxygen K edge}\label{sec:OKedge}\hfill\\
\\
In the TM oxides the O $2p$ states play an important role in the chemical bonding. It is therefore of great interest to investigate their role for electronic ordering phenomena. By performing a scattering experiment at the O $K$ edge, the scattering process involves $1s\rightarrow2p$ transitions, i.e., virtual excitations directly into the $2p$ valence shell. In this way, the resonant scattering at the O $K$ edge becomes very sensitive to spatial  modulations of the O $2p$ states. There is no spin-orbit interaction of the $1s$ core-hole and  the spin-orbit interaction is usually weak in the $2p$ final state also. Resonant scattering at the O $K$ edge is therefore typically rather insensitive towards the ordering of spin moments. However, the hybridisation of the O $2p$ levels with the TM $3d$  states can induce magnetic sensitivity also here as discussed in section\,\ref{sec:Multiferroics}. In addition to this, and in contrast to the TM $L_{2,3}$ and RE $M_{4,5}$ edges described below, the $1s$ core hole does not produce a strong multiplet splitting in the intermediate state. 

Up to date a number of experiments on different materials, including the cuprates and the manganites, have demonstrated a strong resonant enhancement of superlattice reflections at the O $K$ edge, revealing the active role of the oxygen $2p$ states in the electronic order (cf. section\,\ref{sec:ChargeOrder}). Nearly all the resonances observed so far are peaked at the low-energy side of the oxygen $K$ edge, showing that the probed electronic modulations mainly involve the electronic states close to the Fermi level.

\subsubsection{Transition metal $L$ edges}\label{sec:TMLedge}\hfill\\
\\
At the transition metal $L_{2,3}$ edges the resonant scattering process involves $2p\rightarrow 3d$ transitions. This makes it possible to study ordering phenomena related to the $3d$ valence states of the TM. The strong spin-orbit interaction of the $2p$ core-hole splits the x-ray absorption spectra into the so-called $L_3$ edge at lower energies ($2p_{3/2}$ core hole) and the $L_2$ edge at higher energies ($2p_{1/2}$ core hole). This is shown for the Cu $L_{2,3}$-edge in figure\,\ref{fig:KKanalysis}. Furthermore, the coupling of the $2p$ core-hole with a partially filled $3d$-shell in the intermediate state, results in a very pronounced multiplet structure of the $L_{2,3}$ edges not only in absorption, but also in resonant scattering, as can be inferred from the optical theorem.

The spin-orbit interaction causes a greatly enhanced sensitivity towards spin-order\,\cite{DeGroot2008}, which, by means of the optical theorem, is directly related to the magnetic circular and linear dichroism in absorption\,\cite{Lovesey1996}. RSXS at the TM $L_{2,3}$ edges therefore allows the study of  magnetic order of the TM-sites. In addition other electronic degrees of freedom of the $3d$ electrons, like orbital and charge can be probed at these edges in a very direct way, since the virtual transitions depend directly on the configuration of the $3d$ shell. 

\subsubsection{Rare earth $M_{4,5}$ edges}\hfill\\
\\
As will be described in detail in section\,\ref{sec:MagneticStructureOfThinFilms}, RSXS at the RE $M_{4,5}$ edges plays an important role in determining the magnetic structure of thin films containing REs. At the RE $M_{4,5}$ edge, the resonant scattering process involves $3d\rightarrow4f$ transitions. One therefore probes the $4f$ states of the RE elements, which play the most important role for their magnetism. 

In the $3d^{9}4f^{n+1}$  final state of XAS there is a strong spin-orbit coupling that splits the absorption edge into the $M_5$ edge at lower energies ($3d_{5/2}$ core hole) and the $M_4$ edge at higher energies ($3d_{3/2}$ core hole).  Furthermore, the coupling between the core-hole and the valence shell, and the spin-orbit interaction acting on the 4f-electrons, causes a multiplet structure. Both the $M_{4,5}$ splitting and the multiplet structure are also reflected in RSXS as dictated by the optical theorem. Similar to the case of the TM $L_{2,3}$ edge, the 
spin-orbit splitting in the intermediate state of the RSXS process again results a greatly enhanced magnetic sensitivity, which enables the study of the magnetic order in extremely small sample volumes. A particular feature of the RE spectra  is the small crystal field splitting. Hence, the spectral shape at the $M_{4,5}$ resonances is largely determined by the atomic multiplet structure independent of the crystal symmetry. Effects of the local symmetry, as discussed in section\,\ref{sec:Haverkort} are therefore not so pronounced as in case of $3d$ TM, which permits an interpretation of RSXS data without recourse to detailed multiplet calculations\,\cite{Ott2006b,Nagao2008,Schierle2010}.
 
\section{Experimental}

RSXS experiments require photon energies in the range between 200 eV and 2000 eV and substantially benefit from a control of the incident light polarisation, in particular linear in the scattering plane as well as in the perpendicular direction ($\sigma$ and $\pi$ in figure\,\ref{fig:exp_scheme}). But also circular polarisation is important, as will be discussed in connection with cycloidal magnetic structures in section\,\ref{sec:Multiferroics}. These conditions are met by a number of beamlines at third-generation synchrotron radiation sources worldwide. They typically have  elliptical undulators as radiation sources, often of APPLE-II type. Beamlines are  designed using grating monochromators and focusing mirrors operated in grazing incidence. An account of the beamline layout at the Berlin synchrotron BESSY II is given in\,\cite{Englisch2001}.

As pointed out already,  soft x-rays  are subject to strong absorption by ambient atmosphere. Unlike diffractometers for hard x-ray resonant scattering, apparatus for resonant scattering in this energy range must therefore be designed as vacuum instruments, while permitting scattering geometries as shown in figure\,\ref{fig:exp_scheme}. For many applications, even ultra-high vacuum (UHV) conditions are preferred, because the x-ray absorption length in the materials studied can be as short as 200\,\AA \, in the vicinity of soft x-ray resonances\,\cite{Ott2006}, rendering the method quite surface sensitive at small scattering angles. Therefore, cleanliness of the sample surface can become an issue, and particularly thin-film reflectivities at low temperatures will be modified by growing overlayers due to  gas absorption. While resonant diffraction using hard x-rays could make use of conventional diffractometers, instrumentation for soft x-ray diffraction therefore has only recently become available. One of the first soft x-ray diffractometers was developed for studies of thin polymer films and worked in a vacuum of 10$^{-6}$ Torr\,\cite{Jark1988}.
Motivated by the high magnetic sensitivity of the soft x-ray resonances, and following pioneering experiments\,\cite{Kao1990,Tonnerre1995,Durr1999}, the development of vacuum instrumentation for diffraction experiments was initiated. Reflectometers developed at synchrotron radiation sources for the characterization of optical elements were modified for the study of magnetic thin-film materials\,\cite{Laan1999,Roper2001}. As soft x-ray resonances were found to be very useful for the study of complex ordering  in correlated electron systems like charge or orbital order\,\cite{Castleton2000}  dedicated diffractometers were developed that are now operational at various synchrotron radiation sources\,\cite{Johnson1992,Abbamonte2002,Grabis2003,Jaouen2004,Hatton2005,Beutier2007,Lee2008,Staub2008,Bruck2008,Takeuchi2009,Hawthorn2011a}. 

\begin{figure}[t!]
\centering
\includegraphics* [scale=0.8, trim= 0 0 0 0, angle= 0] {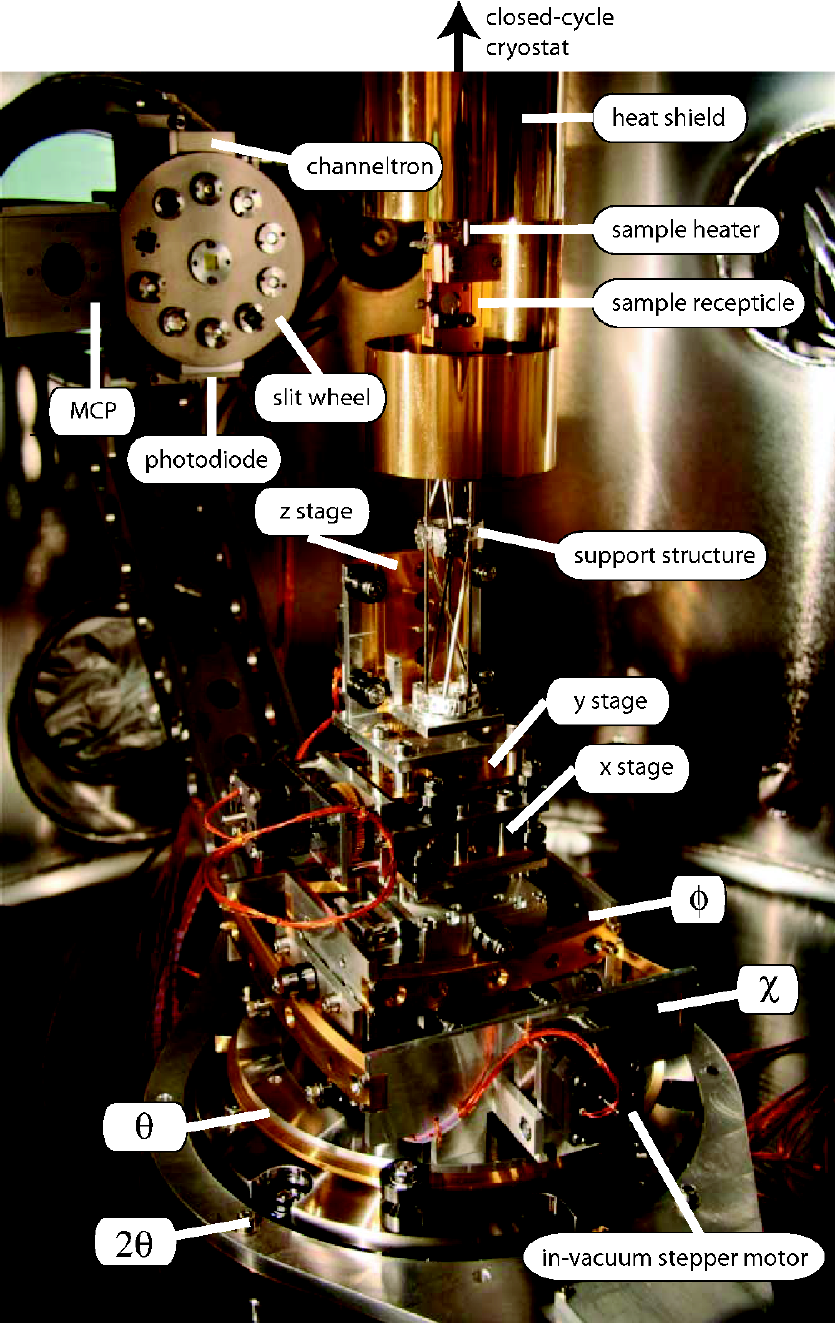}
\caption{\label{Hawthorn2011-1} UHV diffractometer operated at the Canadian Light Source. The instrument is designed as a four-circle diffractometer operated in vacuum. (Reprinted with permission from \cite{Hawthorn2011a}. Copyright 2011, American Institute of Physics.)}
\end{figure}

For the design of these instruments, two approaches are employed. In order to maintain the full flexibility in terms of scattering geometries and sample rotation, diffractometers as known from hard x-ray applications were built from UHV-compatible materials and mounted in a UHV chamber. This requires vacuum-compatible motors and sophisticated mechanical design but maintains the full functionality of a conventional diffractometer. The first instrument of this type was commissioned at the National Synchrotron Light Source\,\cite{Abbamonte2002}. 

Recently, a new diffractometer with a horizontal scattering geometry was launched at beamline 10ID-2 of the Canadian Light Source\,\cite{Hawthorn2011a}. The instrument is shown in figure\,\ref{Hawthorn2011-1} and represents a 4-circle diffractometer, with sample ($\theta$) and detector ($2\theta$) rotation as well as sample tilt ($\chi$) and azimuthal rotation ($\phi$). Sample manipulation is complemented by $x$, $y$, and $z$ stages. Situated inside a vacuum chamber, it operates at a base pressure of $2 \times 10^{-10}$ Torr. Samples can be cooled with a closed-cycle He cryostat connected to the sample holder by flexible Cu braids. Avoiding a rigid mechanical connection to the cryostat has the advantage  that temperature-dependent studies can be carried out without substantial movement of the sample due to the thermal contraction of the sample holder. Sample position was reported to be stable within 100\,$\mu$m\,\cite{Hawthorn2011a}. The base temperature achieved in this instrument is 18 K. The $2\theta$ rotation carries various photon detectors, such as a AXUV 100 silicon photodiode, which works very reliably over a large dynamic range of many orders of magnitude, which is particularly required for reflectivity measurements. For the detection of weak signals a channeltron typically coated with a low workfunction material such as KBr can be used. The instrument  also has a multi channel plate (MCP) with resistive anode readout for the two-dimensional detection of scattering signals.

\begin{figure}[t!]
\centering
\includegraphics* [scale=1, trim= 0 0 0 0, angle= 0] {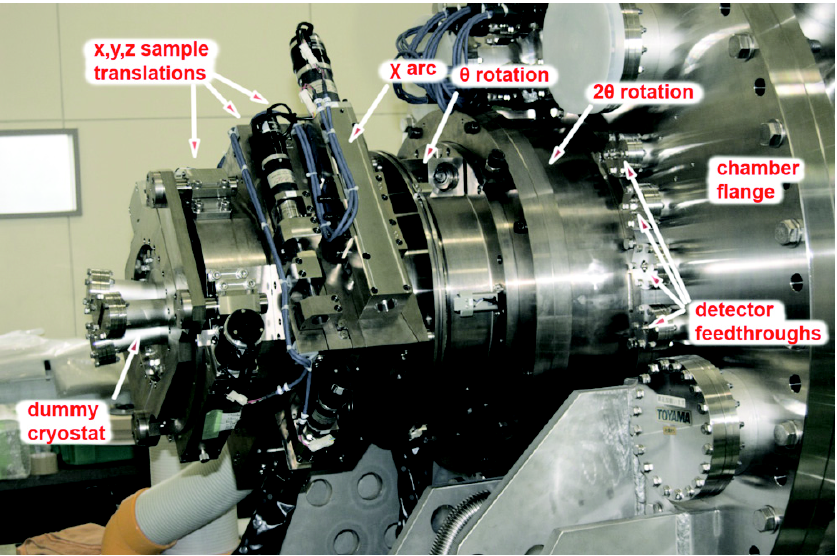}
\caption{\label{Beale2010a-1} External view of the RASOR diffractometer at Diamond Light Source. In this design, sample ($\theta$) and detector (2$\theta$) rotation as well as sample translations are driven by motors and gears outside the vacuum. (Reprinted with permission from \cite{Beale2010a}. Copyright 2010, American Institute of Physics.)}
\end{figure}

Other designs at least partially avoid in-vacuum motors and complex mechanics and have the drives at atmosphere. Figure\,\ref{Beale2010a-1} shows the experimental station RASOR operated with a vertical scattering geometry at beamline I10 of the Diamond Light Source. Here sample and detector rotation are driven by motors and gears outside the vacuum using a double differentially pumped rotary feed-through. Also sample tilt as well as the translational alignment $(x, y, z)$ are actuated from outside the vacuum, while sample azimuth involves an in-vacuum drive\,\cite{Beale2010a}. 

A new in-vacuum spectrometer has been installed on beamline X1A2 at the NSLS, Brookhaven National Laboratory. This spectrometer employs a 6-circle geometry with both horizontal and vertical detector rotations, enabling the detector to be placed anywhere within the hemisphere of the vacuum vessel. This detector configuration dispenses with the need for a $\chi$ rotation of the sample and greatly simplifies the sample rotation hardware enabling a much shorter cryostat to aid in stability and the future use of more complicated sample environments. Unique to this instrument is the capability of performing both surface soft x-ray scattering measurements and, through the use of zone plate optics, soft x-ray nano-diffraction by raster scanning the beam over the sample to measure a real space image of the scattering\,\cite{Wilkins2011}. 

\begin{figure}[t!]
\centering
\includegraphics* [scale=0.5, trim= 0 0 0 0, angle= 0] {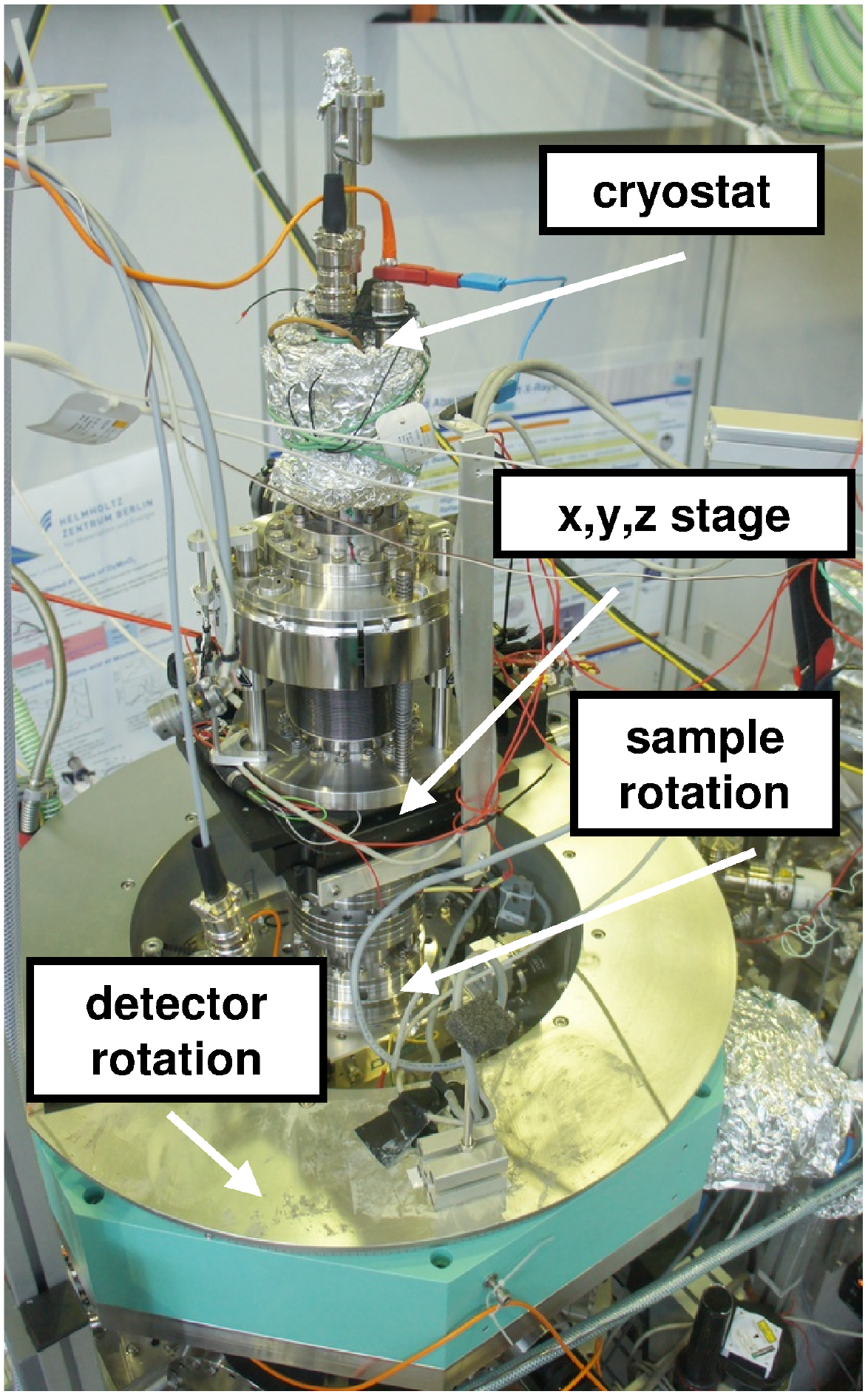}
\caption{\label{Bessydiffractometer-1} RSXS diffractometer operated at BESSY II Berlin. The instrument is based on a large two-stage rotary vacuum feedthrough that provides the basic sample and detector rotation with very high accuracy. }
\end{figure}

\begin{figure}[t!]
\centering
\includegraphics* [scale=0.5, trim= 0 0 0 0, angle= 0] {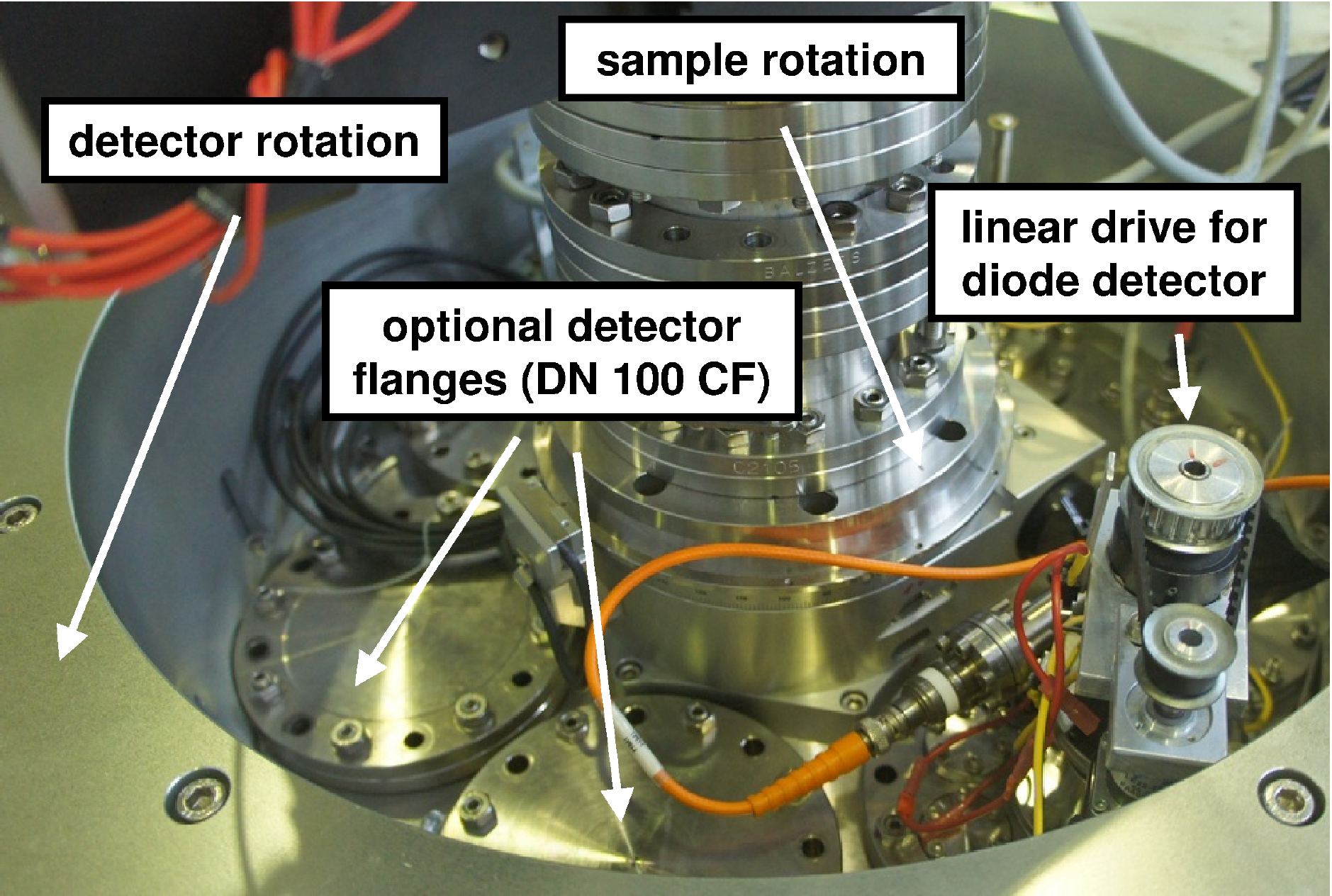}
\caption{\label{Bessydiffractometer-2} Detail of the RSXS diffractometer operated at BESSY II Berlin. Large vacuum flanges on the detector rotation platform provide high flexibility for the application of various detector assemblies. }
\end{figure}

Other designs of soft x-ray diffractometers do not use a conventional four-circle geometry and in this way benefit from avoiding in-vacuum motors and mechanics. Such an instrument is 
operated at beamline UE46-PGM1 of the Berlin synchrotron BESSY~II (see figure\,\ref{Bessydiffractometer-1}). The basis of the diffractometer is a large differentially pumped two-circle rotary feedthrough driven outside the vacuum by two modified Huber rotation stages. Working in horizontal scattering geometry, the instrument combines very high mechanical precision with the capability of carrying high loads, particularly on the detector circle. The latter consists of a large disc that comprises several UHV flanges up to a bore size of 100\,mm diameter. Detectors of various degrees of complexity can be assembled on DN 100 CF flanges and can then be easily transferred and removed from the diffractometer (see figure\,\ref{Bessydiffractometer-2}). Presently, the instrument is equipped with a conventional AXUV 100 Si photodiode, a slit carousel allows in-situ adaption of the resolution. The whole detector is mounted on a linear feed-through, which resembles an $\omega$ rotation of the detector and readily compensates for a missing $\delta$ tilt of the sample. 
The instrument can operate under UHV conditions. Pressures in the low $10^{-10}$~mbar range are reached even without baking the chamber  by operating the sample cryostat together with a liquid nitrogen  cryogenic trap for a few days. 

Various sample holders can be introduced via a CF flange with a bore of 90 mm diameter (see figure\,\ref{Bessydiffractometer-2}). For standard experiments, samples are mounted on a Janis He flow cryostat. Here, base temperatures of $\le 10$\,K are easily achieved and with proper shielding  3~K is reliably obtained. It is to be noted, however, that in this temperature region, heating by the x-ray beam is non-negligible and must be taken into account, e.g., by adequately reducing the photon flux. 

In RSXS experiments, the requirements for sample environment do not only include vacuum conditions and low temperatures. An important parameter for the study of magnetic materials is an external magnetic field. While fields of the order of 100 Gauss can be obtained by small coils in the vicinity of the sample, larger fields require substantial effort. Several apparatus were developed in recent years that are equipped with normal-conducting coil systems, achieving magnetic fields of $< 1$~Tesla. The ALICE instrument operated at BESSY~II has water-cooled coils outside the vacuum, and the magnetic field is guided into the chamber by an iron yoke\,\cite{Grabis2003}. Fields at the sample position can reach 0.8 Tesla. 
The RESOXS instrument now operated at SOLEIL has a four-pole magnet arrangement that can be rotated and allows high flexibility in the direction of the field with respect to the sample\,\cite{Jaouen2004}. Here, $\approx 0.2$~Tesla are achieved.
A recent reflectometer for magnetic measurements developed at the Max-Planck-Institut Stuttgart provides a magnetic field of 0.65 Tesla\,\cite{Bruck2008}. 

Dedicated soft x-ray diffractometers equipped with superconducting coils that provide magnetic fields considerably larger than one  Tesla are presently not available. However, a superconducting magnet that is rotatable inside a vacuum chamber and allows for selected scattering geometries with magnetic fields up to 7\,Tesla is operated at beamline UE46-PGM1 of BESSY II. 

While sample environment is one challenge of soft x-ray diffraction experiments, x-ray detection is now also a focus of instrumental development. Two-dimensional detectors can enhance the efficiency of measurements substantially and some of the instruments are already equipped with vacuum-compatible x-ray CCD cameras that are mounted on the $2\,\theta$-arm of the diffractometer and can be rotated inside the vacuum. Instruments with that option are operated at the Swiss Light Source\,\cite{Staub2008} and the National Synchrotron Light Source\,\cite{Wilkins2011}.

In order to obtain the full information that can be obtained from a soft x-ray diffraction experiment, polarisation analysis of the scattered x-ray beam is also desirable. The method is routinely used in hard x-ray non-resonant and resonant diffraction\,\cite{Paolasini2007} and uses single crystals with a suitable lattice spacing as analyzers (cf. figure\,\ref{fig:exp_scheme}) . With a Bragg diffraction angle corresponding to the Brewster angle ($2\theta=90^\circ$) at the given photon energy, x-rays with $\pi$ linear polarisation are (ideally) completely suppressed. Rotating the analyzer crystal around the x-ray beam scattered from the sample allows to characterize the polarisation of the scattered x-rays. 

\begin{figure}[t!]
\centering
\includegraphics* [scale=1, trim= 0 0 0 0, angle= 0] {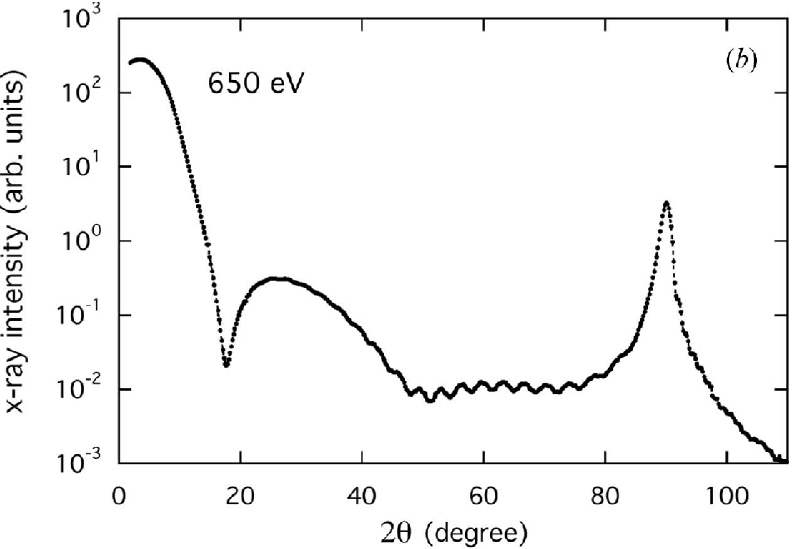}
\caption{\label{Staub2008-1} Reflectivity of a W/C multilayer recorded with $\sigma$ polarised light at a photon energy of 650 eV. The multilayer has a reflectivity of a few percent at the superlattice peak at $2\theta = 90^{\circ}$, which can be used for polarisation analysis at this photon energy. (Reprinted with permission from \cite{Staub2008}. Copyright 2008, International Union of Crystallography.)}
\label{fig:multilayer}
\end{figure}

Using the same method with soft x-rays usually requires artificial multilayer structures with suitable lattice spacings. The technique was initially applied in polarimeters to characterize the polarisation state of synchrotron radiation\,\cite{Kortright1995,Schafers1999}. Recently, it was also integrated in soft x-ray diffractometers. The first instrument to be equipped with a polarisation analysis was reported by Staub et al.\,\cite{Staub2008}. Polarisation analysis is performed by an artificial W/C multilayer with graded layer spacings. By linear translation the layer spacing can be chosen to meet the Bragg condition at the Brewster angle for a given resonance energy. The efficiency of such a multilayer at the first-order diffraction peak is typically of the order of a few percent, as can be inferred from the reflectivity curve shown in figure\,\ref{Staub2008-1}. Despite this substantial loss in intensity, polarisation analysis in soft x-ray diffraction is feasible and can be employed to study complex phenomena such as orbital currents\,\cite{Scagnoli2011}.

\begin{figure}[t!]
\centering
\includegraphics* [scale=1, trim= 0 0 0 0, angle= 0] {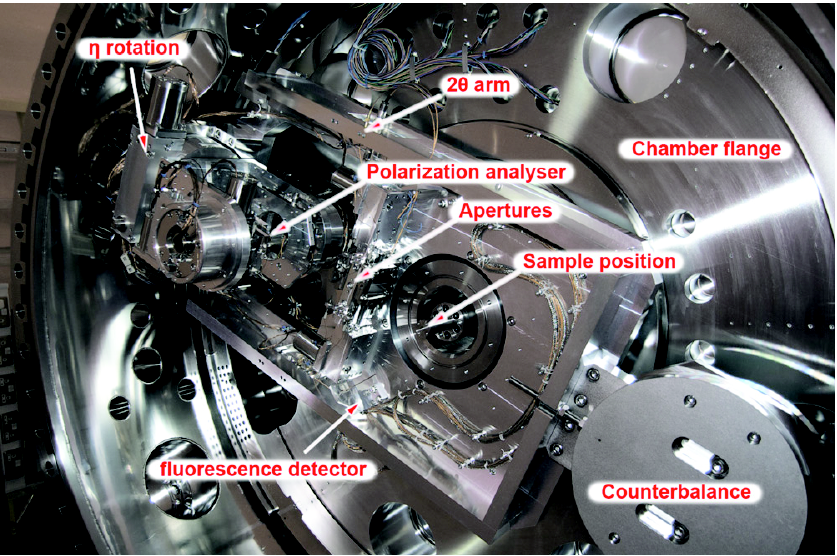}
\caption{\label{Beale2010a-2} $2\theta$-arm of the RASOR diffractometer, equipped with a polarisation analysis.  (Reprinted with permission from \cite{Beale2010a}. Copyright 2010, American Institute of Physics.)}
\end{figure}

Recently, an advanced polarisation analysis has become available with the previously mentioned RASOR instrument\,\cite{Beale2010a}. Here, even a full rotation of the analyzer multilayer around the scattered x-ray beam is possible (see figure\,\ref{Beale2010a-2}).

\section{Case studies}\label{sec:CaseStudies}  

\subsection{Magnetic structure of thin films}\label{sec:MagneticStructureOfThinFilms}

\begin{figure}[t!]
\centering
\includegraphics [width=7cm] {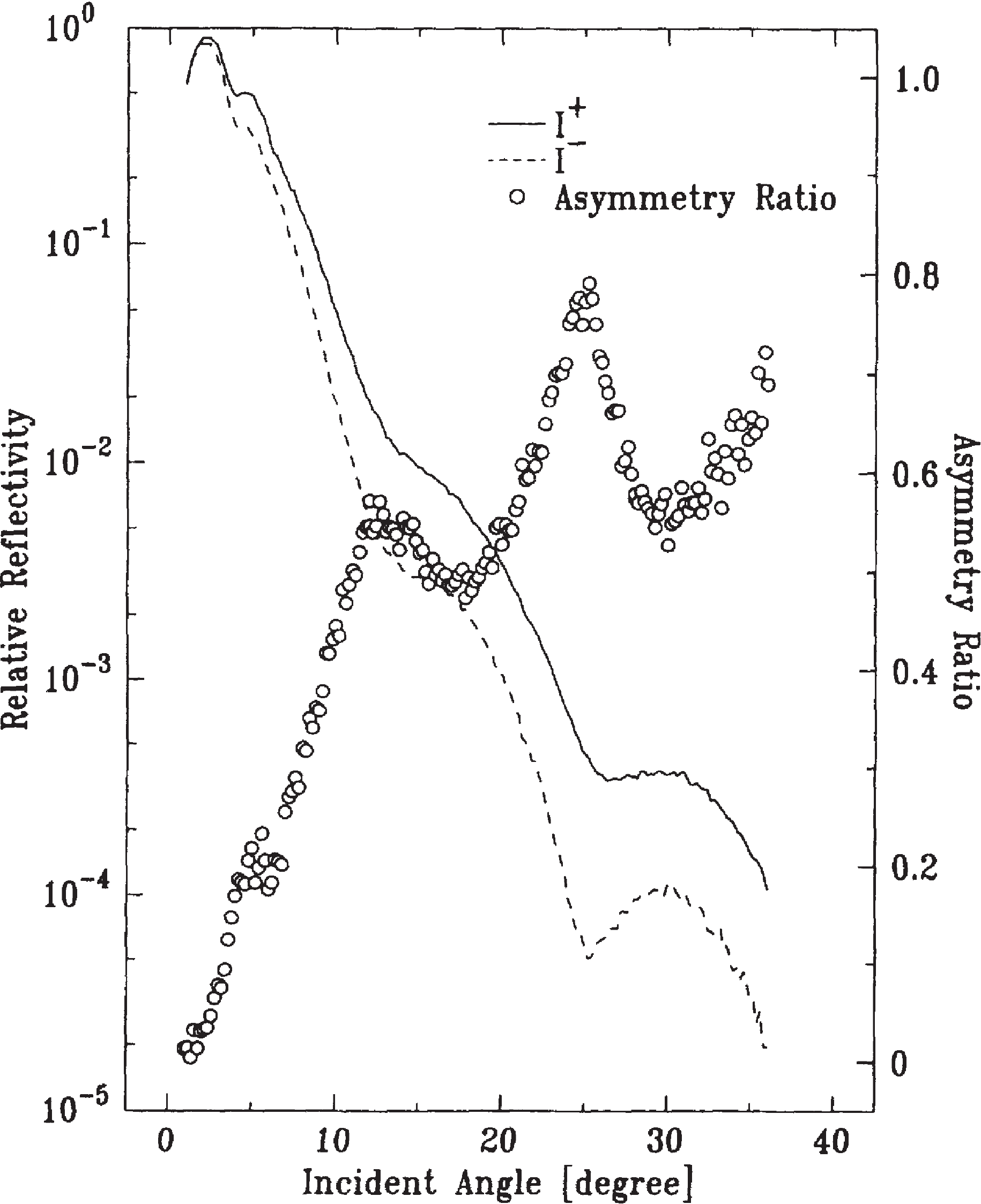}
\caption{\label{p_kao94} Reflectivity of a 37 \AA\,thin Co film as a function of grazing angle measured
at the maximum of the Co $L_{3}$ edge. The solid and dashed
lines denote the two opposite magnetization directions, the dots represent the resulting asymmetry
ratio, which shows a strong angular dependence and amounts up to 80\,\%. (Reprinted with permission from \cite{Kao1994}. Copyright 1994, American Physical Society.)}
\end{figure}

Following the prediction of very strong magnetic scattering cross sections at the Lanthanide $M_{4,5}$ and 3$d$-transition metal $L_{2,3}$ absorption thresholds \cite{Blume1985,Hannon1988} , historically, RSXS  was first employed to study magnetic ordering in thin metallic films and multilayers. The first demonstration of the strength of  magnetic  RSXS  has been reported by Kao et al., who observed changes in the specular reflectivity of a ferromagnetic 35\,\AA-thin single-crystalline Fe film upon magnetization reversal\,\cite{Kao1990}. Exploiting linear polarised x-rays, they found asymmetry ratios $R=(I^{+} - I^{-})/(I^{+} + I^{-})$ as large as 13\,\% at the Fe $L_{2,3}$ resonance  and even stronger effects have been reported for scattering of circular polarised x-rays from a 37\,\AA \,thin Co film with asymmetry ratios up to 80\,\% close to the Co $L_{2,3}$ resonance as shown in figure\,\ref{p_kao94} \cite{Kao1994}. Here $I^{+}$ and $I^{-}$ are the measured specular reflectivity for the two opposite in-plane magnetization directions of this experiment.  The asymmetry ratios strongly depend on the scattering angle, which represents additional structural information on the magnetic order. 

\begin{figure}[t!]
\centering
\includegraphics{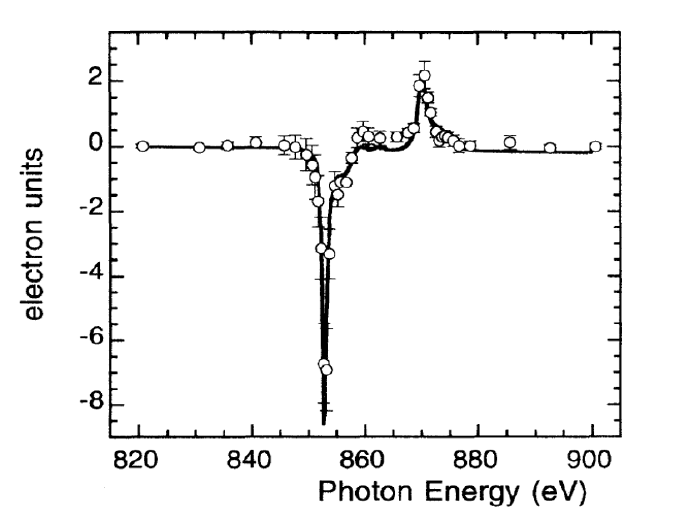}
\caption{\label{p_tonnerre95} Imaginary part of  magnetic RSXS  at the Ni $L_{2,3}$ edge (open circles) obtained from a Ni/Ag multilayer compared to Ni magnetic circular dichroism
data  (solid line)  in electron units. (Reprinted with permission from \cite{Tonnerre1995}. Copyright 1995, American Physical Society.)}
\end{figure}

These two pioneering experiments  revealed the huge potential of RSXS combining the spectroscopic power of x-ray magnetic circular dichroism (XMCD)  with structural information. Even without exploiting the latter aspect, a large number of studies used the strong dichroic effects of the scattered intensities to extract information usually obtained by x-ray absorption techniques \cite{Freeland1997,Tonnerre1998,Hellwig2003,Schulli2004,Radu2009,Tardif2010,Valencia2011}, like monitoring element-specific magnetic hysteresis loops by measuring the field dependence of the circular dichroism of the reflected intensity for fixed photon energies and angular positions. Such studies avoid the experimental difficulties of XMCD performed in total electron yield (TEY) or fluorescence yield (FY) mode, i.e., a  limited probing depth and the influence of external fields and charging  in the TEY mode  or weak signals connected with the FY mode. While these first experiments focused on thin single films, Tonnerre et al.  presented the first demonstration of magnetic RSXS from a multilayer system consisting of  a repetition of Ag(11\,\AA)/Ni(17.5\,\AA) double layers with antiferromagnetic (AFM) or ferromagnetic (FM) coupling of the Ni layers depending on external magnetic field. From this sample either pure half-order magnetic Bragg reflections (AFM coupling) or huge magnetic circular dichroism on the corresponding chemical superstructure reflections (FM coupling) could be observed. By analyzing the magnetic scattering intensities, a magnetic scattering strength at the Ni $L_{2,3}$ edge of the order of ten times the electron  radius $ r_{0}$ was deduced (see figure\,\ref{p_tonnerre95}), i.e., a value very close to the non-resonant charge scattering strength \cite{Tonnerre1995}.

\begin{figure}[t!]
\centering
\includegraphics{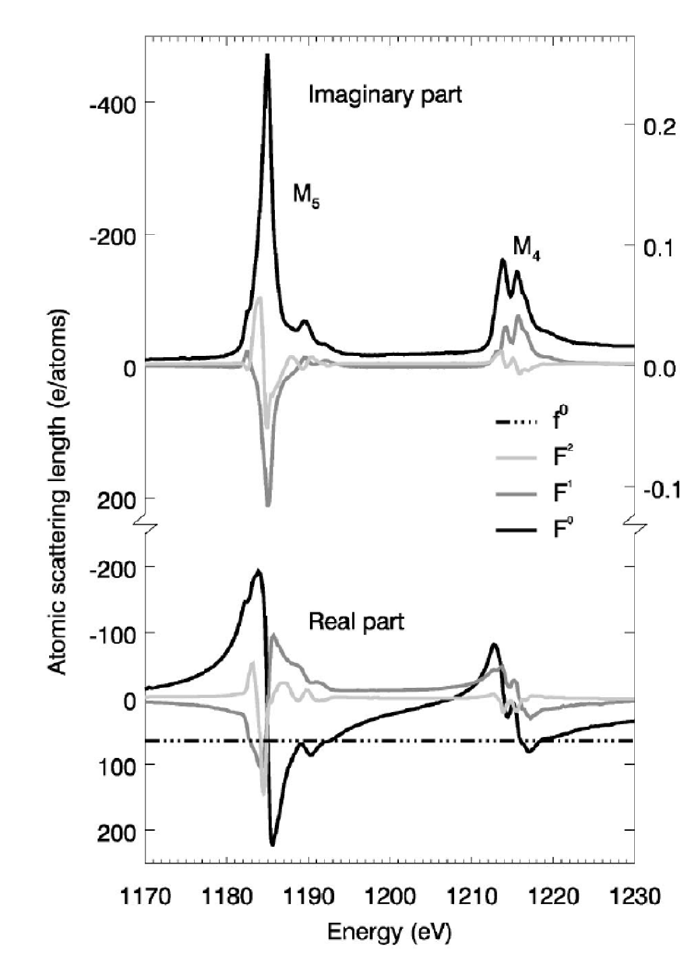}
\caption{\label{p_peters2004} Resonant amplitudes at the Gd $M_{4,5}$ edges. Shown are
the complex charge $F^{0}$, circular magnetic $F^{1}$, and linear magnetic 
$F^{2}$, atomic scattering factors as function of energy in units of $r_{0}$.
Top: imaginary parts, bottom: real parts. Right axis: approximate atomic cross sections
in \AA $^{2}$ using a fixed wavelength for E=1200 eV. Dash-dotted line:
high energy limit of the atomic scattering amplitude Z=64. (Reprinted with permission from \cite{Peters2004}. Copyright 2004, American Physical Society.)}
\end{figure}

The strong enhancement of the scattering cross section at resonance is intrinsically accompanied by a huge change of the optical constants $n=1-\delta+i\beta$ (see section\,\ref{sec:RSXSandXAS}), with  $\beta$ and $\delta$ being of the order of $10^{-2}$. As a consequence, dispersion and absorption effects can no longer be neglected but often have to be included in a quantitative analysis of RSXS data. Consequently, a  large number of studies have been dedicated to a quantitative characterization of soft x-ray resonances, where the optical parameters are typically not well known and the tabulated values \cite{Henke1993} can not be applied. Experimentally, the easiest access to the optical constants is by measuring the absorption coefficient by transmission \cite{Vicentin1995}, TEY, or FY methods and subsequent calculation of the real part of the index of refraction by applying the Kramers-Kronig relation (cf. section\,\ref{sec:RSXSandXAS}). However, the accuracy of the individual methods of measuring the absorption coefficient are often limited by various experimental uncertainties such that it is advantageous to acquire additional information independently as has been done by measuring the Faraday rotation of linear polarised light passing through an Fe/Cr multilayer \cite{Kortright1995} or the energy-dependent Bragg-peak displacement of Fe/V superlattice reflection \cite{Sacchi1998}. In particular, it has been shown that the analysis of polarisation- and energy-dependent position, width and intensity of a superstructure reflection can be used to determine the full set of optical parameters at resonance \cite{Tonnerre1995,Seve1995,Mertins2002}.  While these first studies focused on the $3d$  TM $L_{2,3}$ resonances, detailed characterization of the Lanthanide $M_{4,5}$ resonances has been performed later revealing even stronger resonance effects \cite{Prieto2003,Peters2004,Ott2006b}. Figure \ref{p_peters2004} shows the resonant scattering length through the Gd $M_{4,5}$ resonance calculated on the basis of polarisation dependent absorption obtained in transmission through a thin Gd$_{1-x}$Fe$_{x}$ film. Here, $F^{0}$  represents the resonant charge scattering contribution of the order of 500\,$r_{0}$, which is almost 10 times stronger than off-resonant charge scattering (cf. Eq.\,\ref{eqn:Hannon} in which the photon energy dependent resonant factors are called $R^{(j)}$, i.e., $F^{(j)}$=$R^{(j)}$). Even more impressive, the circular dichroic scattering contribution $F^1$ ($R^{(1)}$ in Eq.\,\ref{eqn:Hannon}) is as large as   200\,$r_{0}$, i.e., a magnetic scattering strength 3 times stronger than off-resonant charge scattering.

With this extraordinarily high magnetic sensitivity, the study of magnetic structures of thin films down to only a few magnetic layers became feasible.

\subsubsection{Magnetic diffraction from long-period antiferromagnets}\label{sec:long-periodAFM}
\begin{figure}[t!]
\centering
\includegraphics{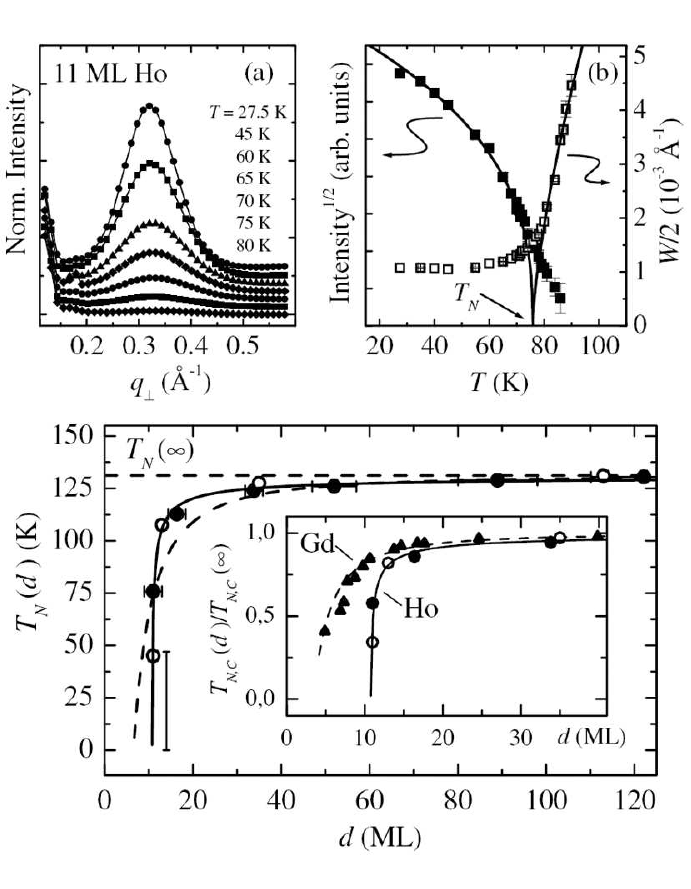}
\caption{\label{p_weschke2004} Top: (a) Magnetic reflection of an 11 ML-thin
Ho film at various temperatures, recorded at the Ho $M_{5}$ resonance.
(b) Square root of the integrated intensity (solid squares) and half widths of the rocking curves of the magnetic reflection (open squares).  
The solid lines represent the result of a simultaneous power-law fit
analysis to determine $T_N$. Bottom: $T_{N}$ as a function of Ho film thickness $d$, including
data from MBE films (solid circles) and films grown in situ on
W(110) (open circles).The inset shows a comparison to the behaviour of  ferromagnet Gd. (Reprinted  from \cite{Weschke2004}. Copyright 2004, American Physical Society.)}
\end{figure}

\hfill\\
\\
The power of magnetic RSXS to study AFM Bragg reflections from thin single films of only a few magnetic layers has been demonstrated for thin Holmium metal films \cite{Schuessler-Langeheine2001,Weschke2004}. Metallic Holmium possesses the largest  magnetic moment of all elements coupled by oscillatory long-range RKKY interactions. Below the bulk ordering temperature of 132\,K, Ho develops ferromagnetic order within the closed-packed planes of the hexagonal closed packed (hcp) structure but with a certain angle between the moments of neighbouring planes resulting in a long-period AFM structure along the [001] direction with an incommensurate temperature-dependent period of about 10 monolayers at 40\,K. In a magnetic diffraction experiment, this magnetic structure causes magnetic superstructure reflections well separated from the crystallographic reflections (0 0 2L~$\pm \tau$) with $\tau \approx 0.2$ at 40\,K as observed from Ho single crystals by neutron scattering\,\cite{Koehler1966, Felcher1976, Pechan1984}. The  potential of magnetic off-resonant and magnetic hard x-ray resonant scattering has been demonstrated  in a series of pioneering magnetic structure studies on Ho by Gibbs et al.\,\cite{Gibbs1985,Gibbs1988}.
In RSXS experiments at the Ho $M_{5}$ soft x-ray resonance, due to the long period of the  magnetic structure, the $(0 0 \tau )$ magnetic reflection of Ho is well within the Ewald sphere. Exploiting the huge magnetic scattering strength of RSXS, this magnetic structure could be readily studied in thin films down to the thickness of a single magnetic period \cite{Schuessler-Langeheine2001, Weschke2004, Ott2006b} as can been seen from the intense and well developed magnetic superstructure reflections obtained from a 11 monolayer thin Ho metal film shown in figure\,\ref{p_weschke2004}.

While for RSXS the relation between magnetic ordering parameters and integrated diffraction peak intensity can be more involved,  the integrated superstucture peak intensity of Ho has been shown to be in good approximation a measure of the helical AFM order parameter\,\cite{Ott2006b} and thus allows to determine the ordering temperatures for various film thicknesses $d$. For small $d$, a finite-size scaling of $T_{N}$ would have been anticipated and was indeed observed. However, this study found a much faster decrease of $T_{N}$ with $d$ (see figure\,\ref{p_weschke2004}) than known from simple antiferromagnets and ferromagnets \cite{Qiu1991,Schneider1990, Li1992,Farle1993}. The observed modified finite-size effect in the ordering temperature is characterized by an offset thickness $d_{0}$ representing a minimum sample thickness below which no helical order can be established. This finding has been related to the long period of the AFM structure\,\cite{Weschke2004}, in line with results obtained for the long-period spin-density wave of Cr in Cr/Fe superlattices \cite{Fullerton1996}. As shown in figure\,\ref{p_weschke2004},  even above $T_{N}$, finite magnetic intensity could be observed characterized by a diverging peak width with increasing temperature.  These remaining broad intensity distributions are caused by short-range magnetic correlations persisting above $T_{N}$, with the peak width being an inverse measure of the magnetic correlation length. This observation shows the huge potential of RSXS for studying  magnetic short-range correlations in films of only a few monolayer thickness, even with the perspective to employ coherent scattering as discussed later.

\begin{figure}[t!]
\centering
\includegraphics{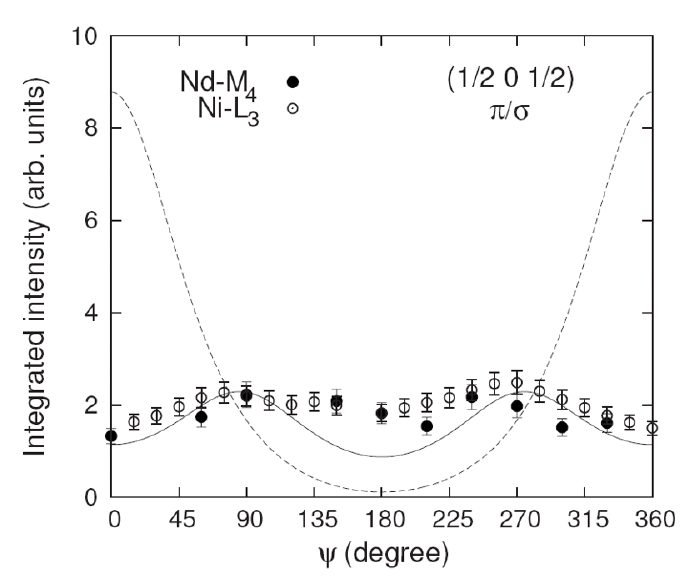}
\caption{\label{p_scagnoli2008} Azimuthal angle dependence of the (0.5 0 0.5) magnetic reflection of NdNiO$_3$ measured at the Ni $L_3$  and Nd $M_{4}$  edge. The solid and
dotted lines correspond to model calculations assuming a non-collinear and a collinear up-up-down-down
magnetic structure, respectively. (Reprinted with permission from \cite{Scagnoli2008}. Copyright 2008, American Physical Society.)}
\end{figure}

The helical magnetic structure of bulk Ho was already well established by neutron scattering. For some materials, however, no large single crystals can be synthesized, which   limits the capability of neutron scattering for magnetic structure determination. In such cases, RSXS can provide  detailed information on the magnetic moment directions exploiting the polarisation dependence of the scattered intensity as discussed in section\,\ref{sec:pol_dep}. Here, the example of thin epitaxial NdNiO$_{3}$ (NNO) films \cite{Scagnoli2006, Scagnoli2008} is presented. In this work  the element specificity of RSXS to study the magnetic structure of Ni and Nd moments in NNO separately is exploited. NNO shows a metal-to-insulator transition at about 210\,K. At the same temperature AFM ordering occurs. From the observed (0.5 0 0.5) magnetic reflection, initially, an unusual collinear up-up-down-down magnetic structure  connected with orbital ordering of the Ni$^{3+}$ $e_{g1}$ electrons has been proposed on the basis of neutron powder diffraction data \cite{Garcia-Munoz1992,Garcia-Munoz1994}. These results have been challenged by the later observation of charge disproportionation without any indication for orbital order using  hard x-ray diffraction\, \cite{Staub2002,Scagnoli2005}. However, off-resonant and resonant scattering at the Ni $K$ edge can suffer from weak signals and only indirect sensitivity to orbital order. Therefore the (0.5 0 0.5) reflection has been studied by RSXS in the vicinity of the Ni $L_{2,3}$ and Nd $M_{4}$ resonance.  Figure \ref{p_scagnoli2008} shows the observed azimuthal dependence of the scattered intensity compared to model calculations based on the formalism developed by Hill and McMorrow\,\cite{Hill1996}. Analogous to the case of the simple bcc antiferromagnet (cf. figure\,\ref{fig:azi}), either a collinear up-up-down-down or a non-collinear magnetic structure was assumed. 

The results clearly reveal a non-collinear magnetic structure of the Ni and Nd moments at low temperatures. In addition, the observed polarisation dependence of scattered intensity, including polarisation analysis of the scattered light at the Ni $L_{2,3}$ resonance, has been found to be inconsistent with orbital order and the energy-dependent line shape of the reflection could only  be reproduced  assuming charge order. Thus these studies fully support a picture of combined charge and non-collinear magnetic order of the Ni sites but are inconsistent with orbital order suggesting that the metal-insulator transition is driven by charge disproportionation in NNO.

At this point, it is appropriate to stress again a caveat: the model calculations for NNO used Eq.\,\ref{eqn:Hannon}, which strictly applies to the case,  in which only the spin breaks the otherwise spherical local symmetry. This approach is commonly used  and yields a reasonable data interpretation in many cases. Nonetheless, it has to be kept in mind that this is an approximation which might not always  be valid in a real material (cf.  section\,\ref{sec:Haverkort}).  In particular, for systems with  symmetry lower than cubic, the standard treatment can yield completely wrong results. This is demonstrated in a study on  NaCu$_2$O$_2$  by Leininger et al.~\cite{Leininger2010}.
NaCu$_2$O$_2$  belongs to the class of edge-sharing copper-oxide chain compounds which have attracted considerable interest in the past years due to their diversity of ground states. The related compound $\mbox{LiCu}_{2}\mbox{O}_{2}$, e.g. has been found to exhibit ferroelectricity. At present it is not clear whether this ferroelectricity is caused by the complex magnetic structure or the strong tendency  of Li and Cu intersubstitution.    A non-collinear spiral antiferromagnetic structure was derived by  neutron diffraction on polycrystalline samples and NMR measurements on NaCu$_2$O$_2$\,\cite{capogna2005,drechsler2006} in line with results from susceptibility measurements ~\cite{Leininger2010}. Serving as an isostructural and isoelectronic reference compound without complicated intersubstitution of alkaline and $3d$ TM ions\,\cite{capogna2005,drechsler2006}, NaCu$_2$O$_2$ is only available in form of tiny single crystals. Therefore, RSXS is the method of choice to study details of the magnetic structure. However, according to the simple treatment of magnetic RSXS data, the azimuthal angle dependence of the magnetic reflection suggested a magnetic structure made of only one magnetic moment direction -- in stark contrast to the results of the susceptibility measurements and neutron powder diffraction data. This  inconsistency could be explained on the basis of the  strong anisotropy observed in  the polarisation dependent absorption probability.  The anisotropy is  caused by a selection rule precluding excitations into the partially occupied planar $3d_{x^{2}-y^{2}}$ orbitals of the $\mbox{Cu}^{2+}$ ions which modifies the polarisation dependent magnetic scattering strength according to the Kramers-Heisenberg formula and hence masks the azimuthal angle dependence of the magnetic signal. These findings are in excellent agreement with the theoretical predictions by Haverkort et al.\,\cite{Haverkort2010}, which shows that in general the local orbital symmetry has to be taken into account for magnetic structure determination by magnetic RSXS.

The first application of RSXS to study long-period AFM structures in bulk materials has been reported by Wilkins et al.\,\cite{Wilkins2003}  showing the high potential of RSXS to study magnetic structures and, even more importantly, other electronic ordering phenomena in bulk oxides. For magnetic structure determination, however, neutron scattering will stay an unrivaled method. On the other hand, despite the limited Ewald sphere,  
the potential of magnetic RSXS for exploring magnetic structure has been clearly shown and its importance will increase with growing interest in thin transition metal oxide films and superstructures that can be obtained with very high quality. Films of $RE$MnO$_3$ perovskites represent a prominent example,  where long period complex magnetic order is strongly coupled to ferroelectricity. Such a thin improper multiferroic has been indeed studied for the first time by RSXS very recently by Wadati et al.\,\cite{Wadati2012}.

\begin{figure}[t!]
\centering
\includegraphics{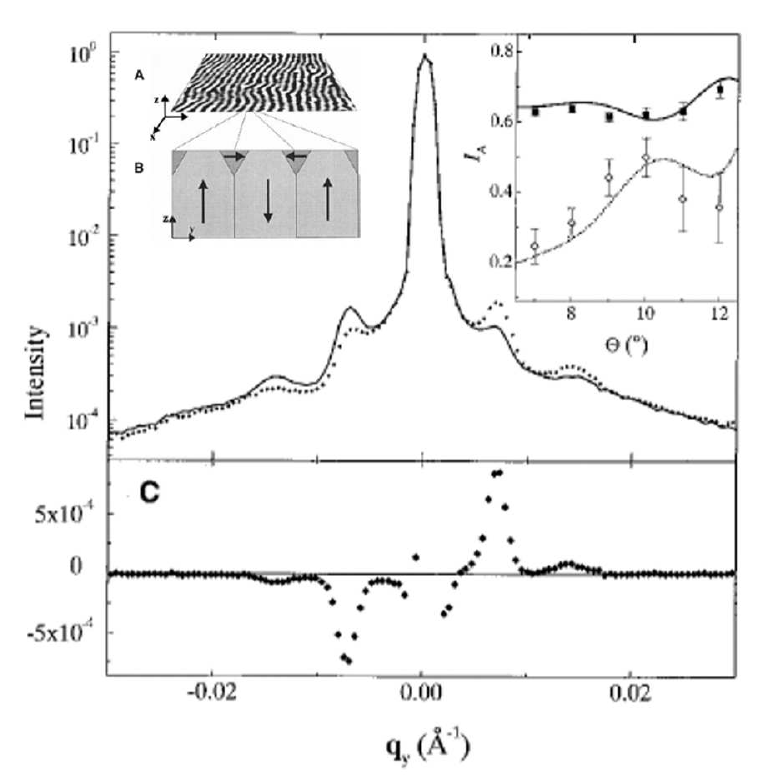}
\caption{\label{p_durr99} Diffraction scans
with momentum transfer perpendicular to the magnetic
stripes in FePd films for left (dotted) and right (solid) circularly polarised x-rays and resulting difference signal below. Insets: right: asymmetry ratios for the first and second order magnetic reflections. Left: magnetic force microscope image of stripe domains and schematic magnetization profile.(Reprinted with permission from \cite{Durr1999}. Copyright 1999, American Association for the Advance of Science.) }
\end{figure}

\subsubsection{Magnetic depth profiles}\label{magneticdepth}
\begin{figure}[t!]
\centering
\includegraphics{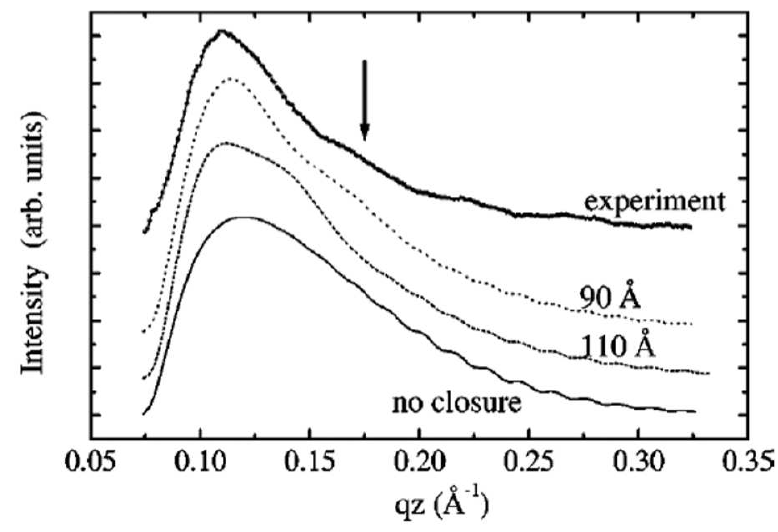}
\caption{\label{p_dudzik00} Rod scan of the magnetic satellite peak from stripe domains in FePd thin films together with model calculations showing the influence of closure domain depth: solid line: no closure domains, dashed line: closure depth
110 \AA , dotted line: closure depth 90 \AA .(Reprinted with permission from \cite{Dudzik2000}. Copyright 2000, American Physical Society.)}
\end{figure}

\hfill\\
\\
While RSXS is particularly useful in case of long-period magnetic structures, RSXS can also provide detailed information on the microscopic magnetic properties of ferromagnets and simple antiferromagnets, in particular when applied to artificial structures that produce an additional periodicity of the order of the wavelength of soft x-rays. Such structures are also naturally obtained  by the formation of domains in chemically homogeneous materials, as has been  observed for thin films of FePd. Ferromagnetic FePd films are characterized by perpendicular magnetic anisotropy leading to the formation of  well ordered stripe domains as shown in figure\,\ref{p_durr99}. In order to reduce the stray field outside the sample, closure domains can be formed at the surface, which  link the magnetizations of the neighbouring stripes, resulting in an overall domain structure of chiral nature. X-ray scattering experiments usually are not sensitive to the phase of the scattered wave and, hence, can not easily distinguish left and right handed magnetic structures. However, for circular polarised incident light, the resonant magnetic structure factor differs for structures of different sense of rotation \cite{Mulders2010}  giving rise to circular dichroism in the scattered intensity not connected with absorption effects but depending on the sense of rotation of a magnetic structure. This property has been exploited by D\"urr et al.\,\cite{Durr1999} to study closure domains in thin FePd films by RSXS. Here, the regular domain pattern at the surface does not cause superstructure reflections in terms of a single peak but rod-like off-specular magnetic scattering intensity with a  maximum intensity at a transverse momentum transfer determined by the inverse of the mean distance of the stripe domains as shown in figure \ref{p_durr99}. The intensity of these off-specular magnetic rods showed pronounced circular dichroism, in this way proving the existence of the above mentioned closure domains. By modeling the Q-dependent magnetic rod intensity, shown in figure\,\ref{p_dudzik00}, the depth of the closure domains could be estimated to be about 8.5\,nm in a 42\,nm thick film \cite{Dudzik2000}. 

As shown  in the latter example, RSXS is capable of studying  ordering phenomena with spatial resolution. The information on spatially  varying ordering phenomena is contained in the dependence of scattered intensity on the chosen incident energy and momentum transfer which can be very sensitive even to tiny modifications of order. Hence, RSXS has been applied very successfully to obtain magnetization depth profiles of thin films and multilayers with very high depth resolution. In the past years, three different ways have been introduced to extract depth dependent information on magnetic ordering, exploiting

\begin{enumerate}
\item the change of the sample volume probed through the resonance due to strong incident energy-dependence of the photon penetration depth \cite{Ott2006},
\item the sensitivity of the shape of a magnetic reflection from very thin samples on the spatial varying magnetic ordering parameter \cite{Schierle2008}, and 
\item the incident energy- and momentum-transfer dependent reflectivity \cite{Tonnerre1998, Grabis2003, Roy2005, Valencia2007, Bruck2008, Tonnerre2008}.
\end{enumerate}
	
\begin{figure}[t!]
\centering
\includegraphics{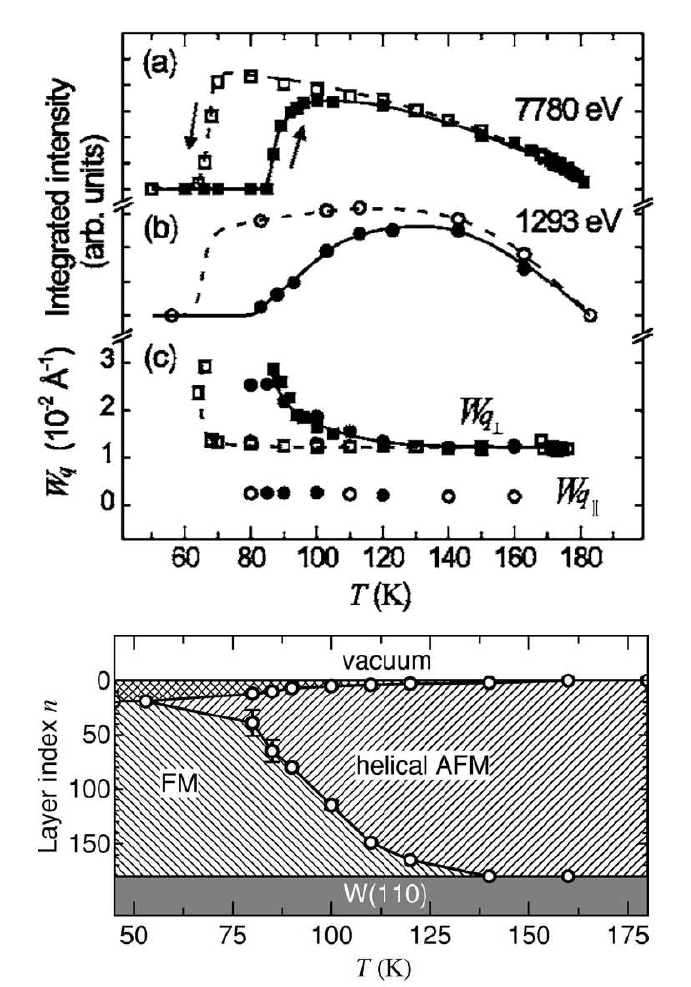}
\caption{\label{p_ott2006} Top: (a) and (b) magnetic diffraction data characterizing the FM/helical AFM phase
transition of 180\,ML Dy/W(110) upon cooling down (open symbols) and
warming up (filled symbols) for two different photon energies; (c) widths of magnetic peaks in
the directions parallel and perpendicular to the film plane recorded
at 7780 eV (squares) and at 1305 eV (circles). Bottom: magnetic depth profile of a 180\,ML thick Dy film during the first order
phase transition from the FM to the helical AFM phase. (Reprinted  from \cite{Ott2006}. Copyright 2006, American Institute of Physics.)}
\end{figure}

While the first two approaches allow to draw  qualitative conclusions  from the raw data without involved data analysis they, however, require a magnetic reflection inside the Ewald sphere at resonance. Approach  (i) has been applied to study depth-dependent AFM ordering in thin Dy metal films grown on W(110). Metallic Dy develops a long-period helical AFM superstructure below $T_{N}$=179 K, which undergoes a first order transition into a ferromagnetic structure at about 80 K. This transition is characterized by a pronounced  thickness dependence of its hysteretic behaviour, i.e.,  the hysteresis smears out with decreasing film thickness. The long period AFM structure gives rise to magnetic superstructure reflections well separated from the structural Bragg reflections similar to those observed from Ho metal described above, which can be easily measured in the vicinity of the Dy $M_{4,5}$ resonance. Through this resonance not only the magnetic scattering strength varies but also the probing depth of the photons due to the energy dependence of the imaginary part of the index of refraction. Figure \ref{p_ott2006} shows the temperature dependent intensity of this AFM reflection through the AFM-FM transition measured for two different photon energies, i.e., with different  probing depths of the photons. Distinctly different hysteretic behaviour was observed, which readily indicates a pronounced depth dependence of the growth of the helical AFM order with temperature. 
From modeling these energy-dependent hysteresis loops, taking the energy-dependent absorption into account, a temperature-dependent depth profile through this first order phase transition could be extracted shown in figure\,\ref{p_ott2006} \cite{Ott2006,Ott2010}. Obviously, upon heating the  helical order starts to grow in the near surface region of the film and develops deeper towards the W interface with temperature. This behaviour has been attributed to pinning of the FM phase at the interface by strain effects. Interestingly, also at the top surface, a phase of modified AFM order characterized by a tendency towards ferromagnetism survives over a large range of temperature.

\begin{figure}[t!]
\centering
\includegraphics[width=7 cm]{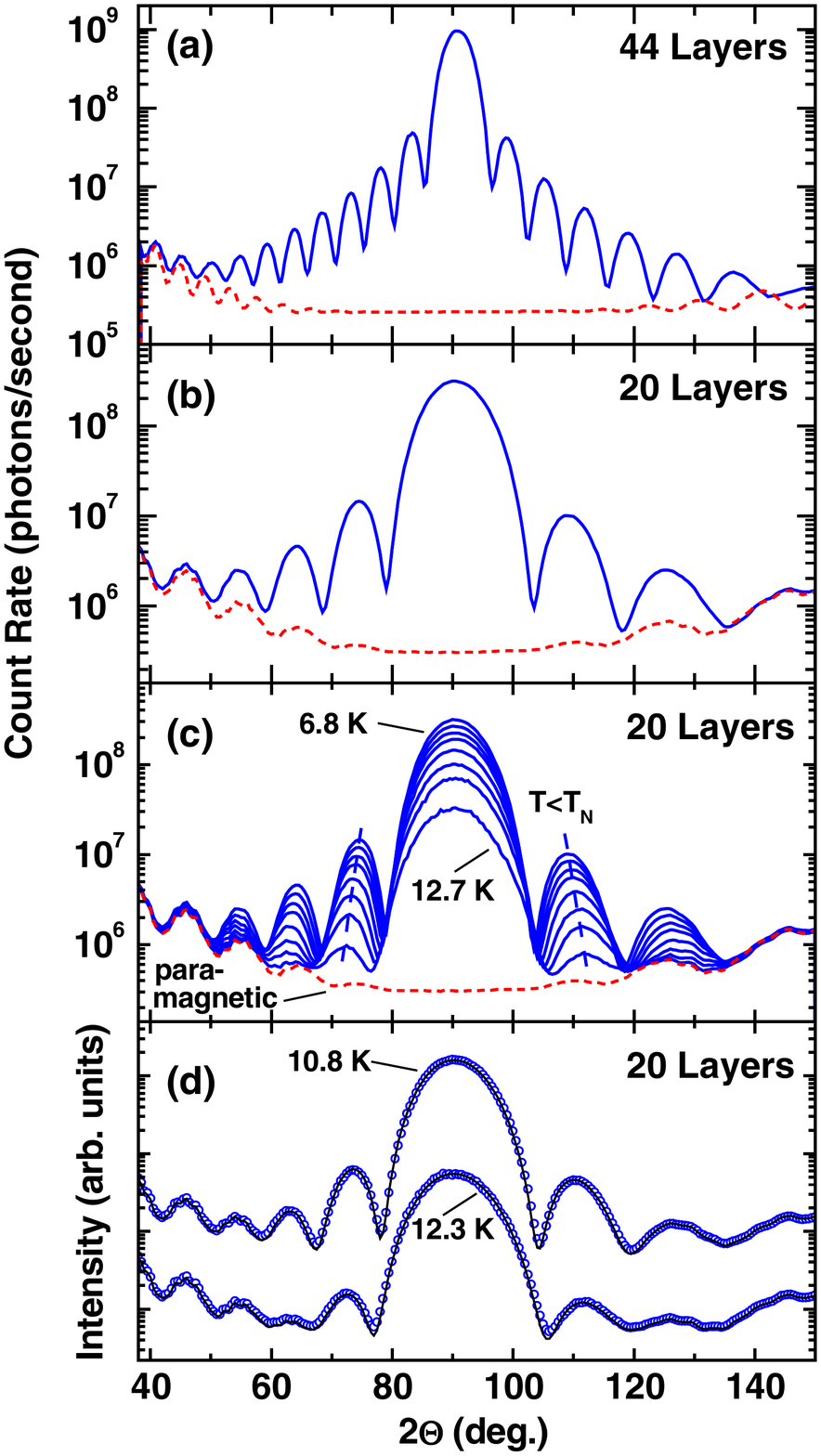}
\caption{\label{p_schierle2008} (Colour online)  Magnetic (0.5 0.5 0.5) Bragg reflections with magnetic Laue oscillations of  thin EuTe films with 44  monolayers (a) and 20 monolayers (b) thickness. (c) Data for a 20 monolayer thick film, recorded for various temperatures below (solid) and above (dashed) $T_{N}$; the photon energy was 1127.5 eV,  the Eu $M_5$ resonance maximum. (d) Two selected experimental diffraction patterns (dots) and results of calculations, taking magnetization depth profiles into account (lines).  (Reprinted  from \cite{Schierle2008}. Copyright 2008, American Physical Society.)}
\end{figure}

This way of achieving depth-dependent information requires a film thickness much larger than the minimum photon penetration depth at resonance and can hence not be applied to ultra-thin samples. In this latter case, however,  according to approach (ii), the shape of a superstructure reflection measured at a fixed energy already contains the information on the spatial modulation of magnetic order, as has been shown for magnetic scattering from the magnetic semiconducter EuTe \cite{Schierle2008}. Below $T_{N}$=9.8K, EuTe develops a simple AFM structure with ferromagnetically ordered (111) planes but alternating magnetization along the [111] direction. By virtue of the strong magnetic Eu $M_{4,5}$ resonance and the very high quality of the epitaxially grown [111] EuTe films, a very well resolved magnetic half order reflection could be observed from a film of only 20 monolayers,  with a peak intensity in the order of almost 1\% of the incident intensity. This magnetic reflection is already comparable to the off-resonant charge scattering from the superlattice peak of figure\,\ref{fig:multilayer}.  At resonance, this magnetic reflections appears very close to the Brewster angle where the charge scattering background for $\pi$ polarised incident light is almost completely suppressed. This ideal combination of magnetic period length and resonance energy in EuTe  therefore leads to a magnetic signal-to-background  ratio of several orders of magnitude, allowing measurements of the magnetic superstructure reflection with unprecedented quality. As can be seen in figure\,\ref{p_schierle2008}, the intensity of this reflection is not contained in a single peak but distributed over several side maxima caused by the finite size of the system (so-called Laue oscillations). The period of these oscillations displays a measure of the inverse   number of contributing magnetic layers and the overall envelope functions depends sensitively on the magnetization depth profile. By virtue of the  high magnetic contrast in this study, the side maxima could be resolved over a large range of  momentum transfer revealing distinct changes with temperature: with increasing temperature the period of the Laue oscillations increases and the intensity of the side maxima compared to the central peak decreases. This behaviour  signals a non-homogeneous decrease of the magnetic order through the film, such that the effective magnetic thickness decreases with increasing temperatures, i.e., the order decays faster at the film boundaries. 

From a kinematical modeling of the observed intensity distributions, temperature-dependent atomic-layer-resolved magnetization profiles across the entire film could be extracted. Even at low temperatures, these profiles displayed reduced order at the film interfaces, which is due to some chemical intermixing. In addition, the profiles reveal a faster decrease of the order close to the film boundaries with temperature, i.e., a layer-dependent temperature dependence of the magnetic order is observed. Due to the half filled 4f shell of Eu$^{2+}$  ions, EuTe exhibits almost pure spin magnetism caused by strongly localised moments and can therefore be regarded as a Heisenberg model system, for which this type of surface-modified order was theoretically predicted\,\cite{Binder1974}.

\begin{figure}[t!]
\centering
\includegraphics{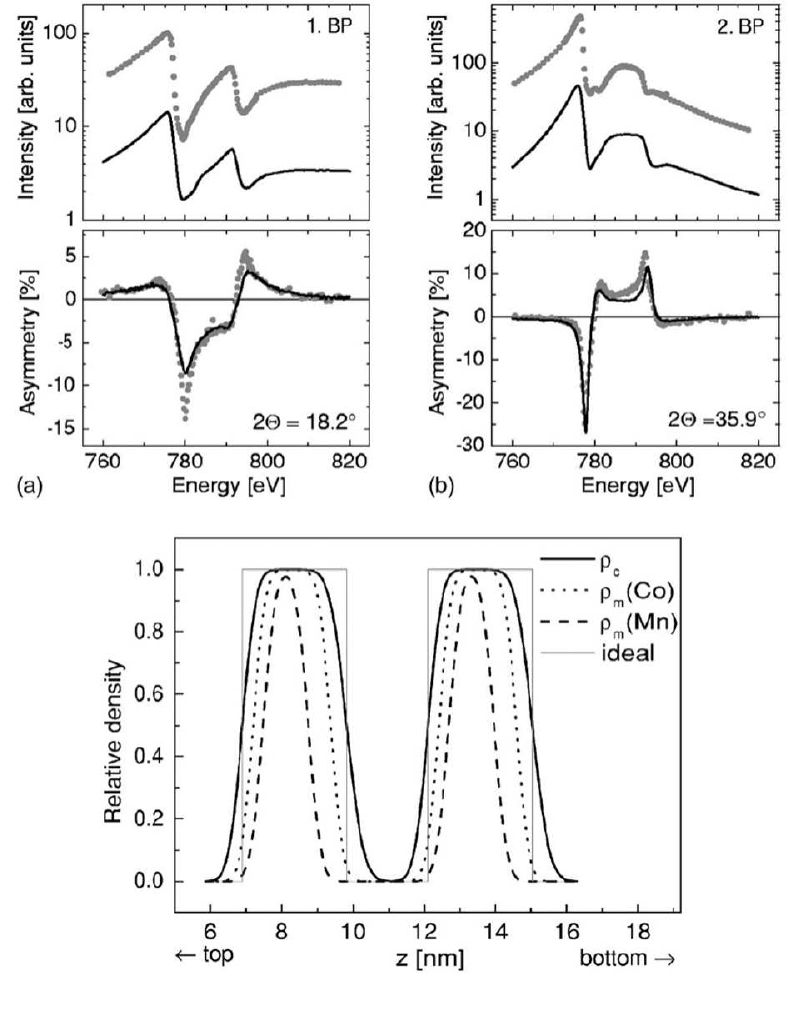}
\caption{\label{p_grabis2005} Top: charge intensities (upper panels) and magnetic asymmetries (lower panels) of the first- and second-order
Bragg peaks recorded at the Co $L_{2,3}$ absorption edges from a Co$_{2}$MnGe/Au multilayer. The dots represent measured data; the lines are model calculations as described in \cite{Grabis2005}. Bottom: structural and magnetic depth profiles of Co and Mn as determined from the model calculations. (Reprinted with permission from \cite{Grabis2005}. Copyright 2005, American Physical Society.)}
\end{figure}

\begin{figure}[t]
\centering
\includegraphics{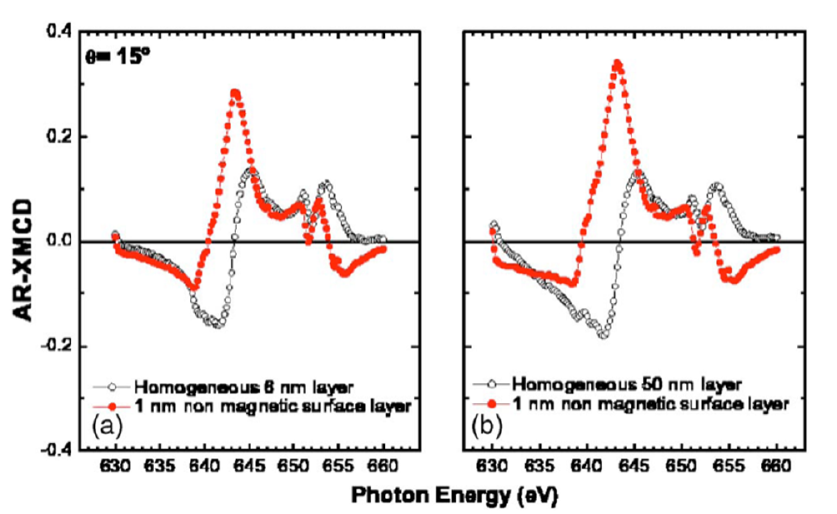}
\caption{\label{p_valencia2007} (Colour online) Simulated XMCD  spectra for two LCMO layers
on strontium-titanate  with thicknesses of (a) 6 and (b) 50 nm. The presence of a 1 nm nonmagnetic surface layer strongly affects the spectral
shape irrespective of the film thickness, making the strong sensitivity of  XMCD in reflection geometry to nonmagnetic layers evident. AR-XMCD here denotes the asymmetry value $\frac{I^{-}-I^{+}}{I^{-}+I^{+}}$, where $I^{+}$ and $I^{-}$ refer to the measured intensities for opposite directions of the external magnetic field. (Reprinted with permission from \cite{Valencia2007}. Copyright 2007, American Institute of Physics.)}
\end{figure}

Modeling of Q-dependent intensity distributions is not restricted to magnetic superstructure reflections but can also be applied in a more general way to the specular reflectivity from systems without any magnetic or chemical reflection inside the Ewald sphere at a soft x-ray resonance, e.g. ferromagnetic films, in this way opening the method to a broader field of applications. This has been  successfully demonstrated for hard x-ray resonant scattering\,\cite{Seve1995,Geissler2001,Jaouen2002}. In the soft x-ray range, in particular at resonance, specular reflectivity typically does not fall off so rapidly and hence can be measured over a large range of momentum transfer due to the strong optical contrast at the interfaces and surfaces. Information on ferromagnetic depth profiles usually requires  to compare reflectivities measured with different polarisations of the incident  x-ray beam or for opposite magnetization directions. Interpretation of such reflected intensities is often less intuitive and  requires detailed theoretical modeling as well as precise knowledge of the energy-dependent optical parameters. Based on an magneto-optical matrix algorithm developed by Zak et al.\,\cite{Zak1991,Zak1992}, several computer codes have been developed within the last decade for this purpose. In the pioneering early experiments, modeling of energy-dependent reflectivity data has been exploited to extract averaged magnetic properties, like the Fe magnetic moment in Fe$_{x}$Mn$_{1-x}$ thin films, which undergo a magneto-structural transition at x=0.75 \cite{Tonnerre1998}. Later studies improved data analysis to incorporate the effects of magnetic and chemical roughness as well as magnetization depth profiles \cite{Grabis2005, Bruck2008, Valencia2007, Valencia2008}. Examples of interest are ferromagnetic half-metals, like La$_{1-x}$Sr$_{x}$MnO$_{3}$ (LSMO) close to $x$=0.3 and a number of Heusler compounds. Half-metals are characterized by a band gap for only one spin direction and therefore are candidates to  show 100\% spin polarisation at the Fermi level, thus being of high interest for spintronic devices. Functionality of such materials crucially depends on the magnetic properties at  interfaces, which can be strongly altered by extrinsic effects like electronic reconstruction, disorder, roughness, or strain, but also by intrinsic changes caused by the broken translational symmetry. Therefore, obtaining depth-dependent information on the ferromagnetic properties of such materials at various interfaces is greatly  needed but challenging.

The interface magnetization depth profiles for Co and Mn moments in Co$_2$MnGe/Au multilayers were obtained from the analysis of element-specific RSXS\,\cite{Grabis2005}. By fitting the XMCD spectra  measured at the angular positions of the first three superlattice reflections, the individual magnetization profiles shown in figure\,\ref{p_grabis2005}  for the Co and Mn moments through the magnetic layers could be reconstructed. They are characterized by non-magnetic interface regions of different extension for the upper and lower boundaries and also differ for the two magnetic species leading to a rather complicated behaviour of the magnetization through a magnetic layer. This observation has been attributed to structural disorder caused by strain at the interfaces which affects Co and Mn spin in a different way since Co on a regular Mn position keeps its ferromagnetic spin orientation and full
moment but Mn on a Co position has an antiparallel spin orientation and a reduced moment \cite{Picozzi2004, Grabis2005}. 

With a similar scope, RSXS was applied to thin manganite films. Manganites show one of the richest set of phase diagrams among the transition metal oxides and  therefore offer a  high potential for application in future functional heterostructures.  Freeland et al.\,\cite{Freeland2005} showed the existence of a  thin non-ferromagnetic and insulating surface layer with a thickness of only one Mn-O bilayer on a single crystal of layered LSMO by analyzing its energy-dependent reflectivity at the Mn $L_{2,3}$ resonance combined with XMCD and tunneling probe techniques\,\cite{Freeland2005}. This finding renders the material a natural magnetic tunnel junction. The sensitivity of RSXS to weak magnetic modifications like a thin non-magnetic layer in manganites is nicely demonstrated in a study of   La$_{1-x}$Ca$_{x}$MnO$_{3}$ (LCMO) films for different thicknesses and substrates by Valencia et al.\,\cite{Valencia2007}. By  analyzing the XMCD spectra,  measured on the specular reflectivity for fixed scattering angles,  detailed knowledge of the chemical and magnetic surface properties could be obtained. In particular, all samples studied developed a thin surface region ranging from 0.5 to 2 nm with depressed magnetic properties. Figure \ref{p_valencia2007} shows simulations of the XMCD spectra  depending on the presence of a magnetic dead layer, revealing distinctly different energy profiles irrespective of the film thickness.  In the same way, Verna et al. very recently reconstructed the magnetization depth profile with high spatial resolution from thin LSMO samples, again, observing a non-ferromagnetic surface region of about 1.5 nm thickness which expands with increasing temperature \cite{Verna2010}. The observation of such non-ferromagnetic dead layers at manganite surfaces for LCMO as well as LSMO single crystals and films of different thickness, independent of the particular epitaxial strain strongly suggests that the observed modified surface magnetization is an intrinsic property of manganite surfaces \cite{Valencia2007}. This conclusion is further supported by an  XAS study exploiting linear dichroism at Mn $L_{2,3}$ resonance from thin LSMO films which observed an orbital reconstruction, i.e., a  rearrangement of the orbital occupation at the surface consistent with a suppression of the double-exchange mechanism responsible for the ferromagnetic properties \cite{Tebano2008}.  

The latter examples of modified, mostly suppressed magnetic properties at surfaces and interfaces, are  important results in view of possible application of materials in heterostructures. However, the proximity of two materials in heterostructures can cause much stronger and more complex changes of magnetic properties at those buried interfaces, which has been studied by RSXS intensively in the last decade.

\subsection{Interfaces}

Interfaces are presently of high interest since here direct control of material properties is possible by several different mechanisms. In particular, the proximity of two different materials at an interface can generate phenomena not present in the individual materials. While this holds in general for a rich variety of electronic phenomena, it is already true for two adjacent magnetic systems. 

\subsubsection{Magnetic Interfaces}\hfill\\
\\
\begin{figure}[t!]
\centering
\includegraphics{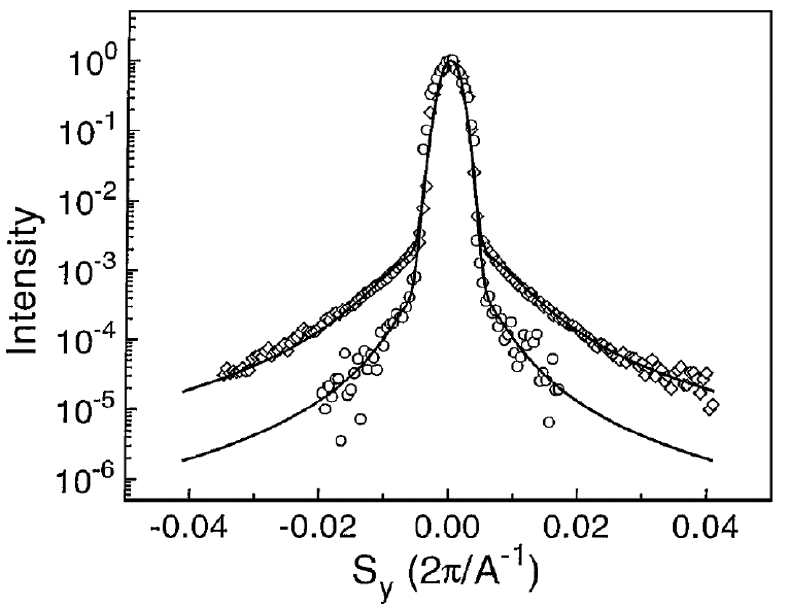}
\caption{\label{p_mckay96} Transverse scans normalized to the maximum of the specular reflectivity  from a 7 nm Co film capped with Al recorded at the Co $L_3$ resonance. Diamonds: average signal from two measurements with opposite magnetization,  representing the correlation of the chemical structure. Circles: difference signal of the two measurements with opposite magnetization, representing the magnetic contribution. The faster decay with $S_y$ in the latter case yields smaller magnetic roughness. (Reprinted with permission from \cite{MacKay1996}. Copyright 1996, American Physical Society.)}
\end{figure} 

\begin{figure}[t!]
\centering
\includegraphics{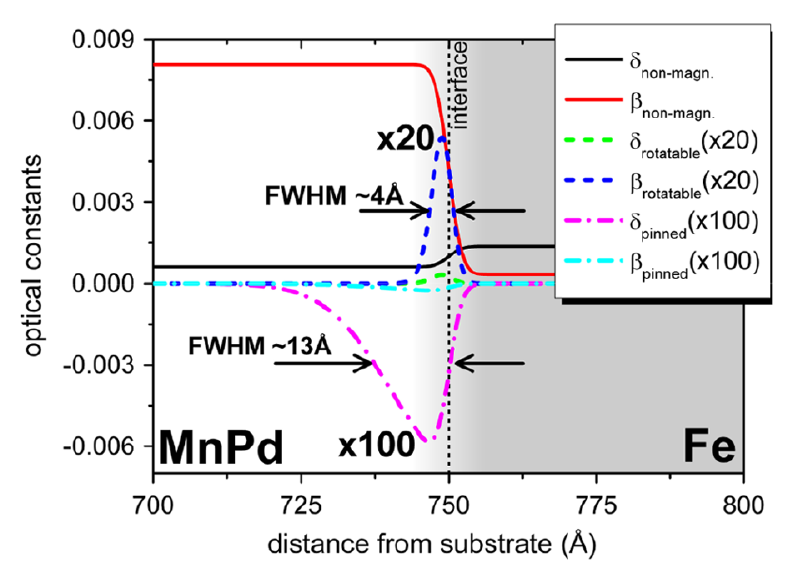}
\caption{\label{p_bruck} (colour online). Profile of the optical parameters $\delta$ and $\beta$ across the interface of the MnPd/Fe exchange bias system, separated into structural non-magnetic (red, black) and various magnetic contributions.  Rotatable Mn moments lead to the
dashed green ($\delta$) and dashed blue ($\beta$) curve. Dashed-dotted magenta and light-blue curves: same for the pinned Mn moments. (Reprinted with permission from \cite{Bruck2008}. Copyright 2008, American Physical Society.)} 
\end{figure}

 Famous examples of interface-generated magnetic phenomena with high technological relevance are giant magneto resistance and exchange bias. The experimental challenge of characterizing such phenomena is due to the extremely small amount of contributing material, the presence of more than one type of magnetic moment, and the buried character of the region of interest. Also here, RSXS has been shown to be a very sensitive tool to study such interfaces since its huge charge and magnetic scattering strength is accompanied by  a moderate penetration depth, element specificity and access to structural information. One crucial parameter that can influence interface-generated functionality is  chemical and/or magnetic roughness, which for a buried interface can be characterized best by scattering techniques measuring the related off-specular scattered intensity distributions. Such diffuse scattering contributions, however, are typically several orders of intensity weaker than the specular reflectivity and require an extraordinary magnetic scattering strength to be detectable. 

The first demonstration of how diffuse RSXS intensities can be exploited to track magnetic roughness was reported by McKay et al. for Co films and Co/Cu/Co trilayers \cite{MacKay1996}. In this study the diffuse intensity around the specular reflectivity was measured at the Co $L_3$ resonance   containing contributions from magnetic and chemical roughness. In order to separate the two quantities, the intensity was measured with circular polarised x-rays for two opposite magnetizations of the sample. In this case the sum of the two measurements represents the chemical roughness while the difference was interpreted as a measure of the magnetic roughness.  From the intensity and width of such broad intensity distributions, properties like the root mean square roughness (intensity) and correlation lengths (width) can be extracted independently for the  magnetic and the chemical interface. Interestingly, the magnetic signal drops much faster with increasing momentum transfer $S_y$ than the chemical intensity as can be seen in figure\,\ref{p_mckay96}. This observation yields a surprising result, namely a much smoother magnetic interface if compared with the chemical roughness. It was explained by a weaker coupling of the Co moments in the intermixing region compared to  that of the homogeneous part of the film.  This means that  the moments connected with chemical roughness do not contribute to the signal from the film magnetization and the magnetic interface therefore appears smoother than the chemical one. Similar results have been obtained by Freeland at al. who studied systematically the relation between magnetic and chemical interfacial roughness by monitoring the diffuse scattering from a series of CoFe alloy films with varying chemical roughness\,\cite{Freeland1998,Freeland1999}. Here, it was also shown that the diffuse intensity can be used to measure magnetic hysteresis loops of bulk and interfacial magnetization independently \cite{Freeland1998}.  It has been shown later, that the difference signal contains charge-magnetic interference contributions, such that it does not represent the pure magnetic roughness \cite{Osgood1999}. The general result, however, that  the  magnetic and chemical roughness can be very different,  is not affected. It was rather supported later by a study of Hase et al., who separated chemical and magnetic roughness in a different way  by comparing the diffuse scattering in the vicinities of either a chemical or half-order pure magnetic reflection from an antiferromagnetically coupled [Co/Cu]$_{n}$ multilayer \cite{Hase2000}. The observation of different roughnesses of the chemical and magnetic interfaces shows that the understanding of macroscopic magnetic phenomena and their  relation to roughness can not be simply obtained from the  knowledge of the chemical roughness alone. Nevertheless,  Grabis et al. observed almost identical chemical and magnetic roughness parameters in a Co$_{2}$MnGe/Au multilayer, showing that different roughness are not a general feature of magnetic interfaces \,\cite{Grabis2005}.  Theory describing diffuse resonant scattering from chemically and magnetically rough interfaces beyond the kinematical approximation has been published by Lee et al.\,\cite{Lee2003}.     

The proximity of two different magnetic systems can cause more complex changes of the magnetic properties than just introducing disorder, as discussed in the previous paragraphs. A prominent example, connected with functionality already applied in technological applications  is the exchange bias effect. 
Exchange bias is the observation of a shift of the  hysteresis along the field axis of a FM material in contact with an antiferromagnet (AF) after field-cooling through the N\'eel temperature. The general mechanism was identified to be the exchange interaction between the spins of the FM and the AF at the interface, which generates an additional unidirectional anisotropy for the FM spins \cite{Meiklejohn1956, Meiklejohn1957}. However, the original models assuming perfect interfaces failed to describe exchange bias and related phenomena in a quantitative way. Rather,  the exchange bias strongly depends on details of the magnetic structure at the interface. Exploiting XMCD techniques, rich knowledge on the magnetic behaviour of the interface was obtained. In particular the presence of a FM component within the AF including a pinned fraction that is   related to the exchange bias was observed \cite{Ohldag2003}. Due to the fact that interfaces are buried, techniques not limited to the top sample surface region are of advantageous. Therefore, a large number of experiments exploited the XMCD on the specular reflectivity to characterize the element specific behaviour of spins in exchange bias  systems \cite{Hase2001,Radu2006, Radu2009, Mishra2009}. Such studies revealed the complex magnetic situation at the interface connected with exchange bias. In particular, several different types of spins are involved, namely the FM spins, the compensated AFM moments as well as uncompensated AFM moments that  can be divided into a pinned and a rotatable fraction.  By exploiting the structural information contained in reflectivity data, detailed characterization of the depth dependence of the different magnetic contributions at the interface could be achieved. By varying the  absorption threshold, magnetization direction, photon polarisation and scattering geometry, RSXS can be tuned to be sensitive to almost all the magnetic ingredients of the magnetic interface. In this way Roy et al. determined the depth profile of FM and unpinned uncompensated AFM spins of the exchange bias system Co/FeF$_{2}$ by analyzing the specular reflectivities measured with constant helicity of the incident photon beam for opposite magnetization directions \cite{Roy2005}. They found the majority of the uncompensated unpinned AFM spins in a region of a few nanometers below the interface, characterized by AFM coupling to the spins of the ferromagnet. In addition to the magnetization direction, elliptical undulators allow to switch the light helicity, providing a detailed picture of the magnetic structure at the interface. In this way, Br\"uck et al. studied the interface  of the MnPd/Fe exchange bias system\,\cite{Bruck2008}. Figure\,\ref{p_bruck} shows the scenario resulting from an analysis  of respective reflectivity difference curves for either opposite external field directions or light polarisations, separating the pinned and unpinned uncompensated spins within the antiferromagnet.  According to this study, the major fraction of the uncompensated AFM spins located very close to the ferromagnet is rotatable and antiferromagnetically coupled to the ferromagnet, while the uncompensated AFM spins deeper in the AFM are predominantly pinned to the AF which finally generates the exchange bias effect. Similar results were obtained by Tonnerre et al. who extended the method to study the interface of a system with perpendicular exchange bias \cite{Tonnerre2008}. 

\begin{figure}[t!]
\centering
\includegraphics[width=15cm]{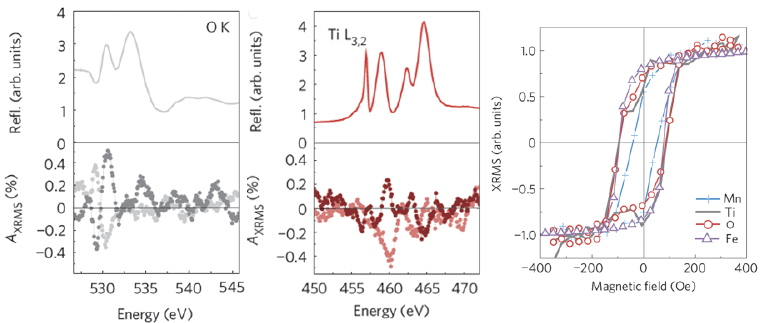}
\caption{\label{p_valencia2011} (Colour online) Absorption signal and circular dichroic asymmetry $A_{XRMS}$ observed in specular reflectivity from Fe/BaTiO$_{3}$ samples through the O $K$ (left) and Ti $L_{2,3}$ resonance as well as corresponding element-selected ferromagnetic hysteresis loops (right). They show a magnetic polarisation of the otherwise non-magnetic O and Ti ions by the Fe at the interface. (Reprinted with permission from \cite{Valencia2011}. Copyright 2011, Nature Publishing Group.) }
\end{figure}
Interesting interface-generated functionality in magnetic heterostructures can  emerge already  from a magnet in contact with a non-magnetic material. In a very recent study, Valencia et al. explored the influence of a ferromagnet in contact with a ferroelectric\,\cite{Valencia2011}. In this experiment, the structural information that can be extracted from scattering experiments was disregarded, still, it can be considered   a text-book example showing the sensitivity of XMCD-like experiments exploiting specular reflectivity rather than TEY or FY methods to study tiny magnetic effects at buried interfaces. In addition, it represents  an important example of interface-generated material properties\,\cite{Valencia2011}. 

In present solid state research, magnetic control of ferroelectric properties and, in particular, electric control of magnetic order are phenomena with very promising perspectives for applications.  Unfortunately, most of the known ferroelectrics are either not magnetically ordered or magnetism and ferroelectricity are weakly coupled. Exceptions are the so called improper ferroelectrics, where ferroelectricity is induced by complex mostly antiferromagnetic structures, which is discussed for bulk materials in more detail in section\,\ref{sec:Multiferroics}. These materials, however,  are often connected with very low ordering temperatures. In the quest for room-temperature multiferroicity, one way to overcome this limitation is to combine a room-temperature ferromagnet and a room-temperature ferroelectric in a heterostructure. Here, the chosen heterostructures consist of Fe or Co on BaTiO$_{3}$ (BTO)\,\cite{Valencia2011}. BTO is a robust diamagnetic room-temperature ferroelectric. The study could show that the tunnel-magnetic resistance of Fe or Co/BTO/LSMO trilayers strongly depends on the direction of the ferroelectric polarisation in the insulating diamagnetic BTO layer. Thus, these heterostructures possess coupled magnetic and ferroelectric, i.e.,  multiferroic, properties. Element-selective XMCD in reflection geometry could identify the key mechanism: the ferromagnet generates ferromagnetism in the topmost layer of BTO by inducing a magnetic moment in the formally non-magnetic O and Ti$^{4+}$ ions as revealed by a seizable asymmetry $A_{XRMS}$ displayed  in figure \ref{p_valencia2011}. The corresponding magnetic interface properties are very weak and hardly detectable by conventional XMCD techniques. Exploiting the dichroism in specular reflectivity through the Ti $L_{2,3}$ and O $K$ resonance, Valencia et al. have been able to unambiguously prove the existence of ferromagnetic moments at the Ti and O ions coupled to the magnetization of the ferromagentic layer. This observation shows the huge advantage of performing XMCD-like experiments exploiting scattered photons to study tiny effects at buried interfaces and, besides these methodological aspects, opens the door for future application of interface-generated room temperature multiferroicity. 
Modified macroscopic and/or microscopic properties at heterostructure interfaces are by far not limited to magnetic behaviour but can be found for other electronic phenomena as well, which offers even more fascinating new perspectives for future nanoscale devices. 

\subsubsection{New Material Properties}\label{sec:newmat}\hfill\\
\\
\begin{figure}[t!]
\centering
\includegraphics{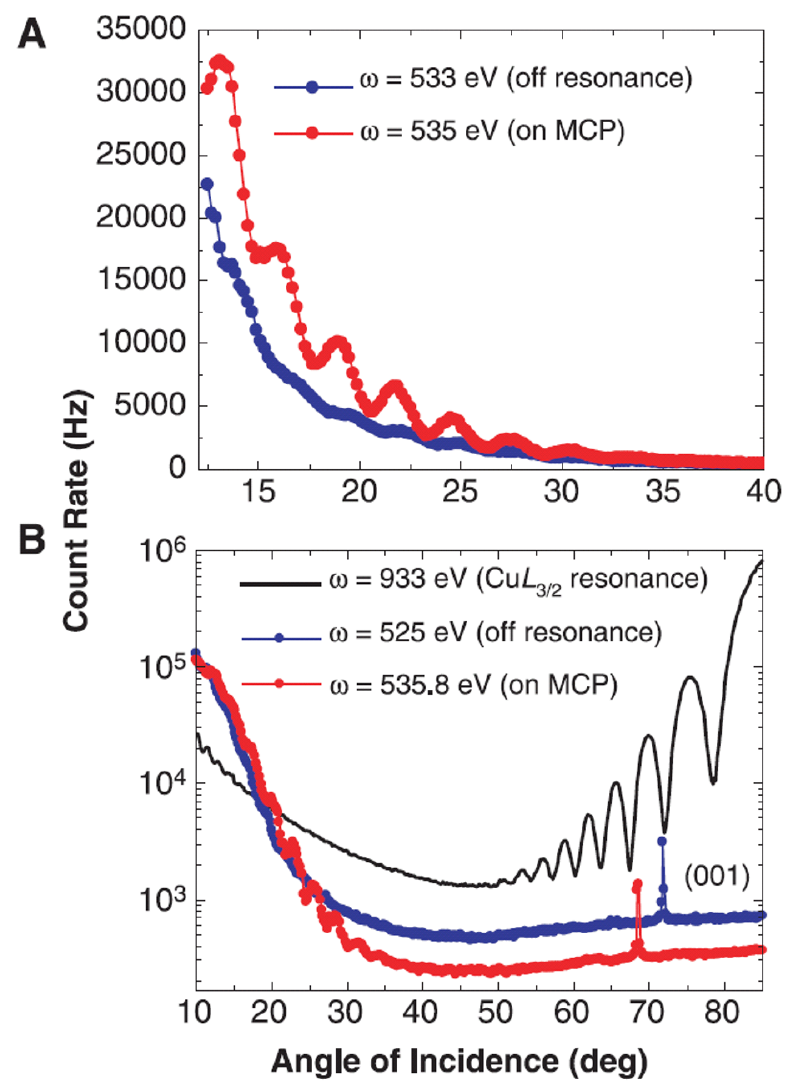}
\caption{\label{p_abbamonte2002} (Colour online) Reflectivity of a 23.2-nm La$_{2}$CuO$_{4+\delta}$ film (A) just below and on the
mobile carrier peak (MCP) at 535 eV , showing the resonant enhancement; and (B) over the full
angular range for three different energies. The faster decrease of the red curve compared to the blue one indicates a smoothing of the charge-carrier density at the interface. (Reprinted with permission from \cite{Abbamonte2002}. Copyright 2002, American  Association for the Advance of Science.) }
\end{figure} 

\begin{figure}[t!]
\centering
\includegraphics{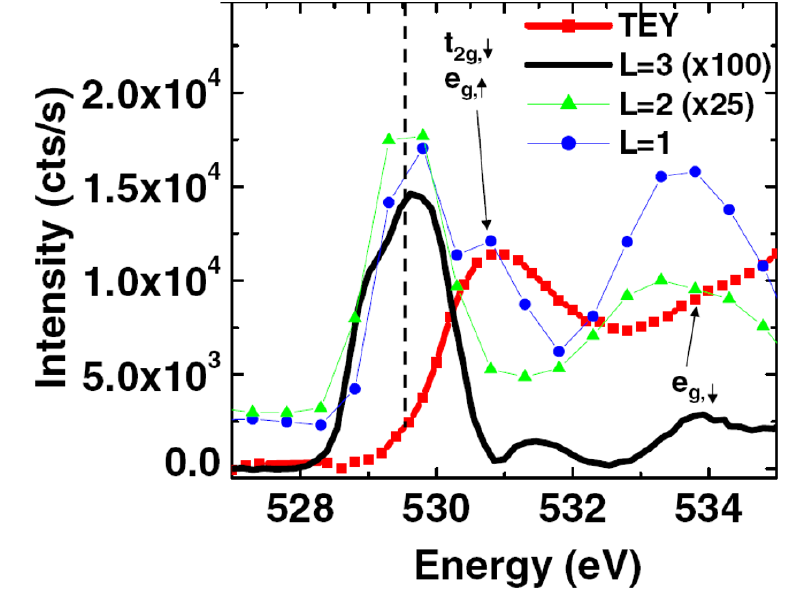}
\caption{\label{p_smadici2007} (Colour online) Energy dependence of (0 0 L)  superlattice reflections of  SMO/LMO multilayers at T=90 K: at L=3 (line), at L=1
(circles), and L=2 (triangles), compared to XAS data (squares,
aligned to zero below the edge). The strong resonant enhancement in the pre-edge region of the 
O $K$ resonance is indicative of an electronic effect. (Reprinted with permission from \cite{Smadici2007a}. Copyright 2007, American Physical Society.)}
\end{figure}

In ultra-thin samples or at heterostructure interfaces completely new phenomena may arise  that cannot be found in the corresponding bulk materials. An important and fast developing field of research was triggered by the discovery that interfaces of transition metal oxides can be grown with very high structural quality down to a flatness on the atomic level. With these structurally perfect interfaces, fundamental studies of new interface properties have become feasible. Examples are  the discovery of a highly mobile electron gas \cite{Ohtomo2004} and even superconductivity \cite{Reyren2007} at the interface of the two insulating perovskite oxides LAO and SrTiO$_{3}$ (STO). A recent example is the observation of insulating and antiferromagnetic properties of LaNiO$_{3}$ (LNO) in  LNO/LaAlO$_{3}$ (LAO) superlattices, as soon as the LNO thickness falls below 3 unit cells, while thicker LNO samples are metallic and paramagnetic \cite{boris2011}. Finally a major reconstruction of the orbital occupation and orbital symmetry in the interfacial CuO$_{2}$ layers at the (Y,Ca)Ba$_{2}$Cu$_{3}$O$_{7}$ / LCMO interface has been observed \cite{Chakhalian2007}. It is a matter of debate to which extent such properties are caused by either electronic reconstruction, strain, oxygen vacancies,  or disorder. Hence, microscopic understanding of macroscopic interface properties may require detailed knowledge on microscopic electronic properties with high spatial resolution. With its strong sensitivity not only to magnetic but also to  charge and orbital order, RSXS is well suited to extract quantitative information on even subtle changes of these electronic degrees of freedom at interfaces.

The general strategy  here  is very similar to the case of magnetic depth profiling: the information on spatially  modulated orbital occupation or charge distribution is contained in the reflectivities, measured as a function of momentum transfer and incident photon energy. This has been demonstrated first by Abbamonte et al. while studying the carrier distribution in thin films of La$_{2}$CuO$_{4+\delta}$ grown on SrTiO$_{3}$ \cite{Abbamonte2002}. The thin film reflectivities of these samples are characterized by well developed thickness oscillations  when measured at the Cu $L_{2,3}$ resonance, proving high sample homogeneity and surface/interface quality (see figure\,\ref{p_abbamonte2002}). In contrast, with a photon energy corresponding to the mobile carrier peak (MCD), the oscillations vanish at higher angles of incidence. This mobile carrier peak in the O $K$ pre-edge region  represents  a characteristic energy for scattering from the doped carriers with an enhancement of the scattering strength of a single hole by about two orders of magnitude. Hence, the reflectivity measured at this photon energy is almost entirely determined by the distribution of the doped carriers. The strong damping of the thickness oscillations at this specific energy is readily explained by assuming a smoothing of the carrier density at the interface towards the substrate.

In a very similar way Smadici et al. studied the distribution of doped holes in superlattices consisting of double layers of insulating La$_{2}$CuO$_{4}$ (LCO) and overdoped La$_{1.64}$Sr$_{0.36}$CuO$_{4}$ \cite{Smadici2009}. While none of these materials is superconducting, the heterostructure shows superconductivity below $T_{C}$=35\,K suggesting charge redistribution at the interface. In RSXS with the photon energy tuned to the La $M_5$ resonance, the superstructure gives rise to a series of  reflections detectable up to the 5th order. These reflections mainly contain information about the distribution of the Sr ions that induce the hole doping.  In contrast, at the mobile carrier peak energy only the first superstructure reflection could be observed. This result directly shows that, although being modulated with the superlattice period, the doped hole distribution does not follow that of the Sr ions. From a quantitative analysis, an average hole density in the insulating LCO layers of 0.18 holes per Cu site has been deduced, suggesting that superconductivity occurs in the formally insulating LCO layers.  

In order to clearly separate interface behaviour, often it is of advantage to design a multilayer such that a specific superstructure reflection is interface sensitive, i.e., its structure factor essentially is given by the difference of the optical properties of an interfacial layer and a bulk-like layer, in that way directly representing the changes at the interface. This approach has been demonstrated first by Smadici et al.\,\cite{Smadici2007a} who studied the interfacial electronic properties of SrMnO$_{3}$ (SMO) / LaMnO$_{3}$ (LMO). While the two single materials are a Mott insulator (LMO) and a band insulator (SMO), repectively,  ferromagnetic and metallic interface properties had been predicted  \cite{lin2006} and macroscopically observed \cite{Koida2002}. In this experiment, (0 0 L) superstructure reflections were studied in a heterostructure consisting   of  $m\times$LMO/$n\times$SMO double layers. $ m$ and $n$ were  chosen such that the L=3  reflection perpendicular to the surface only occurs if scattering from the interfacial and inner MnO$_{2}$ planes of LMO or SMO differ. A pronounced L=3 reflection from this structure could be observed only in the vicinity of the Mn $L_{2,3}$ resonance and in a very narrow energy region at the onset of the O $K$ resonance (see figure\,\ref{p_smadici2007}), i.e., involving electronic states in the vicinity of the Fermi level. The observation of no broken symmetry in the atomic lattice (i.e., no L=3 reflection off-resonance), but an interfacial reflection induced by  the unoccupied density-of-states near $E_{F}$ gives strong evidence that the interface is characterized by electronic reconstruction. The intensity of the L=3 reflection at the Mn $L_{2,3}$ resonance was identified to be of magnetic origin from  a rough azimuthal dependence of the scattered intensity. Comparison of the temperature dependence of the macroscopic properties (conductivity and magnetization) with the observed L=3 intensities at the O $K$ and  Mn $L$ resonances suggests that metallic and ferromagnetic behaviour is indeed interface generated driven by electronic reconstruction. A very similar approach has been used to characterize also the STO/LAO interface in detail very recently by Wadati et al.\,\cite{Wadati2011}.

\begin{figure}[t!]
\centering
\includegraphics{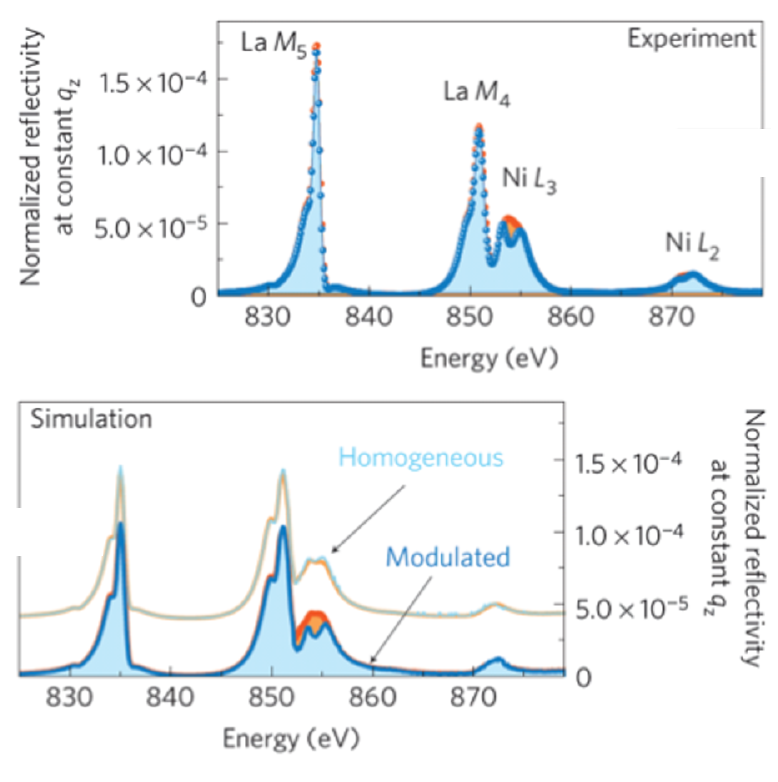}
\caption{\label{p_benckiser2011} (Colour online) Energy scans of the reflectivity from a LNO/LAO superlattice with constant momentum transfer $q_{z}$ close to the (002) superlattice peak using linearly polarised light. 
The top panel shows the
polarisation-dependent experimental data revealing linear dichroism at the Ni$L_3$ resonance. The bottom panel shows the corresponding simulated curves for LNO layers with (1) homogeneous orbital
occupation within the LNO layer stack and (2) modulated orbital occupation. (Reprinted with permission from \cite{Benckiser2011}. Copyright 2011, Nature Publishing Group.)}
\end{figure}

While the latter studies revealed interface-driven modifications in a more qualitative way,  it has been shown very recently that  quantitative information from such superstructure reflections can be derived for each individual atomic plane. This,  however, requires detailed modeling of energy- and polarisation-dependent reflected intensities as in the case of magnetic depth profiling discussed above.  The studied system is a multilayer made of repeated bilayers of four unit cells of LAO and four unit cells of LNO. Ni in cubic symmetry is characterized by a doubly degenerate $e_{g}$ level filled with one electron. According to model calculations, epitaxial strain can be utilized to favour the occupation of the in-plane $x^{2}-y^{2}$ orbital in a superlattice geometry, in this way matching the electronic structure of cuprate  high-$T_c$  superconductors , which may be a route to generate  high-$T_c$ superconductivity in artificial nanostructures \cite{Chaloupka2008,Hansmann2009,Han2010}. While magnetic depth profiling used the contrast obtained by circularly polarised light, light with linear polarisations $\pi$ and $\sigma$  (cf. figure\,\ref{fig:exp_scheme}) was used to measure the linear dichroism of the orbital scattering according to the last term in Eq.\,\ref{eqn:Hannon}.  By changing the light polarisation from in-plane to normal to the superlatice plane, the sensitivity to the $3z^{2}-r^{2}$ orbital could be varied. The observed linear dichroism  in absorption yields the average difference in occupation of the two $e_{g}$ orbitals of 5.5\,\%. In addition, the linear dichroism at the second superlattice reflection, sensitive to the difference of the scattering strength from the inner LNO unit cell and the interfacial LNO unit cell, was monitored. As shown in figure\,\ref{p_benckiser2011}, the observed linear dichroism, here, is much stronger than expected for a homogeneous system with the same orbital occupancy for the inner and interfacial LNO layers. Hence, an inhomogeneity of the orbital occupation through this 4 unit cells thin LNO layer can be readily deduced. Modeling the observed energy dependence for the two different linear polarisations, taking the average dichroism as an fixed input, the data could be explained by a distinct higher occupation of the $x^{2}-y^{2}$ orbitals at the interface (about 7 \%) compared to the inner layers (about 4 \%) , which is explained by the reduced gain of kinetic energy for the electrons in the interface layer by hopping across the interface due to the closed shell of the neighbouring Al$^{3+}$ ions. 

\subsection{Artificial Structures}

\begin{figure}[t]
\centering
\includegraphics{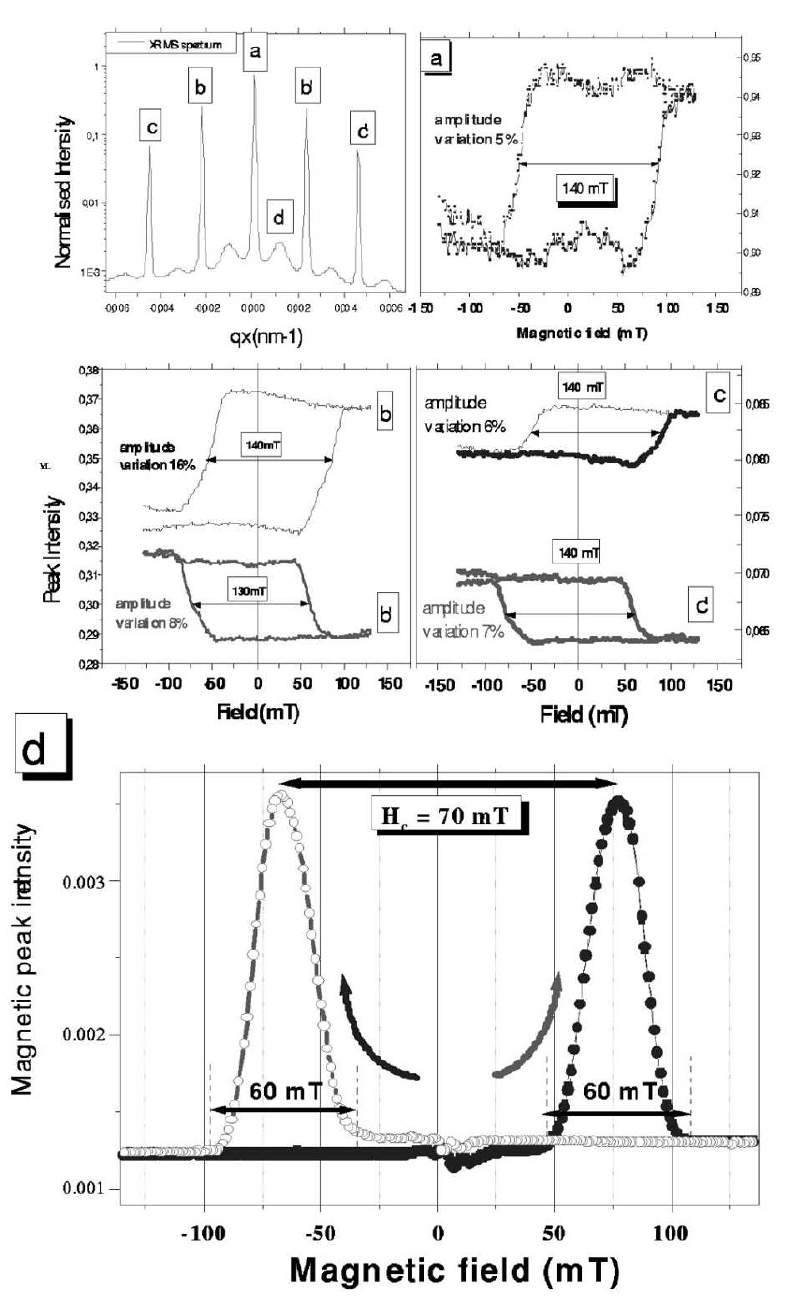}
\caption{\label{p_chesnel2002} Resonant scattering at the Co $L_3$ resonance from an  artificial Co/Pt multilayer structure on Si nanolines separated by 200 nm. Upper left corner: chemical superstructure and half-order magnetic superstructure peaks. Other panels:  evolution of the structural peaks (a, b, c) 
under a perpendicular magnetic field. The bottom panel displays the hysteresis measured on the half-order magnetic peak describing the entire hysteresis loop. (Reprinted with permission from \cite{Chesnel2002}. Copyright 2002, American Physical Society.)}
\end{figure}

The previous sections demonstrated  the capabilities of RSXS for studying ordering phenomena in nanostructures, with particular focus on multilayer samples where interesting physics typically emerge at the interfaces. Besides such depth-structured samples, systems can also be  lateral structured. This has been discussed for stripe-domains in FePd films above, but can be expanded  to artificially designed structures on a nanometer length scale, like regular or irregular line and dot arrays. Such regular patterns of magnetic materials are of interest in connection with  future data storage technologies. The typical length scales (nm) of nanostructures perfectly match the wavelengths of soft x-rays. Interestingly, RSXS has not been widely applied to laterally structured samples so far. The first application was presented  by Chesnel et al. monitoring the magnetization reversal of an regular arrangement of magnetic lines \cite{Chesnel2002}. These samples have been made of Si lines of about 200 nm width and 300 nm height with a line spacing of 75 nm covered by  Co/Pt multilayers. In contrast to the FePd stripe domains discussed above, this structure gives rise to  off-specular intensity rods even without magnetism (see figure\,\ref{p_chesnel2002}). Analyzing the in-plane momentum transfer $q_x$ as well as the width and the relative intensities of the superstructure-rods of different order yields  a detailed characterization of the chemical structure. At the Co $L_{2,3}$ resonance, additional magnetic information can be extracted from the scattered intensities. Here, ferromagnetically aligned neighbouring stripes contribute magnetic intensity to the chemical superstructure rods, while AFM aligned lines give rise to pure magnetic satellites at half order positions. Measuring  hysteresis loops  by recording the scattered intensity at various $q_x$ as a function of an external field  as shown in figure\,\ref{p_chesnel2002}, yields information about the magnetic coupling of such nanolines and   the magnetization reversal process could be modeled in detail. While the hysteresis for $q_{x}$=0 displays the average magnetic behaviour, as also accessible by macroscopic techniques, the off-specular magnetic rod intensity  reflects the magnetic behaviour of the well-ordered part of the sample. The occurrence of significant magnetic intensity at half-order positions for the external field close to the coercive field shows that the demagnetized state of this sample is in fact characterized by an AFM alignment of neighbouring lines with a correlation length of about 4 lines. Similar studies has been performed  on a regular dot array  of ferromagnetic permalloy dots \cite{Spezzani2004} as well as of permalloy rings, where the RSXS intensity distribution has been shown to be sensitive to the possible magnetization states of the single rings, i.e., vortex or so-called onion states \cite{Ogrin2008}.        

\subsection{Charge order}\label{sec:ChargeOrder} 
Ordering of mobile charge carriers is a phenomenon observed in a number of transition metal oxides, such as cuprates, nickelates,  cobaltates, and manganites. Many of the interesting  properties seem to be closely related to this kind of order. 
The most intriguing case is that of the cuprates, as inhomogeneous charge distributions  have been discussed for many years in connection with high-$T_C$ superconductivity (HTSC) in these materials. HTSC has fascinated nearly the whole solid state community, and the interest remained even 25 years after the discovery of HTSC, since there is no consensus on the mechanism for the superconducting pairing of the charge carriers in these compounds. Besides the mechanism of HTSC, it is the physics of electron correlations in general, which makes the cuprates so interesting, and charge order is one manifestation. Since in many cases, charge ordering involves only a small fraction of the total charge, the high sensitivity of RSXS to scattering from particular electronic states renders the method well-suited to study this phenomenon.
 
\subsubsection{\label{2D-cuprates}Two-dimensional cuprates} 
\begin{figure}[t!] 
\centering 
\includegraphics[angle=-90,width=7.5cm]{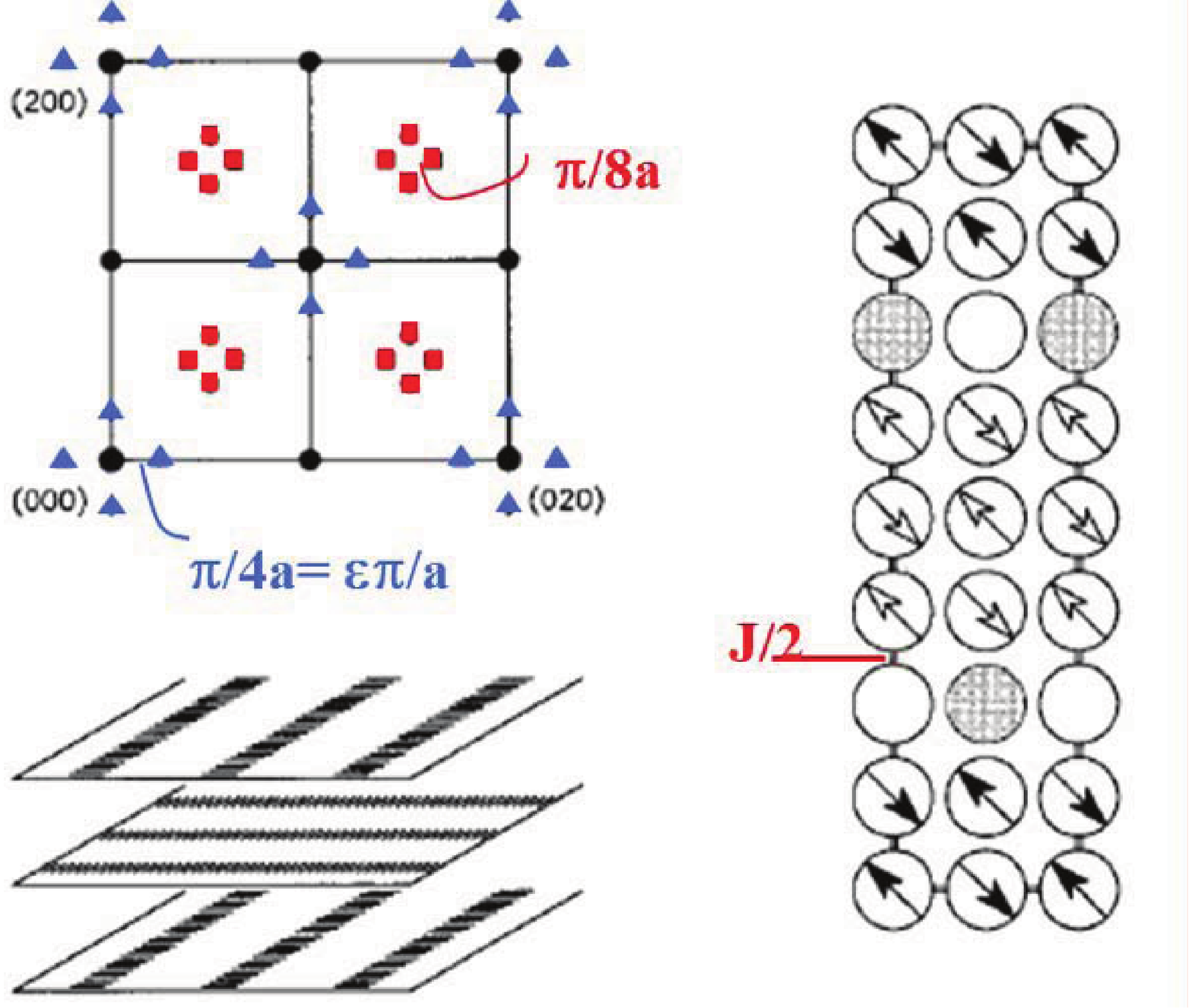}
\caption{(colour online) Upper left panel: diagram of the $(hk0)$ zone in reciprocal space.
Large filled circles, fundamental Bragg peaks; small filled circles,
superlattice peaks of the LTT phase. Red squares, magnetic
superlattice peaks from two different domains of the stripe
structure; blue triangles, charge-order superlattice peaks
from the two stripe domains. Right panel: model for the
stripe order of holes and spins within a CuO$_2$  plane at $n_h=\frac{1}{8}$.
Only the Cu sites are represented. An arrow indicates the presence
of a magnetic moment; shading of arrow heads distinguishes antiphase
domains. A filled circle denotes the presence of one dopant induced
hole centered on a Cu site (hole weight is actually on
oxygen neighbors). Lower panel: sketch showing relative orientation of stripe patterns
in neighbouring planes of the LTT phase.
(Modified versions of figures taken from Tranquada et al.\,\cite{Tranquada1996b}.)}
\label{Tranquada} 
\end{figure}

\hfill\\
\\
 HTSC in cuprates occurs after doping the AFM  Mott-Hubbard insulator\,\cite{Mott1949,Hubbard1964} with holes or electrons.  Since the onsite Coulomb
interaction $U$ on the Cu sites is considerably larger than the charge transfer energy $\Delta$ between Cu $3d$ states and O $2p$ states,  in the undoped system the gap  is essentially
determined by $\Delta$ and the holes are not formed on the Cu sites but on the O sites\,\cite{Zaanen1985}.
The doping-induced mobile holes are shared by four oxygen sites surrounding a divalent Cu site with  one hole in the $3d$ shell, in this way forming  the antiferromagnetically coupled
Zhang-Rice singlet\,\cite{Zhang1988} state. This state may be considered as an effective lower Hubbard band state.
The effective upper Hubbard band is formed predominantly by Cu $3d_{x^2-y^2}$ states
hybridised with some O $2p_{x,y}$ states.
The doping, which leads to a controlled metal-insulator
transition, occurs by block layers between which the CuO$_2$ layer are embedded. In
La$_{2-x}$Sr$_x$CuO$_2$ hole doping of the CuO$_2$ layers occurs via replacing the
trivalent La ions in the LaO block layers by divalent Sr ions. Long-range
AFM order disappears at $x$ = 0.05 and the highest superconducting 
transition temperature $T_c$ is reached at $x$=
0.18. In the overdoped case ($x>$ 0.18) the normal state can be well described by a Fermi liquid. 
For $x<$ 0.15 the normal state properties are far from being understood. 
There is no agreement what should be the minimal model which contains the basic physics of underdoped
CuO$_2$ layers. Several phases are found in this doping regime: an AFM  insulating phase, 
the high-$T_c$ superconducting phase, and a charge and spin ordered phase\,\cite{Vojta2009}. 
Regarding the latter two different phases are discussed in the literature: the checkerboard phase and the stripe-like phase.
In the stripe-like phase,  AFM  antiphase domains are separated 
by periodically spaced domain walls along the Cu-O directions 
in which the holes are situated (see figure\,\ref{Tranquada}). Partially motivated by the detection of incommensurate low-energy spin excitations, the existence of a stripe-like state was  proposed by 
theoretical work on the basis of a Hartree-Fock analysis of the one-band or three-band 
Hubbard model\,\cite{Zaanen1989,Machida1989,Schulz1990,Poilblanc1989}.
The proposed ground state resembles a soliton state in doped conjugated polymers\,\cite{Su1979}.
The theoretical predictions of a stripe phase were based on a mean-field approximation
and the importance of the long range Coulomb interaction, not taken into account in the
Hubbard model, was pointed out\,\cite{Emery1994}.
The phase separation, which seems that the AFM  background expels holes,
was supposed to have a strong influence on the physical properties of
doped cuprates: (i) a central ingredient to the pairing mechanism in HTSC, (ii) the non-Fermi-liquid behaviour
around optimal doping, and (iii) the existence of the pseudogap in the underdoped region.

The stripe-like structure  consists of three concomitant
modulations of the spin density, of the charge density, and of the lattice which is coupled to the charge modulation.  
The spin density modulation  has twice the wavelength of the
charge and lattice modulation, and in a scattering experiment, these two modulations give rise to superstructure reflections around the AFM Bragg peaks at  (0.5$\pm\frac{1}{2}\epsilon$,0.5,L) and around the structural reflection at ($\pm\epsilon$,0,L),  respectively, and symmetry related positions (see figure\,\ref{Tranquada}). Assuming half-doped stripes, independent of the average doping concentration, the propagation vector $\epsilon$ is determined by $1/n$ where $n$ is the average stripe distance given in $d$-spacings of the lattice. Using the same assumption about the constant doping concentration in the stripes, $\epsilon$ should be equal to 0.5$n_h$. Actually this is observed for $n_h\leq\frac{1}{8}$\,\cite{Yamada1998}, indicating  that in real space incommensurate stripes exist,  where the stripe distance varies on a local scale and where only the mean distance between the stripes is determined by $n_h$.  

In several doped cuprates such as La$_{2-x}$Ba$_x$CuO$_2$ (LBCO),  (La,Nd)$_{2-x}$Sr$_x$CuO$_2$, and (La,Eu)$_{2-x}$Sr$_x$CuO$_2$,
the stripe-like order is stabilized near $x=n_h=\frac{1}{8}$,
concomitant with a suppression of superconductivity. 
It occurs in the so-called low-temperature tetragonal (LTT) phase, which is characterized by a corrugated pattern caused by rotation of the CuO$_6$ octahedra.
Evidence for static stripes was  first detected in an elastic neutron scattering study on the system  
(La,Nd)$_{2-x}$Sr$_x$CuO$_2$, where the superstructure reflection for the spin and the lattice modulation
have been detected\,\cite{Tranquada1995}.

\begin{figure}[t!] 
\centering 
\includegraphics[width=7.5cm]{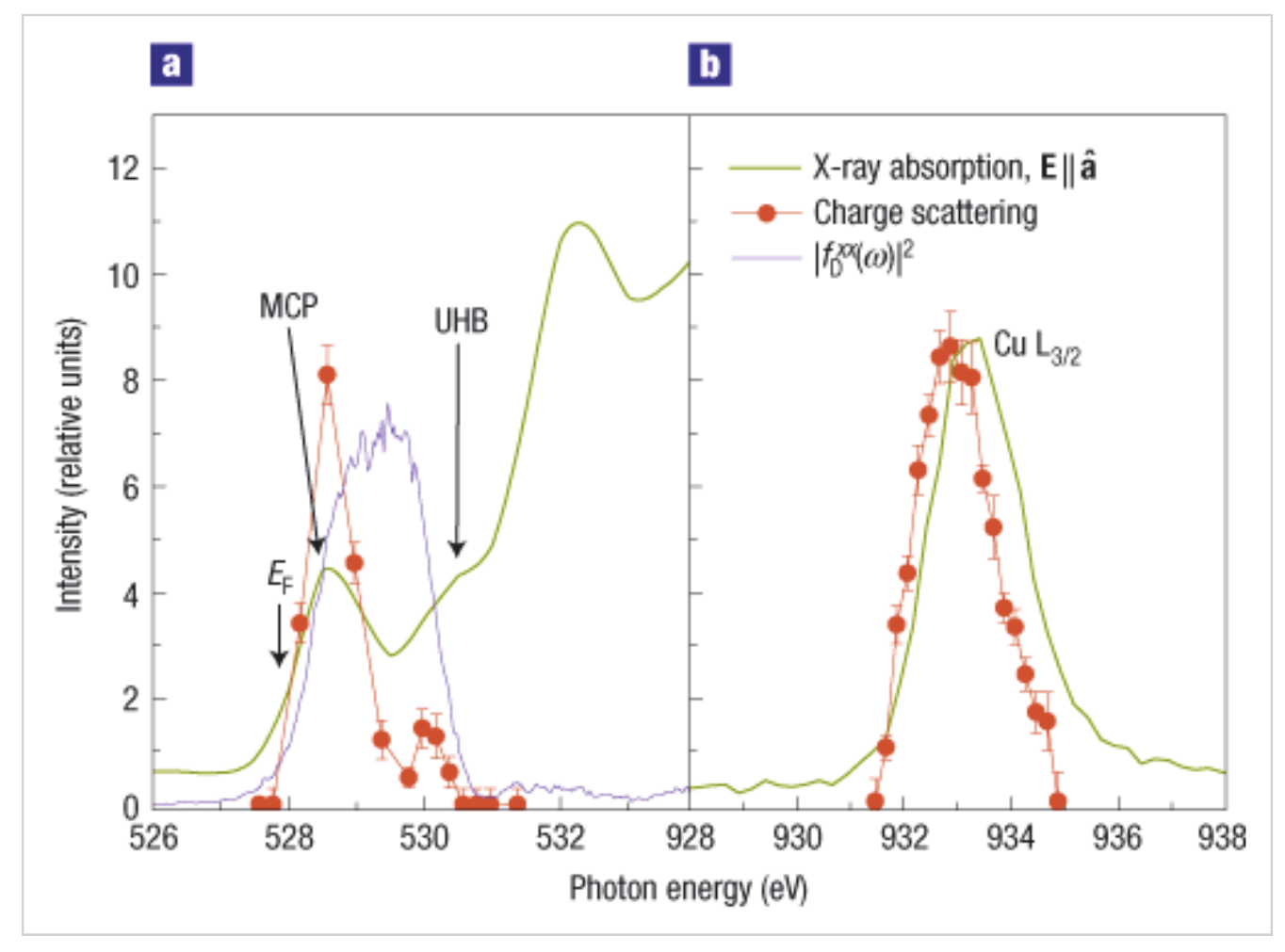}
\caption{(colour online) Comparison of RSXS and XAS data of LBCO.  (a) Data near the O $K$ edge. Green line: XAS
spectrum for photons polarised parallel to the CuO bond
direction. Red circles: intensity of the ($\frac{1}{4}$,0,L) superstructure reflection at L=0.72.
The enhancement at the mobile charge carrier peak (MCP) and the upper Hubbard band (UHB)  demonstrates a significant modulation of the
doped hole density. Blue line: form factor for scattering from doped holes calculated from XAS data. (b) Data near the
Cu $L_3$ edge for L=1.47.
(Reprinted with permission from \cite{Abbamonte2004}. Copyright  2005, Nature Publishing Group.)}
\label{LBCO_Abba} 
\end{figure}

Neutron scattering\,\cite{Tranquada1995} and non-resonant hard x-ray
scattering\,\cite{VonZimmermann1998} monitor the ordering of the charges 
indirectly by the associated lattice distortion. The reason
for this is that these techniques are mainly
sensitive to the nuclear scattering and the core electron scattering,
respectively. Detection of charge order by these methods is hence not fully conclusive, since lattice distortions may also occur with very small or even  no charge order of the valence electrons:
they may be caused as a result of a bond-length mismatch between different
units of a solid (e.g. planes or chains) or by a spin density wave alone like, e.g. in chromium metal. 
On the other hand, RSXS is a method which can directly probe the existence of the charge modulation of the
conduction electrons. Furthermore, this method enables one
to study the wave length of the modulation, the coherence length, the temperature dependence of the order parameter, and in principle also the momentum dependence of the form factor which would give the detailed spatial dependence of the charge modulation. This has been demonstrated 
for the doped cuprates by RSXS at the O $K$ and the Cu $L_{2,3}$
edges by Abbamonte et al.\,\cite{Abbamonte2002,Abbamonte2005}. In  figure\,\ref{LBCO_Abba} 
we show data from this work on LBCO\,\cite{Abbamonte2005}, comparing x-ray absorption data,
measured with the fluorescence method,  with
the resonance profile, i.e., the photon energy dependence of the intensity of the charge-order superstructure reflection at energies near a core excitation. 

Absorption edges at the O $K$ and Cu $ L_3$ edges in cuprates  have been extensively studied by
electron energy--loss spectroscopy (EELS)\,\cite{Nucker1988,Romberg1990,Fink1994}
and XAS\,\cite{Yarmoff1987,Chen1991,Pellegrin1993,Fink1994}.
The O $K$ edge  spectra of  hole-doped cuprates show two pre-peaks: 
the lower one at 528.6 eV  is assigned to  transitions into empty
O $2p$ states related to the Zhang-Rice singlet states.  The intensity of this pre-peak, also called the mobile
carrier peak (MCP)
is directly related to the number of the mobile charge carriers. 
The second pre-peak (UHB) at 530.4 eV is related to the upper Hubbard band.
With increasing doping concentration there is a spectral weight transfer from the upper
UHB peak to the MCP peak which is a classical signature of a doped correlated system\,\cite{Eskes1991}, 
not observed in doped semiconductors. 
The resonance profile for the ($\approx\frac{1}{4}$,0,L) superstructure reflection shown in figure\,\ref{LBCO_Abba}(a) for the O $K$ edge
exhibits a prominent resonance at the MCP, in this way providing a direct link to the doped holes and hence identifying the charge order. Interestingly, the reflection also resonates at the UHB energy, which means that also the 'Mottness' is spatially modulated. 
'Mottness' in this context means that the system acts like a Mott insulator. 
The resonance profile  for the
($\approx\frac{1}{4}$,0,L) superstructure reflection shown in figure\,\ref{LBCO_Abba}(b) shows also a strong enhancement
near the  Cu $L_3$ absorption peak. This resonance probably arises from the modulation of the Cu lattice\,\cite{Abbamonte2006}.
The charge-order reflections  are sharp along H but rod-like along L,  indicating quasi-long-range order in the CuO$_2$ plane but weak coupling between planes. 

In the work by Abbamonte et al.\,\cite{Abbamonte2005} not only the existence of a charge  modulation in LBCO has been directly demonstrated but also a quantitative analysis has been presented.
Analogous to the procedure outlined in section\, \ref{sec:RSXSandXAS}, the imaginary part of the form factor can be derived from x-ray absorption data and finally the real part can be derived via a  Kramers-Kronig transformation. The  form factor for that part which is related
to doped holes  is  shown in  figure\,\ref{LBCO_Abba}(a) by the blue curve.
 One realizes a strong enhancement of the calculated form factor near the two pre-peaks although the line shape does not perfectly agree with
the measured resonance profile. According to this analysis, the form factor for the scattering of a doped  hole at resonance has a value of 82, which means that a single hole in this compound scatters like a Pb atom off resonance. Since the scattered intensity is determined
by the form factor squared, this also means that at resonance the scattered intensity of a hole is amplified by a factor of  about 7000.
This fact demonstrates the enormous sensitivity of RSXS to the charge modulation of valence electrons which renders this method an ideal tool to study the charge modulation of doped holes in transition metal oxides.
The amplitude of the charge modulation was estimated to be 0.063 holes. In a one-band model, which only contains Cu sites, the  charge modulation for $x=\frac{1}{8}$ should be close to 0.5 holes. As outlined above, this value is expected for this doping concentration  in a simple model in which three Cu-O lines are not doped and the fourth line is 50\,\% doped. The relatively small experimental value of 0.063 hole modulation can be explained by the fact that the holes are distributed among many different (oxygen) sites.  

In figure\,\ref{LBCO_Abba}(b) XAS data at the Cu $L_3$ edge together with a resonance profile is presented. Here the enhancement of the scattering length was estimated to be about 300. 

Finally, the temperature dependence of the charge ordering in LBCO has been studied yielding a transition temperature $T_{CO}$ = 60 K which is the same as the transition temperature $T_{LTT}$ for a lattice transition from the low-temperature orthorhombic (LTO) to the low-temperature tetragonal (LTT) phase. In the latter structure, the CuO$_6$ octahedra are tilted along the Cu-O bond direction stabilizing the stripe formation.

A similar RSXS study has been performed  on the system La$_{1.8-x}$Eu$_{0.2}$Sr$_x$CuO$_4$ (LE§SCO)\,\cite{Fink2009,Fink2011}. Generally,
with decreasing ionic radius of the substitutes of La (here Eu), the tilt angle $\Phi$ in the LTT phase increases leading to a higher lattice transition temperature T$_{LTT}$. This stabilizes the stripe order and leads to a stronger suppression of high-$T_c$ superconductivity\,\cite{Buchner1994}. Charge-order diffraction peaks for LESCO could be observed over a range of doping levels $x$. Besides information about the phase diagram discussed  below, data for different  $x$ allow to extract more detailed information about the amplitude of the  hole doping modulation from the photon energy dependences according to the following considerations.  

\begin{figure}[t!] 
\centering 
\includegraphics[width=7.5 cm]{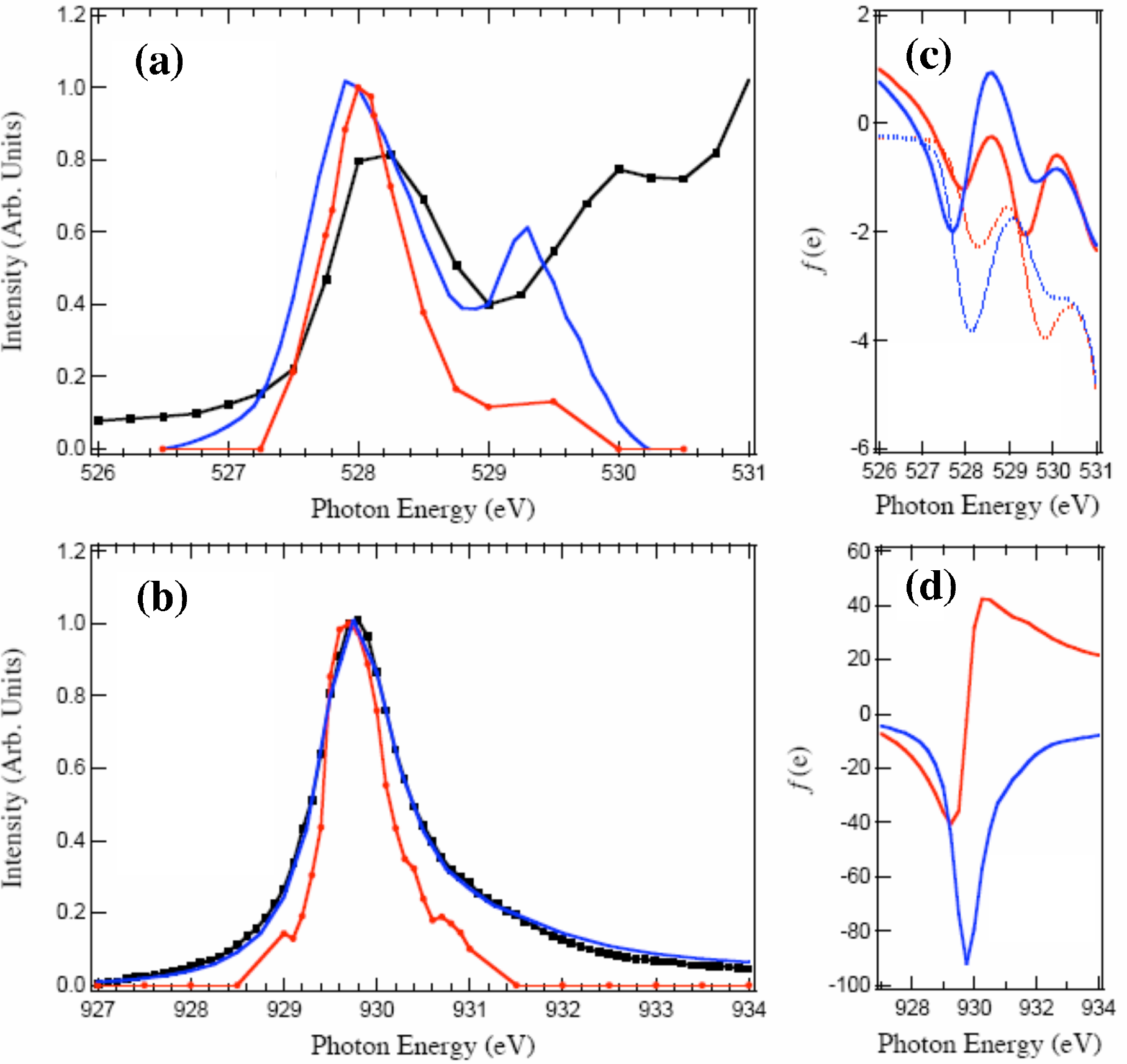}
\caption{(colour online)(a) [(b)] The resonant scattering intensity of LESCO in red as a function of photon energy through the O $ K$
[Cu $ L_3$] absorption edge for the stripe superstructure peak. Data were taken near the O $K$ [Cu $L_3$] edges for L=0.75 [L=1.6].
 Also shown is the x-ray absorption spectrum in black and the calculated scattering intensity in dark blue. (c) The real (solid line) and the imaginary (dotted
line)  parts of the atomic form factor $f$ at the O $K$ edge of LSCO for $x$=0.07 depicted in red and $x$=0.15 depicted in 
blue. (d) The real part in  red and the imaginary part in blue of the atomic form factor $f$ at the Cu $ L_3$ edge of LSCO for $x$=0.125.
(Reprinted  from \cite{Fink2009}. Copyright  2005, American Physical Society.)}
\label{LESCO1_Fink} 
\end{figure}

In  figure\,\ref{LESCO1_Fink} (a) and (b) LESCO data for $x$=0.15 are reproduced, showing XAS results, calculated intensities,  and a RSXS scattering profile. The calculated intensities were derived from the calculated form factors presented in figure\,\ref{LESCO1_Fink} (d) and (f) which in turn were evaluated from doping dependent XAS data as described above.
Assuming a linear variation of the atomic form factor $f_j$ as a function of the doping concentration $x$ one can expand 
 $f_j(E,x+\delta x_j)$= $f(E,x)+(\Delta f(E,x)/\Delta x)\delta x_j$. Then the intensity ratio between the lower Hubbard band (LHB) and the upper
Hubbard band (UHB) is $I_{LHB}/I_{UHB} = |[\Delta f(E,x)/\Delta x]_{LHB}/[\Delta f(E,x)/\Delta x]_{UHB}|^2$. A remarkable result,
typical of a correlated electron system,  is that due to a spectral weight transfer,  in the XAS data the intensity of the UHB decreases proportional to $1-x$
while the intensity of the MCP peak increases proportional to $2x$\,\cite{Eskes1991}. Since the absorption is related to the form factor, in first approximation,
the ratio for the form factor of the UHB to the LHB should be near 2, i.e., the intensity ratio in the resonant diffraction profile should be four, in agreement with
the calculated ratio [see figure\,\ref{LESCO1_Fink} (a)], but in clear disagreement with the measured resonance profile presented in the same figure.
This discrepancy was explained in terms of a non-linear change of the absorption at the UHB and the MCP observed for doping 
concentrations $x>0.2$\,\cite{Romberg1990,Chen1991,Chen1992,Pellegrin1993,Peets2009}. From this analysis it was concluded that the hole doping modulation per 
Cu site was larger than 20 $\%$ in agreement  with the modulation of about 50 $\%$ derived for LBCO mentioned above\,\cite{Abbamonte2005}. 

\begin{figure}[t!]
\centering
\includegraphics[width=7.5cm]{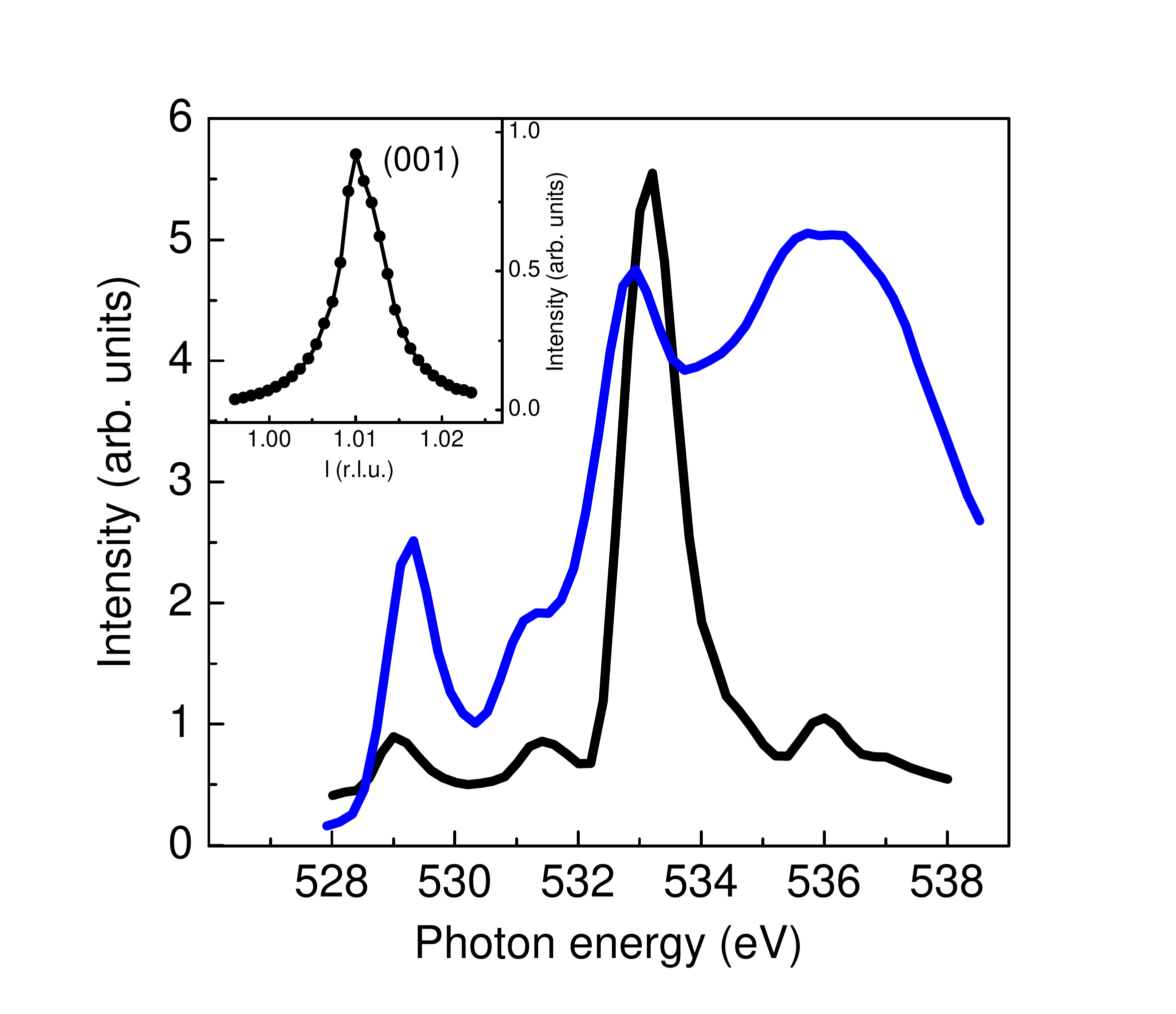}
\caption{(colour online)  X-ray absorption spectrum of La$_{1.65}$Eu$_{0.2}$Sr$_{0.15}$CuO$_4$ near the O $K$ edge measured in the
fluorescence yield mode  in blue. Also shown in black is the (001) superstructure reflection intensity as a function of
the photon energy. The inset shows the (001) reflection measured with a photon energy of 533.2  eV at $T$ = 6 K.
(Reprinted  from \cite{Fink2011}. Copyright  2011, American Physical Society.)}
\label{LESCO2_Fink}
\end{figure}

\begin{figure}[t!]
\centering
\includegraphics[width=7.5cm]{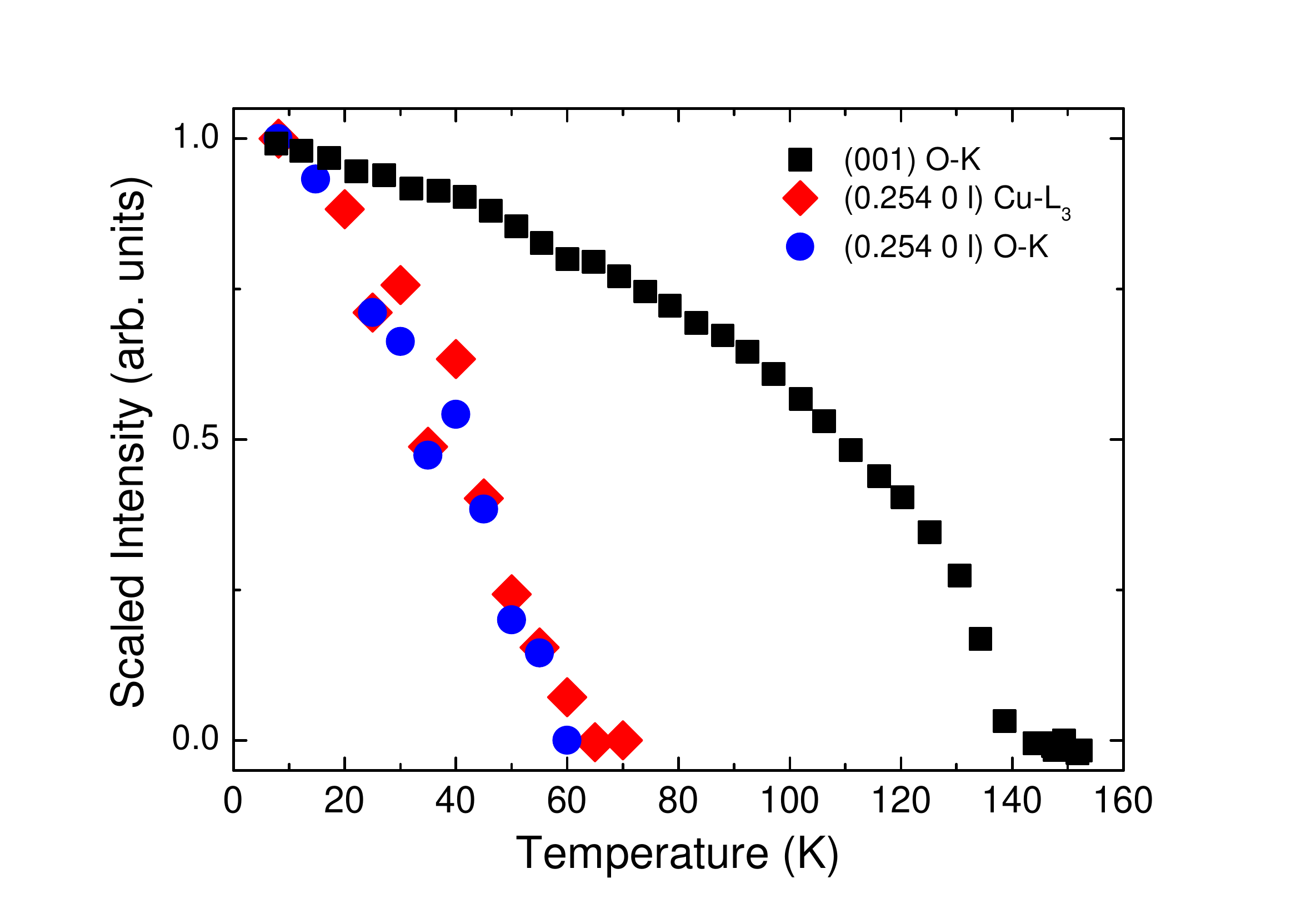}
\caption{(Colour online) Temperature dependence of the intensities
of the superstructure reflections of La$_{1.65}$Eu$_{0.2}$Sr$_{0.15}$CuO$_4$  normalized
to the intensity at $T$= 6 K. Squares: (001) reflection measured with
photon energies 533.2 eV near the O $K$ edge. Circles: (0.254 0
0.75) reflection measured with photon energies  529.2 eV near
the O $ K$ edge. Diamonds: (0.254 0 1.6) reflection measured with
photon energies 929.8 eV near the Cu $ L_3$ edge.
(Reprinted  from \cite{Fink2011}. Copyright  2011, American Physical Society.)}
\label{LESCO3_Fink} 
\end{figure}

In figure\,\ref{LESCO2_Fink} we illustrate that other resonances  near the O $K$ edge can be used to obtain information
on the structure of cuprates. There, besides the first two pre-peaks in the XAS spectrum,  discussed already before,
near 533.2  eV  a further peak appears due to a hybridisation of O $2p$ states with rare earth ($RE$ = La
and Eu) $5d$ and/or $4f$ states\,\cite{Nucker1988}. In figure\,\ref{LESCO2_Fink} we also show the
photon energy dependence of the (001) reflection measured in the LTT phase at $T$=6 K. In this phase, neighbouring CuO$_2$
planes are rotated by 90$^\circ$, yielding O sites with different (rotated) local environments. As a result, the (001) reflection
becomes allowed at resonance\,\cite{Dmitrienko1983}. The strong resonance at 533.2 eV is due to octahedral tilts, which cause different
local environments and affect the hybridisation between the apical O and the $RE$ orbitals. In the LTO phase, neighbouring
CuO$_2$ planes are just shifted, not rotated, with respect to each other. In this case the (001) reflection remains forbidden even
at resonance. Using the resonance at 533.2 eV it is possible the detect the LTO-LTT phase transition with soft x-rays. This
transition is not detectable  by the orthorhombic strain (a-b splitting),  because the (100)/(010)
reflections cannot be reached due to the limited range in momentum space in RSXS.

\begin{figure}[t!] 
\centering 
\includegraphics[width=7.5 cm]{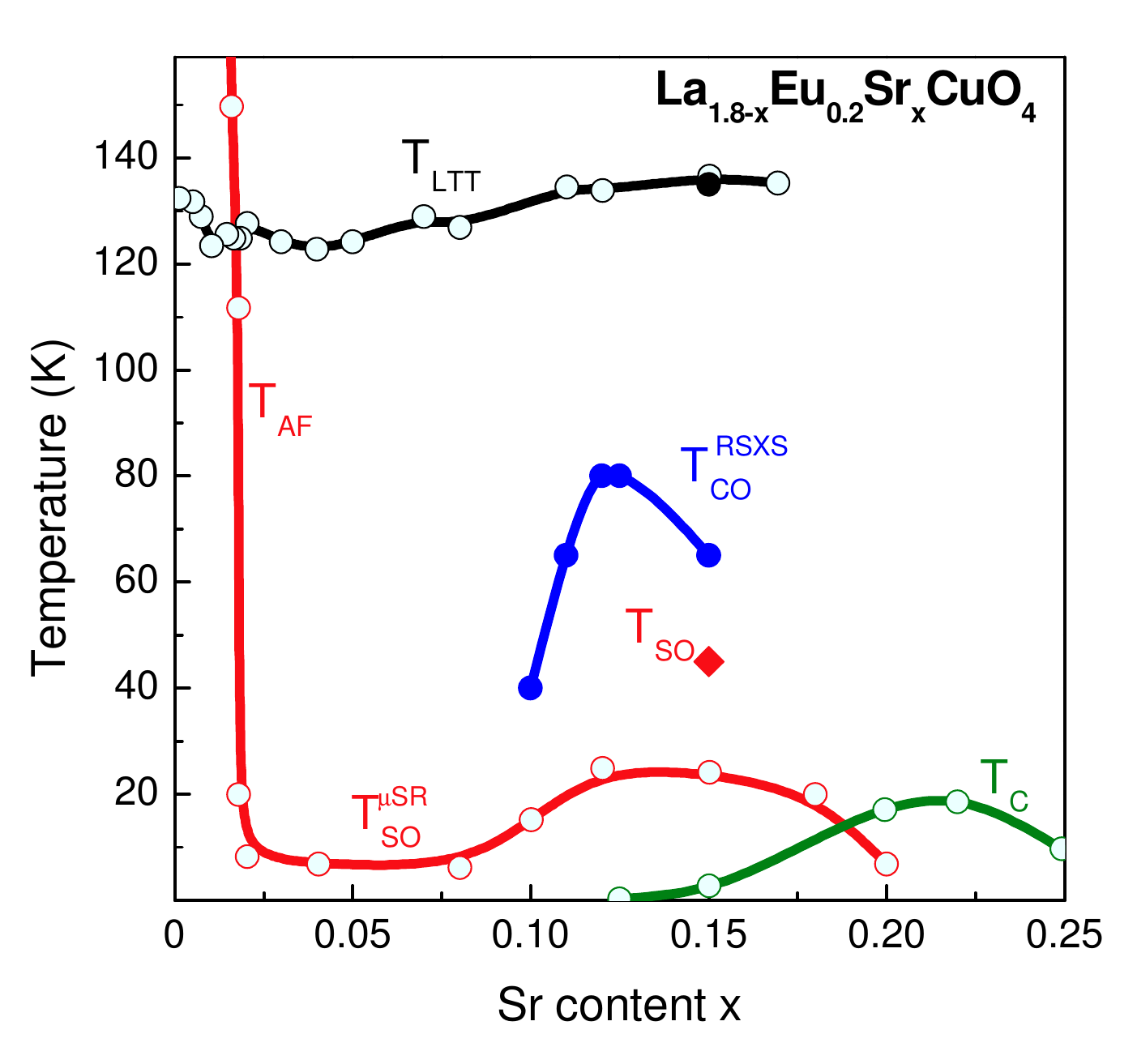}
\caption{(Colour online) Phase diagram of La$_{1.8−x}$Eu$_{0.2}$Sr$_x$CuO$_4$
showing transition temperatures for the LTT phase $T_{LTT}$, the antiferromagnetic
structure $T_{AF}$, the magnetic stripe order $T_{SO}$, the stripe
like charge order $T_{CO}$, and the superconducting transition temperature
$T_c$. Closed circles from  RSXS experiments\,\cite{Fink2011}.  Open circles
from Ref.\,\cite{Klauss2000}. Closed diamond from neutron diffraction data presented
in Ref.\,\cite{Hucker2007}.
(Reprinted  from \cite{Fink2011}. Copyright  2011, American Physical Society.)}
\label{LESCO4_Fink} 
\end{figure}

With the help of this resonance feature, it is possible to study the temperature dependence of the structural LTT order and the charge order in one experiment, as shown in figure\,\ref{LESCO3_Fink}. Here the   temperature dependent RSXS data of LESCO $x$=0.15 of the stripe superstructure reflection and the 
(001) reflection are depicted. These data   indicate a first order LTO-LTT transition at $T_{LTT}$=135 K and a charge order 
transition at $T_{CO}$=65 K,  clearly showing that the two phase transition are well separated. A systematic RSXS study of the doping dependence of charge order yielded the phase diagram for LESCO shown in  figure\,\ref{LESCO4_Fink}.

Different from  LBCO, the LESCO results represent the first example
in which the lattice transition temperature $T_{LTT}$ is so high that
the charge order can no longer be stabilized at this temperature
and therefore a gap of 55 K exists between $T_{LTT}$ and $T_{CO}$. This
is a remarkable result since it demonstrates that the charge
order is at least partially  electronic in origin.

The width of the superstructure reflections, being about five times larger than the experimental resolution,
indicate disorder effects and/or glassy behaviour. From the widths the coherence
lengths of the charge order have been determined. These lengths are of the order of about 100 lattice constants  and increase, at least for smaller
$x$, linearly with increasing Sr concentration. As a function of doping no maximum of the coherence lengths has been detected for integer numbers of stripe separation in units of the in-plane lattice spacing  corresponding to $n_h=\frac{1}{2n}$ with n being integer. This is different from doped nickelates where a clear maxinum was detected for $n_h=\frac{1}{3}$\,\cite{Yoshizawa2000}. The results on the cuprates indicate that in these systems the coherence length is not only related to the commensurability of the stripe lattice with the underlying CuO lattice. Furthermore the results point to the fact that the coherence lengths in these systems are not determined by the impurity potential of the doping atoms but is probably related to an increasing tilt angle with increasing Sr content.

Finally the wave vector of the superstructure reflection increases with increasing Sr content, at least up to $x$=0.125.
Since the nesting vector between parallel segments in the Fermi surface decreases with increasing doping\,\cite{Kordyuk2002},  this  indicates that
stripes are not  conventional charge density waves caused
by nesting. Thus these results point to more strong coupling scenarios for the stripe formation.

In a more recent work the stripe modulations in LBCO and La$_{1.48}$Nd$_{0.4}$Sr$_{0.12}$CuO$_4$ (LNSCO)
were compared using RSXS and hard x-ray scattering\,\cite{Wilkins2011}. Making use of a two-dimensional detector for RSXS an isotropic coherence length of the charge modulation of the hole density in the CuO$_2$ planes has been detected. Also the strain modulation of the lattice shows an isotropic coherence length. These results are  surprising given that the stripes in a single CuO$_2$ layer are highly anisotropic.
In the direction perpendicular to the CuO$_2$ planes, the stripes are weakly correlated giving rise to the uniform streak of intensity along L. Both the charge and the strain modulation is better correlated in LBCO than in LNSCO.
The in-plane  hole correlation lengths for LBCO and LNSCO are $255 \pm 5 \,\AA$ and $111 \pm 7 \,\AA$, respectively.
Similar to LESCO, in LNSCO the hole density modulation sets in well below the transition into the LTT phase while for LBCO $T_{CO}$ is equal to $T_{LTT}$ and below $T_{CO}$ the amplitude of the modulation is independent of the temperature. This suggests that  the LTT transition occurred at a higher temperature in
LBCO then it is likely that both the electronic and structural modulations would also have persisted to higher temperatures. A further result from this investigation is that the electronic charge stripe modulation in LBCO is 10 times larger in amplitude than in LNSCO. This result is surprising since from various other experiments, for LBCO, LNSCO, and LESCO a similar amplitude for the charge modulation is expected. Further work is required to solve this puzzle.

Until now we have reviewed RSXS studies on static stripe-like CDW order in LBCO, LNSCO and LESCO. In all these compounds the  CDW is stabilized by an LTT  lattice distortion. An important issue in this context is whether static stripes also exist in other two-dimensional cuprates. Checkerboard-like static charge order has been reported  in Ca$_{2-x}$Na$_x$CuO$_2$Cl$_2$ (NCCOC) using surface sensitive scanning tunneling experiments\,\cite{Hanaguri2004}.
CDW order  could be related to the electron pockets detected in underdoped YBCO in quantum oscillation measurements\,\cite{Doiron-Leyraud2007}. The observation of a  complete wipe-out of the Cu nuclear quadrupole resonance signal at low temperatures in Ni substituted NdBa$_2$Cu$_3$O$_{6+\delta}$\,\cite{Grafe2008}, similar to that of stripe ordered underdoped cuprates, suggests  the existence of static stripes in this compound.
Thus one may think that static stripes are generic to the cuprates and it is very important to look at stripe-like CDW order in other systems which are not stabilized by an LTT lattice distortion. 

A  RSXS study on a tetragonal compound, already discussed in section\,\ref{sec:newmat},  has been performed on an excess O doped La$_2$CuO$_{4+\delta}$  layer ($n_h=2\delta$) epitaxially grown on a SrTiO$_3$ crystal\,\cite{Abbamonte2002}.  In this work the resonance effects at the pre-edge of the O $K$ shell  excitation and near the Cu $L$ edge have been used for the first time to exploit charge modulation in the CuO$_2$ layers. It was there where it was demonstrated that at the pre-edge of the O $K$ shell excitations, the scattering amplitude for mobile holes is enhanced by a factor of 82 which leads to an amplification of diffraction peaks related to these holes by more than 10$^3$. Extensive non-successful search for superstructure reflections due to a stripe-like CDW lead to the suggestion of an absence of static stripes  in this high-$T_c$ superconductor.

A further RSXS study on a possible charge modulation in a tetragonal cuprate has been performed on NCCOC\,\cite{Smadici2007}. From experiments at the MCP peak at the O $K$ edge or at the Cu $L_{3}$ edge  no evidence for a checkerboard-like  charge modulation has been detected. From this null experiment the authors have concluded that the checkerboard order observed in the STS experiments is either glassy or nucleated by the surface. A further RSXS experiment on Ni substituted NdBa$_2$Cu$_3$O$_{6+\delta}$ was stimulated by the above mentioned Cu nuclear quadrupole resonance experiment\,\cite{Grafe2008}. No superstructure reflection due to stripe-like charge order could be detected at low temperatures in this compound\,\cite{Fink2010}.

More  RSXS experiments  on charge order have been performed on the most studied high-$T_c$ superconductor YBCO which contains a   CuO$_2$ double layer and in addition  CuO$_3$ chains along the $b$ axis. In a first paper on the
ortho-II YBa$_2$Cu$_3$O$_{6.5}$  phase, in which the CuO$_3$ chain layers are ordered into  alternating  full and empty chains, charge order in the planes and in the chains were reported\,\cite{Feng2004}. 
This observation was later questioned, partially by the same authors, in a more refined RSXS study on ortho-II  and ortho-VIII oxygen ordered YBCO\,\cite{Hawthorn2011}. However, charge order in underdoped YBCO was postulated on the basis of NMR studies in high magnetic fields~\cite{Wu2011} and was eventually detected in very recent experiments by RSXS~\cite{Ghiringhelli2012} and also with hard x-ray diffraction~\cite{Chang2012}. The former study showed by the resonance profile near the Cu $L_3$ resonance, that the charge order in fact takes place in the CuO$_2$ planes rather than the chains. The temperature dependence of the charge order found in both studies clearly revealed competition with superconductivity. This result has important implications for the understanding of the material and the cuprates in general. It readily explains the anomalously low superconducting temperature in underdoped YBCO and shows that charge order seems to be a generic feature of CuO$_2$ planes in layered cuprates. 

\subsubsection{\label{1D-cuprates}(Quasi) one-dimensional cuprates} \hfill\\
\\
Among the various  mechanisms discussed for high-$T_c$ superconductivity in cuprates, the electronic model based on ladder-like structures in which
Cu and O atoms are ordered in two chains, coupled by rungs, has  been  heavily discussed\,\cite{Dagotto1992,Sigrist1994}.
In this model, depending on the exchange coupling along the chains and that along the rungs, singlets  can exist on the rungs of a doped ladder compound
and exchange-driven superconductivity can be formed. Depending on the size of parameters or doping concentration,  an insulating hole crystal in which the carriers crystallize into a static Wigner crystal may also  form the ground state. 
Note, this would be an electronic charge density wave (CDW), driven by the Coulomb interaction and not by a coupling to the lattice (Peierls transition).
 The competition of the two phases is similar to that believed to occur
between ordered stripes and high-$T_c$ superconductivity in two-dimensional cuprates.

Indeed the only known doped ladder compound Sr$_{14-x}$Ca$_x$Cu$_{24}$O$_{41}$ (SCCO) exhibits both phases: superconductivity has been detected for $x$= 13.6 below $T_c$=12 K at hydrostatic pressures
larger than 3.5 GPa\,\cite{Nagata1998}, while for $x$=0 SCCO exhibits a CDW. Therefore and because the interplay between charge and 
spin degrees of freedom can be easier studied theoretically in quasi one-dimensional systems,  experimental studies of charge density modulations in SCCO by RSXS are 
extremely important\,\cite{Abbamonte2004,Rusydi2006,Rusydi2007,Rusydi2008}.

\begin{figure}[t!]
\centering
\includegraphics[angle=-90,width=7.5cm]{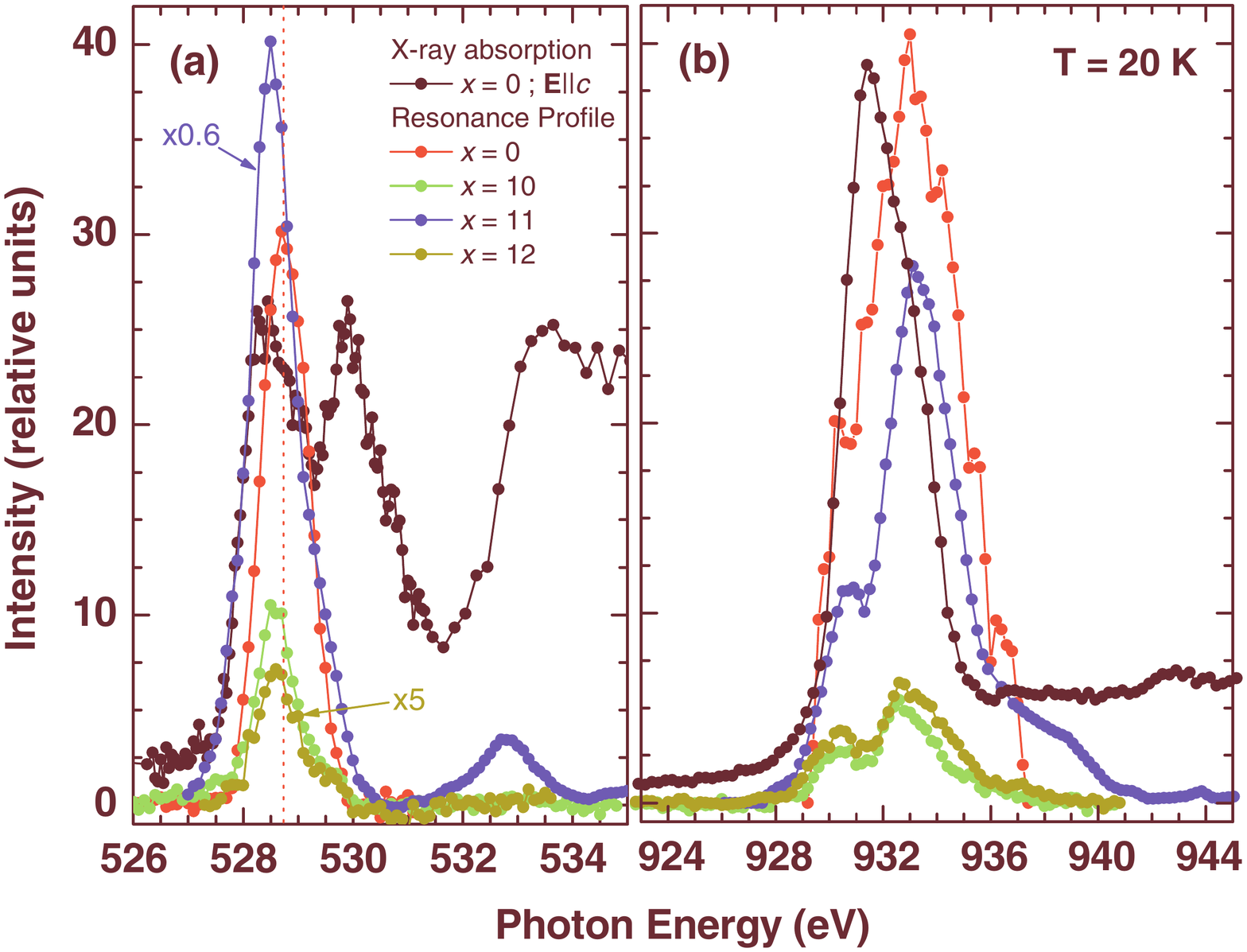}
\caption{(colour online)  Scattering profiles of Sr$_{14-x}$Ca$_x$Cu$_{24}$O$_{41}$, compared to XAS data (black points) of the $x$ = 0 system
($E\parallel c$) near (a) the O $ K$ and (b) the Cu $L_3$ edge.
Red, green, blue, and brown symbols correspond to scattering profiles for $x$ = 0, 10, 11, and 12, respectively.
(Reprinted with permission from \cite{Rusydi2006}. Copyright  2006, American Physical Society.)}
\label{SCCO_Rusyd1}
\end{figure}

\begin{figure}[t!]
\centering
\includegraphics[angle=-90,width=7.5cm]{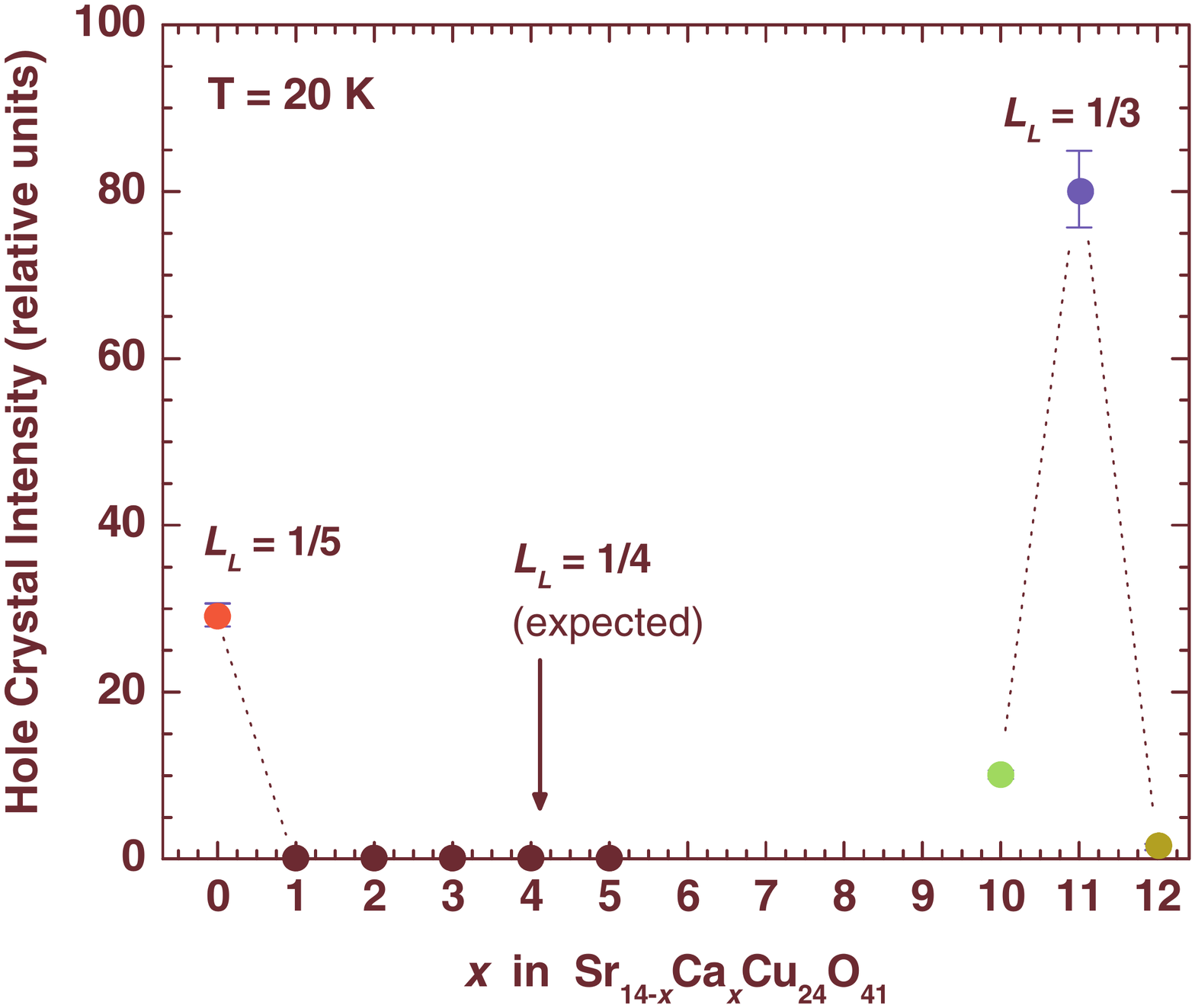}
\caption{(Colour online) Intensity of hole crystal scattering at the
O $K$ mobile carrier peak (MCP, 528.6 eV) of Sr$_{14-x}$Ca$_x$Cu$_{24}$O$_{41}$ for various chemical
compositions. The dashed lines are a guide to the eye. Hole crystallization
occurs only on the rational wave vectors $L$=5 and
$L$=3. No reflection with $L$=4 was observed.
(Reprinted with permission from \cite{Rusydi2006}. Copyright © 2006, American Physical Society.)}
\label{SCCO_Rusyd2} 
\end{figure}

SCCO consists of two different alternating types of copper oxide sheets perpendicular to the $b$ axis. One with chains,
formed out of edge-sharing CuO$_4$ plaquettes and one with weakly coupled two-leg ladders, i.e., two parallel adjacent chains formed out of corner-sharing CuO$_4$ plaquettes.  Both the chains and the ladders are aligned
along the $c$  axis.
The sheets are separated by Sr/Ca ions. Since the CuO$_4$ plaquettes in the chains and those in the ladders are rotated by 45$^\circ$ with respect to each other, the ratio of the  lattice constant for the ladder $c_L$ to that of the chain $c_C$ is about $10/7  \approx \sqrt 2$. This means that one unit cell is composed out of ten CuO$_4$ plaquettes forming the chains and
two times seven equal to 14 CuO$_4$ plaquettes forming the two-leg ladders. This leads to a "misfit compound" which is structurally incommensurate and has a large unit cell along the c axis 
with the length $c$=27.3\,\AA $\approx 7c_L \approx 10c_C$. 
According to the chemical formula and assuming that Cu is divalent, there are 6 holes per formula unit.
The distribution of the holes among chains and ladders is still under debate. Earlier polarisation dependent XAS measurements for $x$=0\,\cite{Nucker2000} suggested
that  0.8 holes are on the ladder and the rest is on the chains. Replacing the Sr ions by the smaller isovalent Ca ions, i.e., by chemical pressure, there is a hole transfer from the chains to the ladders
which according to the earlier XAS results\,\cite{Nucker2000},  for $x$=12 increases the number of holes on the ladders to 1.1, i.e., 0.08  holes per Cu site. Optical spectroscopy\,\cite{Osafune1997}, ARPES experiments\,\cite{Koitzsch2010} and recent XAS experiments\,\cite{Rusydi2007} showed for high Ca replacements values of 0.2, 0.15-0.2, and 0.31 holes per Cu site, respectively, with no sign of convergence with time.

In  figure\,\ref{SCCO_Rusyd1} we reproduce SCCO data showing XAS spectra for $x$=0 and a photon polarisation parallel to $c$ together with resonance profiles for various Ca concentrations\,\cite{Rusydi2006}.
In agreement with previous XAS data near the O $K$ edge\,\cite{Nucker2000} the first pre-peak at 528.4 eV and the shoulder at 
528.7 eV were assigned to hole states at the chains and the ladders, respectively. The next peak at 530 eV was
interpreted in terms of the UHB. Like in other doped cuprates, the first peak at 931.5 eV in the Cu $L_3$ XAS spectrum is assigned to a transition into empty Cu $3d$ states while the shoulder at 933 eV is related to ligand (O) hole states. For $x$=0, a resonance profile measured at a photon energy of 528.7 eV and  a wave vector $Q=(0, 0, \frac{2\pi}{c_L}L_L)$ shows a clear resonance for $L_L$=1/5 (see figures\,\ref{SCCO_Rusyd1} and 
\,\ref{SCCO_Rusyd2}). This observation together with the fact that away from the ladder hole shoulder no scattering intensity  has been detected, signals the existence of a commensurate hole crystal in the ladders with a wave length $\lambda=2\pi/|Q|=c_L/L_L=5c_L=19.5$ \AA.
At higher Ca concentrations (10$\leq x \leq12$) a resonance is observed for $L_L$=1/3 (see figures\,\ref{SCCO_Rusyd1} and 
\,\ref{SCCO_Rusyd2}).  On the other hand,  no resonance has been detected for $L_L$=1/4. This could indicate that the hole crystal is stable for odd, though not even, multiples of the ladder period.
The resonance profiles at the Cu $L_3$ edge shows, in particular for higher $x$ values two peaks, at 930 eV and at 933 eV. The latter peak coincidences with the ligand-hole shoulder and thus indicates the presence of
a hole modulation. The former indicates a lattice modulation\,\cite{Abbamonte2006} which is more pronounced for larger hole
(Ca) concentrations. The hole crystal intensities show both for $L_L$ = 1/5 and 1/3 a strong temperature dependence which points to melting of the hole crystal at higher temperatures and also signals that the modulation is not caused by the misfit between the chains and the ladders. The non-existence of hole crystals with non-integer L values signals that incommensurate hole crystals melt even at very low temperatures.

Hole crystallization in ladder compounds was predicted by a $t-J$ model\,\cite{Dagotto1992} and density matrix renormalization group calculations\,\cite{White2002}. In particular it was found that holes like to pair up on the rungs of the ladder. Thus doped rungs are separated by the wave length $\lambda$ and the doping concentration $p$ per Cu site should be then $p=c_L/\lambda=L_L$ or the number of holes on the ladder per unit cell should be $n_L=14L_L$. From these considerations we expect for zero Ca concentration
$x=0$ $p_L=0.2$ and $n_L=2.8$ and for $x=11$ $p_L=0.33$ and $n_L=4.7$. These values are in good agreement with the values derived in the recent XAS evaluation\,\cite{Rusydi2007} and thus offer strong support for a pairing of holes along the rungs and thus also support models on the mechanism of high-$T_c$ superconductivity in cuprates based on ladder systems.  On the other hand, these results are in striking disagreement with other results on the number of holes an the ladders mentioned above.

 The absence of a $L_L$=1/4 periodicity was explained in terms of resonant valence bond calculations indicating that the Wigner hole crystal is stable for odd, but not for even multiples of the ladder period. Further theoretical work showed that  the widely used ladder $t-J$ model  is not sufficient and has to be supplemented by Coulomb repulsion between the holes on neighbouring ladders to explain the existence of a hole crystallization on the ladders. A mean field calculation of the extended model could explain a charge density wave with an odd period\,\cite{Wohlfeld2010}.

At the end of the paragraph reviewing charge order in the system SCCO, we report on RSXS results on the chains of these compounds\,\cite{Rusydi2007}. For the undoped compound in which divalent Sr was replaced by trivalent La, a superstructure reflection for $L_C=0.30$ has been detected for photons with an energy near the Cu $L_3$ and  $L_2$ edges. Since in this compound, no holes can form a charge density wave and since the superstructure  reflection was independent of the temperature, it has been interpreted in terms of a pure strain wave formed by the misfit between the chains and the ladders. For doped  SCCO an incommensurate superstructure reflection at $L_C=0.318$ has been detected. It shows a remarkable temperature dependence and appears mainly near the ligand-hole shoulder. From this it was concluded that in SCCO a strain-stabilized charge density wave is formed.

\subsubsection{\label{Nickelates}Nickelates} \hfill\\ 
\\
The first transition metal oxide system in which static stripe-like order was detected was La$_{2-x}$Sr$_x$NiO$_{4+y}$ (LSNO). In contrast to LSCO, upon doping, i.e., replacing the trivalent La ions by divalent Sr ions or by adding oxygen, the nickelates remain insulating except for very high doping  concentrations, and superconductivity has not been detected so far. Both in the electron diffraction \,\cite{Chen1993} and neutron diffraction\,\cite{Tranquada1994} data,  superstructure reflections were observed, which were interpreted in terms of coupled charge and spin-density modulations in the NiO$_2$ planes, similar to the case depicted for the cuprates in figure\,\ref{Tranquada}. Different to cuprates, however, the stripes are rotated by 45$^\circ$.  In an orthorhombic unit cell with axes rotated by 45$^\circ$ with respect to the Ni-O bonds,  the magnetic peaks are split about the antiferromagnetic position (1,0,0) along the [100] and [010] directions by $\epsilon$, while the charge order are split about the fundamental Bragg peaks by 2$\epsilon$, as in the case of cuprates. Also  similar to the cuprates, for hole concentrations
$n_h<\frac{1}{8}$, in the nickelates $\epsilon\approx 0.5n_h=0.5(x+2y)$. Stripe-like charge and spin order in LSNO is  observed for 0.15 $\leqq n_h\leqq$ 0.5. The stripe order is most stable at a doping level
$n_h=\frac{1}{3}$ where it shows the highest charge and spin ordering
temperatures and the longest correlation length. Different from the cuprates  in La$_2$NiO$_4$ there are two intrinsic holes on each Ni site in  the $3d_{x^2-y^2}$  and  the $3d_{3z^2-r^2}$ state\,\cite{Pellegrin1996}. 

\begin{figure}[t!] 
\centering 
\includegraphics[width=7.5 cm]{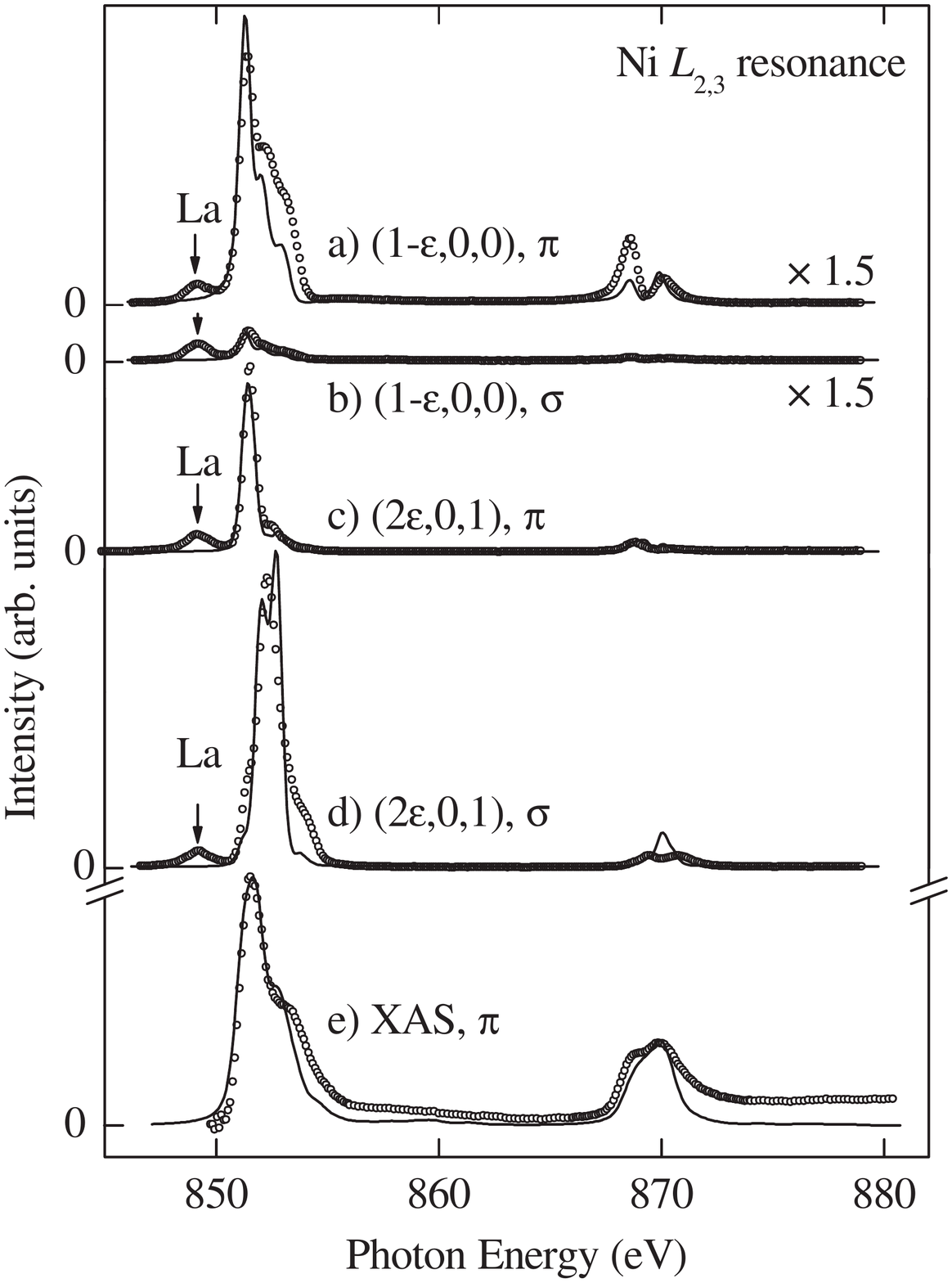}
\caption{(colour online) The resonant scattering profile at the Ni $L_{2,3}$ edges of La$_{1.8}$Sr$_{0.2}$NiO$_4$
for the spin and charge modulation superstructure reflections recorded with $\pi$ and $\sigma$ polarised photons. Also shown are XAS data. The experimental results (points) are compared to simulations (solid line).
(Reprinted  from\,\cite{Schussler-Langeheine2005}. Copyright  2005, American Physical Society.)}
\label{LSNO_Schuessler} 
\end{figure}

RSXS experiments were performed on a La$_{2-x}$Sr$_x$NiO$_4$ single crystal\,\cite{Schussler-Langeheine2005}. 
The energy  profiles at the Ni $L_{2,3}$ edges, shown in figure\,\ref{LSNO_Schuessler}, exhibit resonances at the
charge-order wave vector (2$\epsilon$,0,1) and at the spin-order wave vector (1-$\epsilon$,0,0) with $\epsilon$=0.196 in the above  notation.
A  resonance enhancement  is also observed at the La $M_4$ edge at 849.2 eV, however, this enhancement  is much weaker than at the Ni $L_{2,3}$ edges   indicating that the modulation occurs mainly in the NiO$_2$ planes. 
The resonances show a strong polarisation dependence, e.g. the intensity of the magnetic 
superstructure peak for $\sigma$ polarisation) is only 10 $\%$ of 
that for $\pi$ polarisation. Since only a magnetic moment perpendicular to the polarisation of the incoming light is probed and 
because in the chosen experimental geometry the $\pi$ polarisation is perpendicular to the stripes, one can conclude that the Ni spins are 
essentially collinear with the stripes. Further conclusions can be drawn from a comparison with calculations based on a configuration-interaction model (see  figure\,\ref{LSNO_Schuessler}). 
The difference of about 1 eV  for  the charge modulation 
resonances for $\pi$ and $\sigma$ polarisation  indicates a large energy splitting between the $3d_{x^2-y^2}$  and the $3d_{3z^2-r^2}$  levels. 
The comparison also signals that the holes are going mainly to the O sites on a ligand molecular orbital with $x^2-y^2$ symmetry. 
Due to the strong on-site Coulomb repulsion of two holes on the Ni sites, the Ni 3$d$ count is not strongly modulated in the stripe structure, ruling out a Ni$^{2+}$/Ni$^{3+}$ charge order scenario. On the contrary, the 
the situation is very close to that of stripe structures in cuprates, except that in the nickelates there is an additional intrinsic hole in the Ni 3$d_{3z^2-r^2}$ states.

In summary, the comparison with the RSXS data with the cluster calculations yields interesting results on the stripe structure and the electronic structure of nickelates: (i) both the charge order 
and the spin order resides in the NiO$_2$ layers, (ii) the doped holes are mainly located on O sites and the spin of these holes are 
coupled anti-parallel to the spins of the intrinsic hole on the Ni 3$d_{x^2-y^2}$ states in close analogy to the  Zhang-Rice singlets in the cuprates.

\subsubsection{\label{Magnetite}Magnetite} 
\begin{figure}[t!] 
\centering 
\includegraphics[width=7.5 cm]{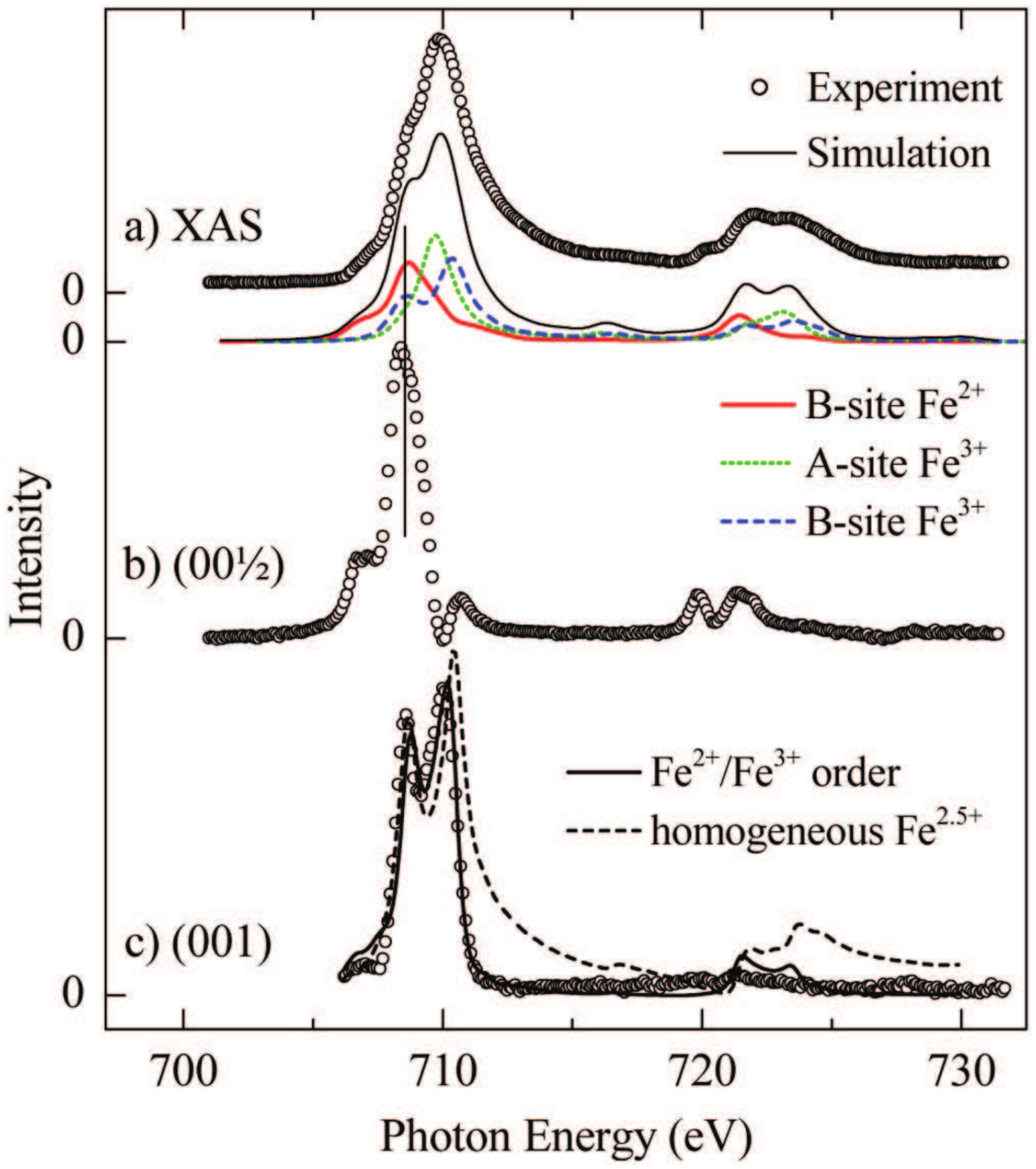}
\caption{(colour online) Experimental Fe $L_{2,3}$  XAS spectra of Fe$_3$O$_4$ compared to resonance profiles of the (0,0,$\frac{1}{2}$) and the (0,0,1) superstructure reflections. The solid lines are simulations for the charge order scenario while the dashed line is related to the homogeneously mixed-valence scenario. 
(Reprinted  from\,\cite{Schlappa2008}. Copyright © 2008, American Physical Society.)}
\label{FeO_Schlappa} 
\end{figure}

\hfill\\ 
\\
Magnetite, Fe$_3$O$_4$, is another prototype correlated transition metal oxide, in which a subtle interplay of lattice, charge, spin, and orbital degrees of freedom determines the physical properties\,\cite{Walz2002}. It was the first magnetic material known to mankind.
Verwey discovered in the late 1930 that upon lowering the temperature below the Verwey temperature $T_V$=123 K,  Fe$_3$O$_4$ undergoes a first-order transition connected with a conductivity decrease by two orders of magnitude. Magnetite has been considered as a mixed valence system. In this compound at high temperatures, the tetrahedral  $A$ sites are occupied by  Fe$^{3+}$ ions while the octahedral $B$ sites are occupied by an equal number of randomly distributed  Fe$^{2+}$ and  Fe$^{3+}$ cations. According to Verwey the transition into the low conducting state is caused by an ordering of the Fe valency on the $B$ sites accompanied by a structural transition. Although numerous experimental\,\cite{Wright2001}  and theoretical studies\,\cite{Leonov2004,Jeng2004}  have been performed on magnetite, no full consensus over all aspects of the order transition exists. With its particular sensitivity to the Fe $3d$ electronic structure, recent RSXS studies at the Fe $L_{2,3}$ edges contributed significantly to the understanding of
the Verwey transition\,\cite{Schlappa2008}. XAS data of Fe$_3$O$_4$ at the Fe $L_{2,3}$ edge together with resonace profiles of the (0,0,$\frac{1}{2})$ and the (0,0,1) superstructure reflections are displayed in figure\,\ref{FeO_Schlappa}.

Based on previous interpretations of the  Fe $L_{2,3}$ absorption spectra and on simulations it was possible to decompose the absorption  structure and to assign the different peaks to specific Fe sites [see figure\,\ref{FeO_Schlappa} (a)]. The low-energy features (red solid line) were assigned to Fe$^{2+}$. Remarkably,   the (0,0,$\frac{1}{2}$) shows a resonance exactly at these energies which  means that the (0,0,$\frac{1}{2}$) peak is due to an order, which involves only $B$-site  Fe$^{2+}$ ions. This  implies further that there is an orbital order of the $t_{2g}$ electrons  that distinguishes different
$B$-site Fe$^{2+}$  ions. This experimental result on the orbital ordering of the $t_{2g}$ electrons of  the Fe$^{2+}$ ions agrees with the theoretical predictions on the basis of LDA+$U$ calculations\,\cite{Leonov2004,Jeng2004}.
The (0,0,1) superstructure reflection shoes  maxima at two energies corresponding  to the $B$-site 
 Fe$^{2+}$ and the  $B$-site  Fe$^{3+}$ ions. This is exactly expected for a charge order involving the two $B$-site Fe ions. In figure\,\ref{FeO_Schlappa}(c) the measured resonance profile is compared with simulations for a charge ordered state (solid line) and a homogeneous Fe$^{2.5+}$ configuration (dashed). The very good agreement with the charge order state and the disagreement with the homogeneous state clearly supports  charge order on the $B$ sites in magnetite.
In this context, one should mention that the charge order of the Fe $3d$ electrons on the $B$-site is small and that one should rather speak of a modulation of the $t_{2g}$ count  which is partially screened by a charge transfer from the oxygen neighbors to the empty $e_{g}$ states\,\cite{Leonov2004}.

\subsection{Orbital ordering}
In transition metal compounds the degeneracy of partially occupied electronic states is very often lifted, resulting in a  long-range orbital ordering of the occupied
orbitals below a transition temperature. Therefore orbital ordering phenomena play a fundamental role in determining  the electronic and magnetic properties of many transition metal oxides. The origin of the orbital ordering is still under discussion. 
One scenario explains orbital ordering in terms of a purely electronic mechanism, i.e., by a super-exchange between the transition metal ions\,\cite{Kugel1973}.  
The other model explains orbital ordering by a lifting of the degeneracy of the atomic level by a cooperative Jahn-Teller effect, i.e., by a net energy gain due to lowering of the electronic state and an energy loss by a local distortion of the ligands around the transition metal, similar to  the Peierls transition mentioned above. In this second scenario lattice effects are important.  Possibly in real systems  a combination of the two models is realized.

\subsubsection{\label{sec:Manganites}Manganites} \hfill\\ 
\\
Classical examples of orbitally ordered systems are KCuF$_3$~\cite{Paolasini2002} and LaMnO$_3$. In case of manganites, in particular, orbital order has  attracted considerable attention  and controversy since they exhibit a wide diversity of  ground states including phases with colossal magnetic resistance\,\cite{Searle1970,Tokura2006} or charge and orbitally ordered ground states. In many cases small  changes in some parameters such as the charge carrier doping or temperature can lead to transitions between disparate ground states. The origin of this rich physics is widely believed to be due to a complex interplay of charge, magnetic, orbital, and lattice degrees of freedom, which results in different strongly competing electronic phases. 

Two prototypical examples are the manganites $RE_{1-x}A_x$MnO$_3$  and $RE_{1-x}A_{x+1}$MnO$_4$  ($RE$=trivalent rare earth ions, $A$= divalent alkaline-earth ions), which belong to the famous Ruddlesden-Popper series. In all these compounds each Mn ion is surrounded by an O octahedron. The interaction with the O ions leads to a partial lifting of the degeneracy of the $3d$ states into the lower $t_{2g}$ states and the twofold degenerate $e_{g}$ states at higher energies. In the undoped case ($x$=0), the Mn ions have a formal valence of 3+ with four electrons in the $3d$ shell.  Since the Hund's rule energy is much larger than the crystal field splitting, the Mn $3d^4$ is in a very stable high-spin state that corresponds to three spin-up electrons in the $t_{2g\uparrow}$ states and a single spin-up electron in the $e_{g\uparrow}$ state, which results in a $S$=2 state of Mn. 
The $t_{2g}$ states, being less hybridised with the O ions than the $e_g$ states, are assumed to be more localised, also by strong correlation effects. The $e_{g}$ electrons are supposed to be more delocalised due to the stronger hybridisation with the O ions. On the other hand, for a cubic MnO$_6$ octahedron, a single electron in the degenerate $e_g$ levels can occupy any linear combination of the  $x^2-y^2$ and $3z^2-r^2$ orbitals. This is referred to as the orbital degree of freedom, which plays a very prominent role for the physics of manganites. The ground state of a cubic octahedron is therefore degenerate and, according to the theorem of Jahn and Teller, this degeneracy will always be lifted by a symmetry reduction, i.e., a distortion of the octahedron. The Mn $3d^4$ is therefore strongly Jahn-Teller active causing local lattice distortions.

When the filling of the $e_{g}$ states is close to 1 or close to a commensurate value, the distortions of the interconnected octahedra can occur in a cooperative way and one is then talking about a collective Jahn-Teller distortion. This collective ordering of Jahn-Teller distorted octahedra also corresponds to an orbital ordering. In other words the collective Jahn-Teller effect is one way to stabilize an orbitally ordered state. There is however another mechanism that can stabilize orbital order, even in the absence of electron-phonon coupling, which is given by the so-called Kugel-Khomskii exchange\,\cite{Goodenough1955,Kugel1973}. It was discussed controversially which of the two mechanisms is the main driving force for orbital order in the manganites.

A particular example for this controversy was given by the half-doped manganite La$_{1-x}$Sr$_x$MnO$_4$, which is composed of MnO$_2$ planes separated by a cubic (LaSr)O layer. Upon hole doping, more and more Mn ions with a formal valence of 4+ ($3d^3$) are created, which suppresses the collective Jahn-Teller effect and causes an increased itineracy of the $e_{g}$ electrons. The $e_g$ electrons thus play the role of conduction electrons coupled to the localised $S$=3/2  $t_{2g\uparrow}$ electrons. As discussed first by Zener\,\cite{Zener1951}, the mobility of $e_{g}$ conduction electrons is strongly affected by the spin degrees of freedom: due to the so-called double exchange mechanism,  the  $e_{g}$  electrons can delocalise only for ferromagnetic alignment of neighbouring spins.  Upon increasing the temperature above the ferromagnetic transition temperature the configuration of the spins gets disordered, which causes a strong spin-charge scattering  and thus an enhancement of the resistivity close to the Curie temperature. Applying an external magnetic field at this temperature can easily align the local spins leading to a strong magneto-resistance. This is a simple  qualitative explanation of the colossal magneto-resistance observed e.g. in La$_{1-x}$Sr$_x$MnO$_3$ (x=0.3)\,\cite{Tokura2006} which is not correct on a quantitative level, since it neglects the orbital and lattice degrees of freedom. 

The relevance of the latter becomes particular evident near $x$=0.5 where an electron ordering takes place which has  attracted much attention and controversy. Several studies have been performed on the compound La$_{0.5}$Sr$_{1.5}$MnO$_4$ which is composed of MnO$_2$ planes separated by a cubic (LaSr)O layer. At room temperature the Mn sites in this system are all crystallographically equivalent and have an average valency of +3.5. At $T_{CO}\approx$ 240\,K a charge disproportionation occurs creating two inequivalent Mn sites. Originally the two sites were assigned to Mn$^{3+}$ and Mn$^{4+}$ ions,  but later on it turned out that the charge difference of the two sites is much less than 1. Below $T_N$=120 K a long range antiferromagnetic ordering of the Mn ions into a CE-Type structure is realized, originally detected in the related compound  La$_{1-x}$Ca$_x$MnO$_3$  by neutron scattering\,\cite{Wollan1955}. 
It is believed that this complex electronic order is stabilized by several interactions, but it was discussed controversially whether orbital order really exists in this system and, if it does, whether it is driven by a collective Jahn-Teller distortions or by superexchange interactions.

Synchrotron-based investigations of orbital order in the manganites were initiated by a seminal study of   La$_{0.5}$Sr$_{1.5}$MnO$_4$ using hard x-rays at the Mn $K$ edge\,\cite{Murakami1998}. Superstructure reflections were detected,  which could not be explained on the basis of the high-temperature crystal structure and were therefore explained in terms of an orbital ordering of the Mn $3d$ states. While there was consensus about the orbital origin of these reflections, the direct observation by resonant hard x-ray scattering was challenged, as the M $K$ edge is dominated by dipole-allowed transitions to the 4p states, which are not sensitive to orbital order but rather lattice distortions, as pointed out in  a theoretical study~\cite{Benfatto1999}. Another theoretical paper\,\cite{Castleton2000} thus suggested the use of RSXS experiments at the Mn $L$ edges, which are directly related to virtual excitations into Mn $3d$ states and thus to orbital ordering. The calculations predicted  different photon energy dependencies of the scattered intensity for  orbital ordering being stabilized by superexchange interaction  compared to Jahn-Teller effect.
Such RSXS studies  on single-layered manganite  La$_{0.5}$Sr$_{1.5}$MnO$_4$ at the Mn $L$ edges were subsequently reported in Refs. \cite{Wilkins2003a,Wilkins2005,Wilkins2005a,Dhesi2004,Staub2005,Staub2006}.

\begin{figure}[t!]
\centering
\includegraphics[width=7 cm]{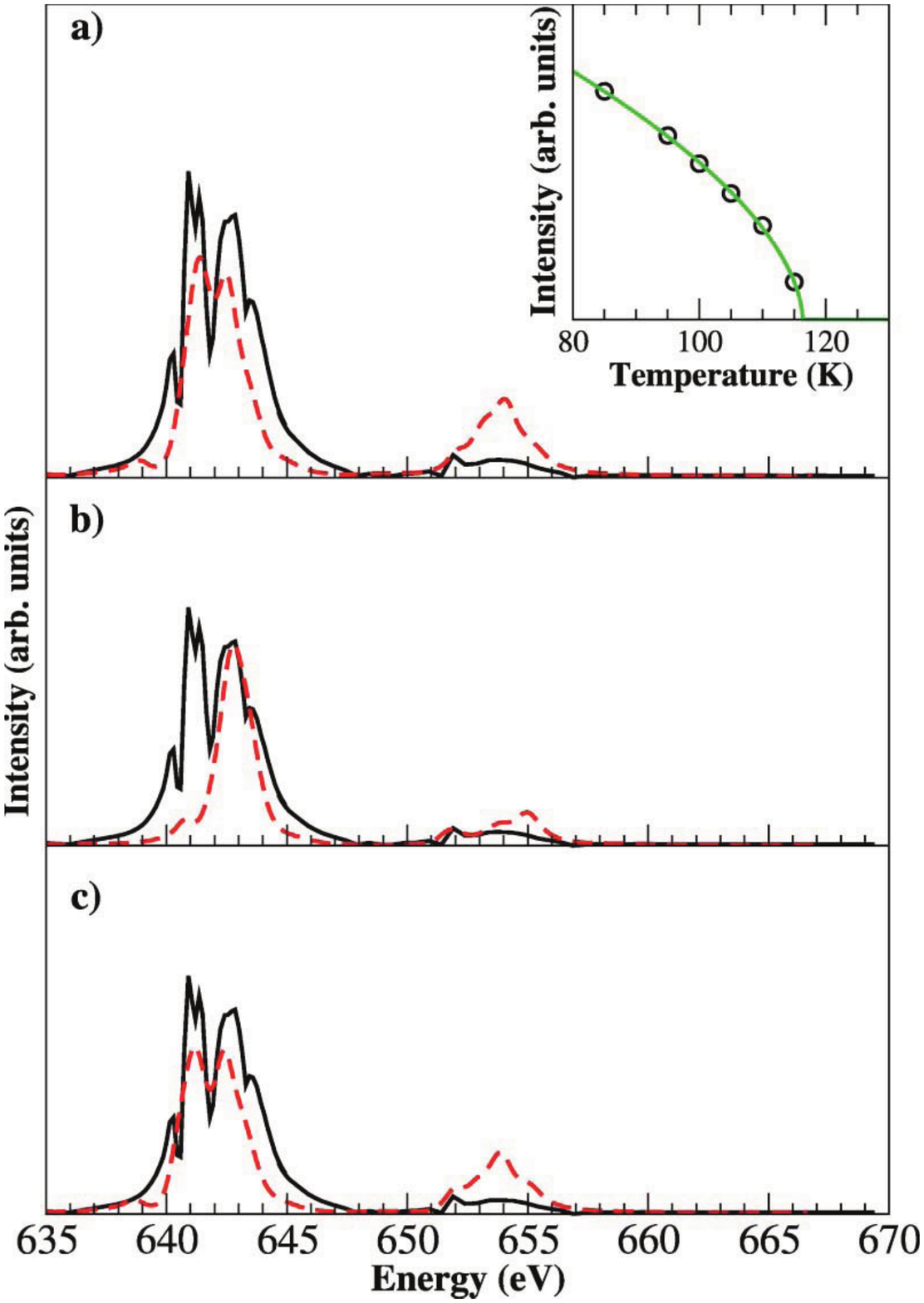}
\caption{(colour online) The resonant scattering profile at the Mn $L_{2,3}$ edges of La$_{0.5}$Sr$_{1.5}$MnO$_4$ at the magnetic superstructure reflection $(\frac{1}{4},-\frac{1}{4},\frac{1}{2})$ measured at 63 K (full black line) compared with theoretical fits (dashed red lines) for (a) $d_{x^2-z^2}$ and (b) $d_{3x^2-r^2}$ types of orbital ordering. In panel (c) a fit for an orthorhombic crystal field is presented. The inset shows the temperature dependence of the magnetic order parameter. (Reprinted with permission from \cite{Wilkins2005a}. Copyright © 2005, American Physical Society.)}
\label{LSMOm_Wilkins}
\end{figure}

The first report was given by Wilkins et al.\,\cite{Wilkins2003a}, who studied  the  orbital and magnetic reflections at  wave vectors of $\vec q_{OO}=(\frac{1}{4},\frac{1}{4},0)$ and $(\vec q_{AF}=\frac{1}{4},\frac{1}{4},\frac{1}{2})$, respectively. We note that the occurrence of orbital order causes the unit cell to quadruple in the $a-b$ plane.
\begin{figure}[t!]
\centering
\includegraphics[width=7 cm]{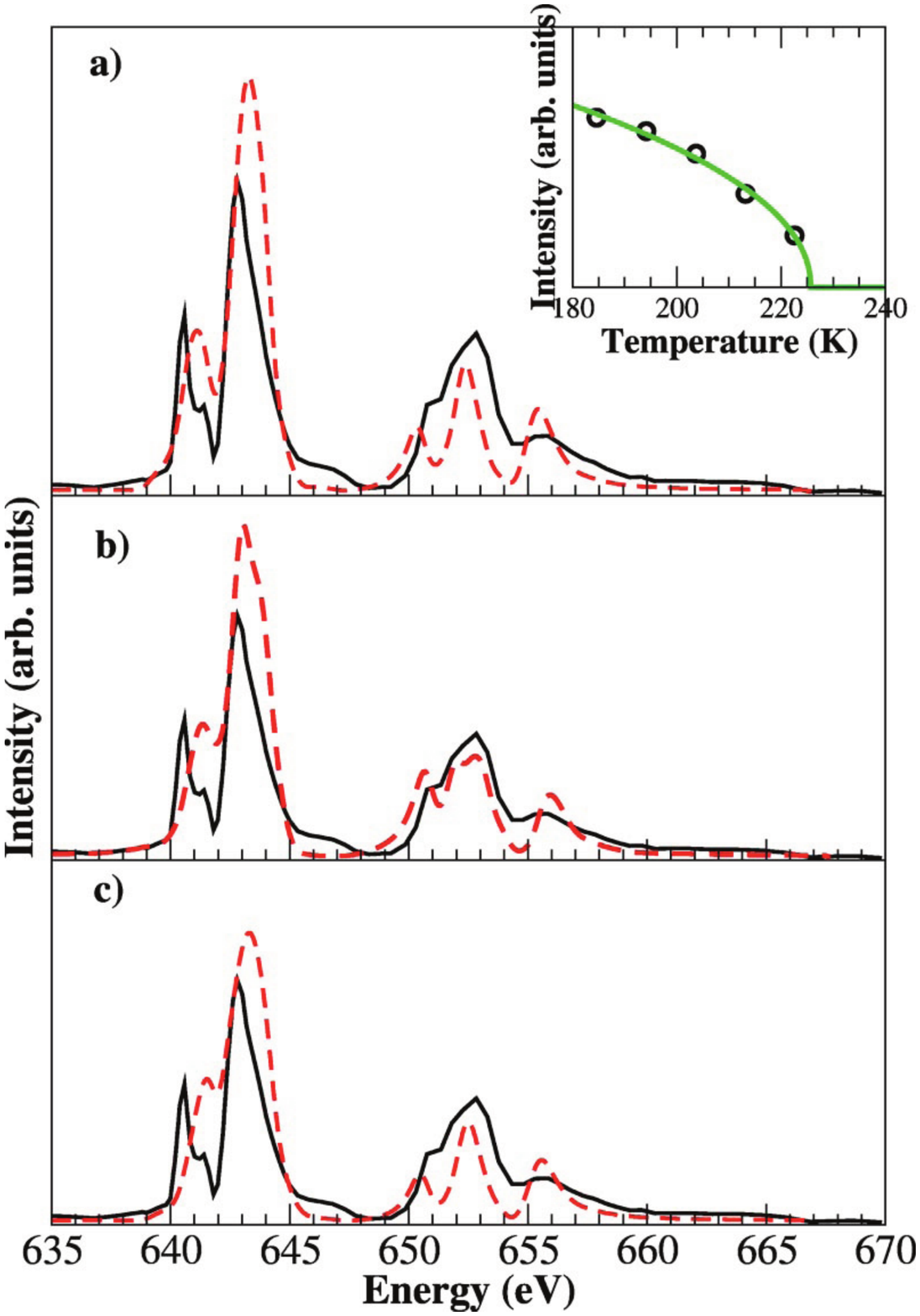}
\caption{(colour online)  The resonant scattering profile at the Mn $L_{2,3}$ edges of La$_{0.5}$Sr$_{1.5}$MnO$_4$ at the orbital-order superstructure reflection $(\frac{1}{4},\frac{1}{4},0)$ measured at 63\,K (full black line) compared with theoretical fits (dashed red lines) for (a) $d_{x^2-z^2}$ and (b) $d_{3x^2-r^2}$ types of orbital ordering. In panel (c) a fit for an orthorhombic crystal field is presented. The inset shows the temperature dependence of the orbital order parameter. (Reprinted with permission from \cite{Wilkins2005a}. Copyright © 2005, American Physical Society.)}
\label{LSMOo_Wilkins}
\end{figure}
Figure\,\ref{LSMOo_Wilkins} shows the energy dependence of the scattered intensity at the orbital order wave vector $\vec q_{OO}$ through the Mn $L_3$ (near 640 eV) and $L_2$ edges (near 650 eV) which are related to virtual electric-dipole transitions between the Mn $2p_{3/2}$ and $2p_{1/2}$ core levels to the unoccupied  $3d$  states. Figure\,\ref{LSMOm_Wilkins} shows analogous data for the AFM order reflection $\vec q_{AF}$. It can be clearly observed in these figures that the orbital and magnetic reflection exhibit strong resonances at the Mn $L_{2,3}$ edges. We note that these measurements provide the most direct prove of orbital ordering in these materials.
 In figures\,\ref{LSMOo_Wilkins} and \ref{LSMOm_Wilkins}  the insets show the temperature dependences of the orbital and magnetic order parameters, respectively. The orbital order parameter decreases continuously with increasing temperature and disappears at $T_{OO}$=230 K. The intensity of the antiferromagnetic reflection shows a similar behaviour but disappears at a lower temperature of $T_N$=120 K.

In both figures the experimental data are compared with theoretical calculations of the RSXS spectra based on atomic multiplet calculations in a crystal field. In the calculations the crystal fields were modified in such a way as to fit the experimental data. In the panels (a) and (b) the cubic and the tetragonal crystal field parameters were adjusted for a  $d_{x^2-z^2}$ and $d_{3x^2-r^2}$ type of orbital ordering, respectively. In the panel (c) for the $d_{x^2-z^2}$  type ordering a small orthorhombic crystal field component was added. From the comparison of the experimental data with the theoretical calculations the authors concluded that the orbital ordering below $T_{OO}$  is predominantly of 
$d_{x^2-z^2}$ type. The inclusion of a small orthorhombic crystal field component moderately improved the fits, although many details of the complicated experimental lineshape could not be reproduced by the calculation.

\begin{figure}[t!]
\centering
\includegraphics[width=7 cm]{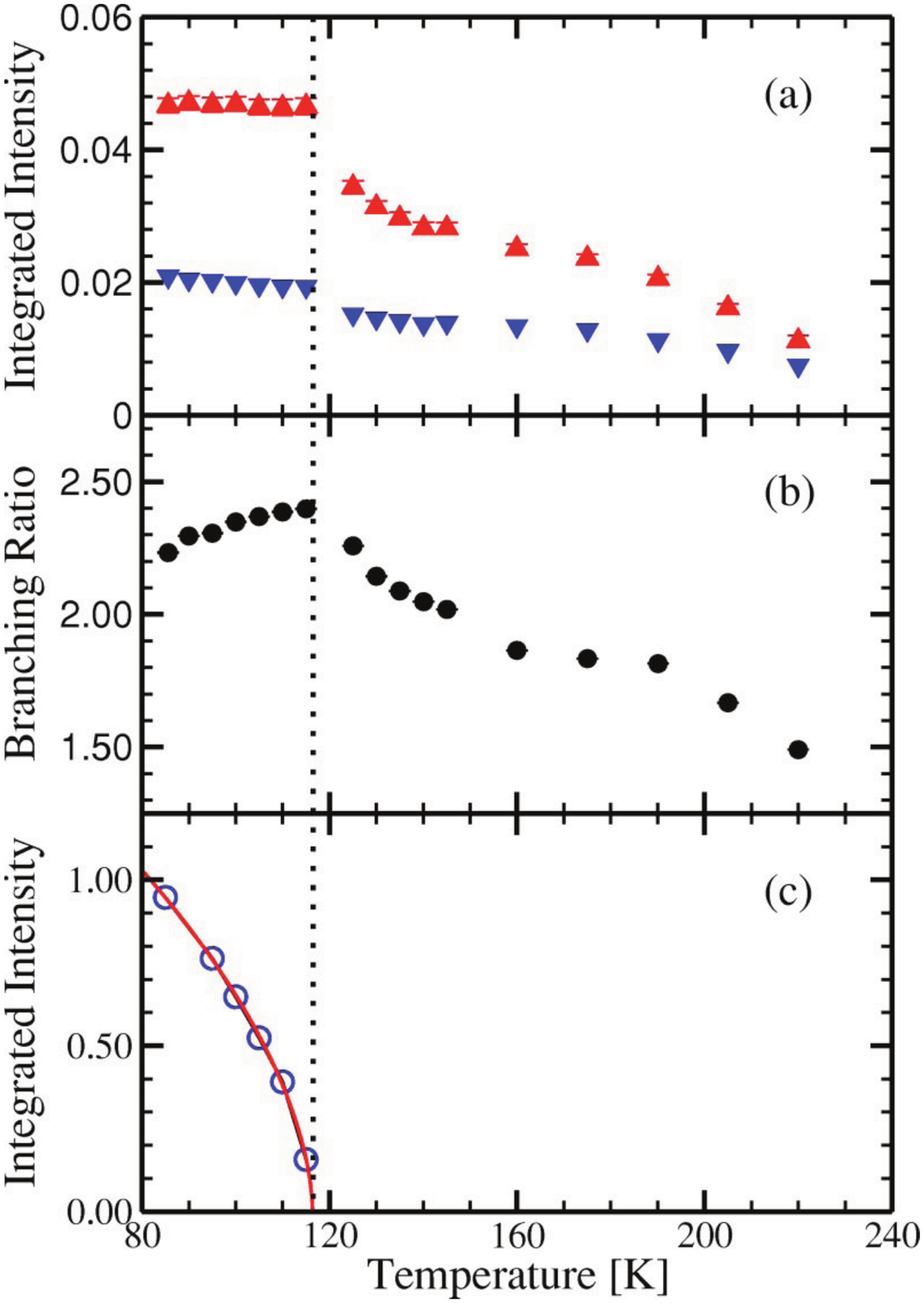}
\caption{(colour online) Temperature dependence of RSXS data at the Mn $ L$ edges of La$_{0.5}$Sr$_{1.5}$MnO$_4$. (a) Integrated intensities of the main features of the orbital ordering $(\frac{1}{4},\frac{1}{4},0)$  reflection at  at the $L_3$ edge (triangles) and at the $L_2$ edge (inverted triangles). (b) Intensity ratio of the integrated intensities $L_3/L_2$. (c) Temperature dependence of the integrated intensity of the magnetic  $(\frac{1}{4},-\frac{1}{4},\frac{1}{2})$ reflection at 643 eV.  (Reprinted with permission from \cite{Wilkins2005a}. Copyright © 2005, American Physical Society.)}
\label{LSMOt_Wilkins}
\end{figure}

Notwithstanding these difficulties, calculations of the resonance profiles predicted a drastic reduction of the $L_3/L_2$ intensity ratio with  decreasing tetragonal crystal field.  This theoretical prediction was used for the interpretation  of experimental results. The experimental temperature dependent RSXS data shown in figure\,\ref{LSMOt_Wilkins} (a) and (b) indicate that below the orbital ordering temperature $T_{OO}$  the $L_3/L_2$  ratio increases, suggesting an increase of the tetragonal crystal field, i.e., an increase of the Jahn-Teller distortion with cooling.  Approaching $T_N$ the  ratio $L_3/L_2$  increases further and saturates below $T_N$.

 From these findings the authors concluded that there are two separate contributions causing the orbital ordering: a dominant mechanism related to the  cooperative Jahn-Teller distortion of the O ions around the Mn$^{3+}$ ions and the direct magnetic Kugel-Khomskii superexchange  mechanism which further strengthens orbital ordering by short-range AFM order and another increase by long-range AFM order below $T_N$. Furthermore the authors concluded a strong interaction between orbital and magnetic correlations.

Nearly at the same time another group has performed similar RSXS experiments on the  same compound La$_{0.5}$Sr$_{1.5}$MnO$_4$\,\cite{Dhesi2004}. The diffraction profile at the Mn $L_{2,3}$  edges for the the orbital order $(\frac{1}{4},\frac{1}{4},0)$ superstructure reflection  were very close to those shown in figure\,\ref{LSMOo_Wilkins}.  In a following paper\,\cite{Mirone2006},  the experimental data were  compared to calculations of a small planar cluster consisting of a central active Mn site with a first neighbor shell comprising O and Mn sites. Thus by allowing a hopping between the Mn and the O ions an integer charge ordering of the Mn ions could be abandoned. In addition to the Jahn-Teller distortion of the O ions, the spin magnetization of the inactive Mn$^{4+}$ sites were explored as adjustable parameters. The calculations based on a $3z^2-r^2/3x^2-r^2$ orbital ordering of the $e_g$ electrons  reproduced all spectral features of the resonance profile while calculations on a $x^2-y^2/z^2-y^2$ order did not, although the differences were not dramatic enough to provide a definite answer. A further result of the RSXS studies of Dhesi  et al.\,\cite{Dhesi2004} was the detection of a complicated temperature dependence of the individual features of the $L_3$ edge indicating that the complete $L_3$ edge intensities cannot be assigned to one order parameter (e.g. orbital ordering by a collective Jahn-Teller effect). 

These findings were supported by a following RSXS study of the same compound by Staub et al.\,\cite{Staub2005,Staub2006}. They realized that the features observed in the resonance profiles of the magnetic reflection show the same temperature dependence while those of the orbital order reflection show a strong energy and temperature dependence. The high-energy features of the $L_{2,3}$  edges (in figure\,\ref{LSMOo_Wilkins} at 647 and 656 eV) which were assigned to a Jahn-Teller ordering parameter saturate below $T_N$. The other features assigned to a direct super-exchange ordering parameter  still increases below $T_N$. From this experimental result together with a polarisation dependence of the resonance profiles they came to the conclusion that there are two interactions  leading to the orbital order. 
 It was concluded that the orbital super-exchange  interaction dominates over the Jahn-Teller distortion strain field interaction and drives the transitions.

The RSXS results on the manganites described above,  demonstrate that the  lineshapes at the Mn $L_{2,3}$ edges display rich spectral features, which contain a lot of information about the underlying order phenomena. To extract this information and to interpret the spectra is, however, challenging as it requires extensive theoretical modeling. 
Here it should be mentioned  that in addition various RSXS studies have been performed on double and multi layer manganites\,\cite{Wilkins2003,Thomas2004a,Herrero-Martin2006,Wilkins2006,Grenier2007,Beale2009,Zhou2011,Lee2011}.

Although largely discussed in connection with 3d transition metal compounds, orbital order is a phenomenon that also occurs in other materials, such as $RE$ compounds. Here,  the shielding of the $4f$ shell by the outer $5d$ electrons causes a weaker coupling of the electronic degrees of freedom of the $4f$  electrons to the lattice and hence, allows studying orbital order mechanisms in a less complex situation. In a series of publications on orbital ordering of $4f$ electrons in $\mathrm{RE}{\mathrm{B}}_{2}{\mathrm{C}}_{2}$ compounds Mulders et al. could show that RSXS performed at the $RE-M_{4,5}$ resonances is not just able to observe orbital order of the $4f$ electrons but can in addition quantify the contributions of the different higher order multipole moments to the electronic ordering\,\cite{Mulders2006a,Mulders2007,Princep2011,Princep2012}.

\subsection{\label{sec:Multiferroics}Multiferroics}
 
In recent years, the interest in materials with coupled magnetic and ferroelectric order has seen an extraordinary revival\,\cite{Fiebig2005,Cheong2007}. The incentive for numerous applied as well as fundamental studies of these materials is the perspective to manipulate electric polarisation by a magnetic field\,\cite{Kimura2003,Hur2004} or magnetic order by an electric field\,\cite{Chu2008}. In this context, the notion multiferroics has been coined, which is, of course, more general and applies to a broader range of materials with coupled order parameters. The renewed interest in magnetoelectric materials was initiated by the discovery of strong coupling between magnetic and electric order in TbMnO$_3$ for which one can flip the electric polarisation by application of a magnetic field\,\cite{Kimura2003}. The ferroelectric order in TbMnO$_3$ is connected to the occurrence of a cycloidal magnetic arrangement of the Mn spins\,\cite{Kenzelmann2005,Katsura2005,Mostovoy2006}, a property that was also found for other members of the perovskite $RE$MnO$_3$ series, where $RE$ denotes a rare earth Gd, Tb, or Dy\,\cite{Kimura2005}. Different theoretical models were used to explain the strong coupling of magnetic and electric order, however, arriving at the same expression that links the direction of the electric polarisation $\mathbf{P}$ to the chirality of the spin structure\,\cite{Katsura2005,Mostovoy2006}: 

\begin{equation}\label{eqnpolarisation}
\mathbf{P} \propto \sum_{i,j}{\mathbf{e}_{i,j}\times (\mathbf{S}_{i}\times \mathbf{S}_{j})}.
\end{equation}

In this equation, $\mathbf{S}_{i}$ and $\mathbf{S}_{j}$ denote spins at the sites $i$ and $j$,and $\mathbf{e}_{i,j}$ is the unit vector connecting the two sites.

Other rare earth manganites of the form $RE$Mn$_2$O$_5$ also exhibit multiferroic behaviour, with reversible switching of the ferroelectric polarisation by a magnetic field, as found  in TbMn$_2$O$_5$\,\cite{Hur2004}. The coupling of magnetic and electric order in these latter materials is believed to be of a different origin, involving magnetoelastic coupling, ionic displacements or electronic ferroelectric polarisation. The majority of the recently studied magnetoelectrics are manganites, however, also cuprates, among them cupric oxide CuO itself have been found to be multiferroic\,\cite{Kimura2008}, with a particularly high ordering temperature of $\approx 230$~K.

As the multiferroic materials are transition metal oxides with complex order of the magnetic moments, neutrons have initially been the method of choice for their magnetic characterization\,\cite{Kenzelmann2005,Yamasaki2008,Chapon2004}. 
RSXS, on the other hand, is perfectly matching these materials 
as both the $3d$ and $4f$ electronic states of the transition metals as well as the $2p$ states of oxygen can be addressed via resonant dipole transitions\,\cite{Koo2007,Okamoto2007,Forrest2008,Bodenthin2008,Wilkins2009,Staub2009,Staub2010b,Schierle2010,Partzsch2011,
Souza2011,Huang2008,Wu2010,Beale2010b,Scagnoli2011,Jang2011,Wadati2012}. 
Thus, the method has provided substantial contributions  to the understanding of the materials, also because circularly polarised x-rays are readily available at synchrotron radiation sources. By the magnetic structure factor, this provides direct access to a possible handedness, as present in helical\,\cite{Mulders2010} or cycloidal arrangements\,\cite{Schierle2010,Jang2011}. In addition, frustrated magnetic interactions result in rather long-period superstructures that can thus be accessed at soft x-ray wavelengths.

\begin{figure}[t!]
\centering 
\includegraphics* [scale=0.5, trim= 0 0 0 0, angle= 0] {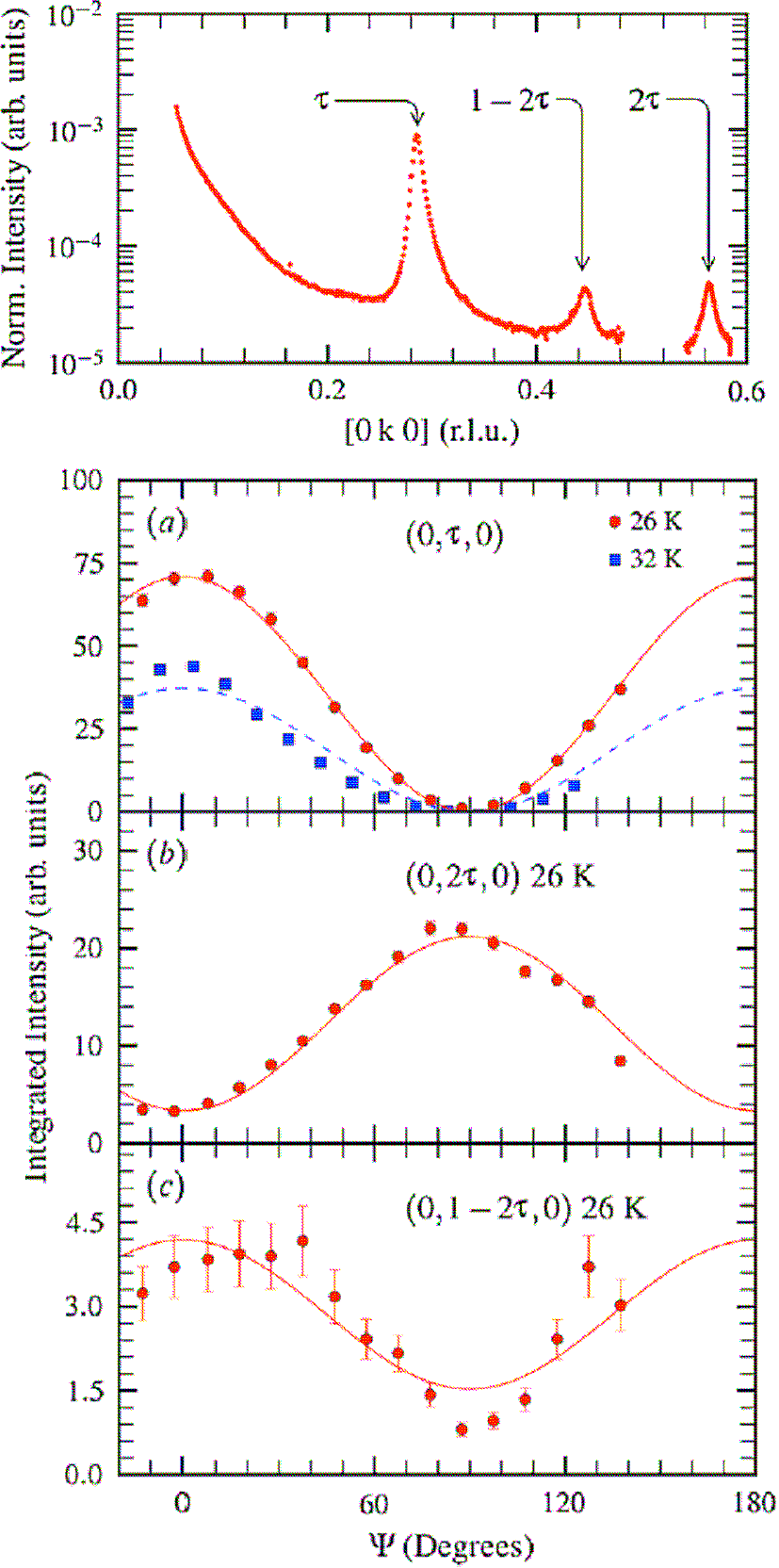}
\caption{\label{Wilkins2009-1} Magnetic diffraction from TbMnO$_3$ , recorded at the Mn $L_3$ resonance. Top: scan along [0~K~0] with three diffraction peaks due to a long-period magnetic modulation of the Mn spins. Bottom: azimuthal dependencies of the diffraction peaks, providing information on the magnetic-moment directions along the [0~K~0] direction. (Reprinted with permission from \cite{Wilkins2009a}. Copyright © 2009, American Physical Society.)}
\end{figure}

TbMnO$_3$ is the most studied material of the $RE$MnO$_3$ series. For this material, magnetic neutron diffraction provided the first experimental observation of a cycloidal magnetic structure in connection with ferroelectric polarisation\,\cite{Kenzelmann2005}. 
From the observation of  components of the Mn $3d$  magnetization along both the $b$ and the $c$ crystallographic axis in the ferroelectric phase a non-collinear structure was concluded that together with a propagation along the $b$ axis  has the proper chirality to induce a ferroelectric polarisation along the $c$ axis according to Eq.\,\ref{eqnpolarisation}. The Tb $4f$ structure, on the contrary, remained collinear, as only a component along the $a$ axis was found, pointing out the essential role of the Mn ordering for the ferrolectric properties of the material. 
The Mn spin cycloid leads to substantial circular dichroism in x-ray scattering as first found in a non-resonant hard x-ray study using circularly polarised light, in this way directly linking to the handedness of the magnetic structure\,\cite{Fabrizi2009}. While it was generally agreed that ferroelectricity is essentially due to a Mn $bc$ cycloid, further details of the magnetic structure could be elucidated taking advantage of the very high sensitivity of resonant magnetic soft x-ray diffraction.

A refinement of the Mn magnetic structure was obtained from resonant soft x-ray diffraction at the Mn $L_2$ resonance\,\cite{Wilkins2009}. Here, a magnetic F-type reflection along [0~K~0] was observed  that would not be allowed for strictly A-type magnetic ordering in the perovskite structure of these materials\,\cite{Goto2004}. From modeling the azimuthal dependences of the diffraction intensities measured with linearly polarised x-rays (Figure\,\ref{Wilkins2009-1}) it was concluded that a canted magnetic component along the $c$ axis must be present in all ordered phases of the material, rendering also the previously assumed sinusoidal phase non-collinear. 

\begin{figure}[t!]
\centering 
\includegraphics* [scale=0.8, trim= 0 0 0 0, angle= 0] {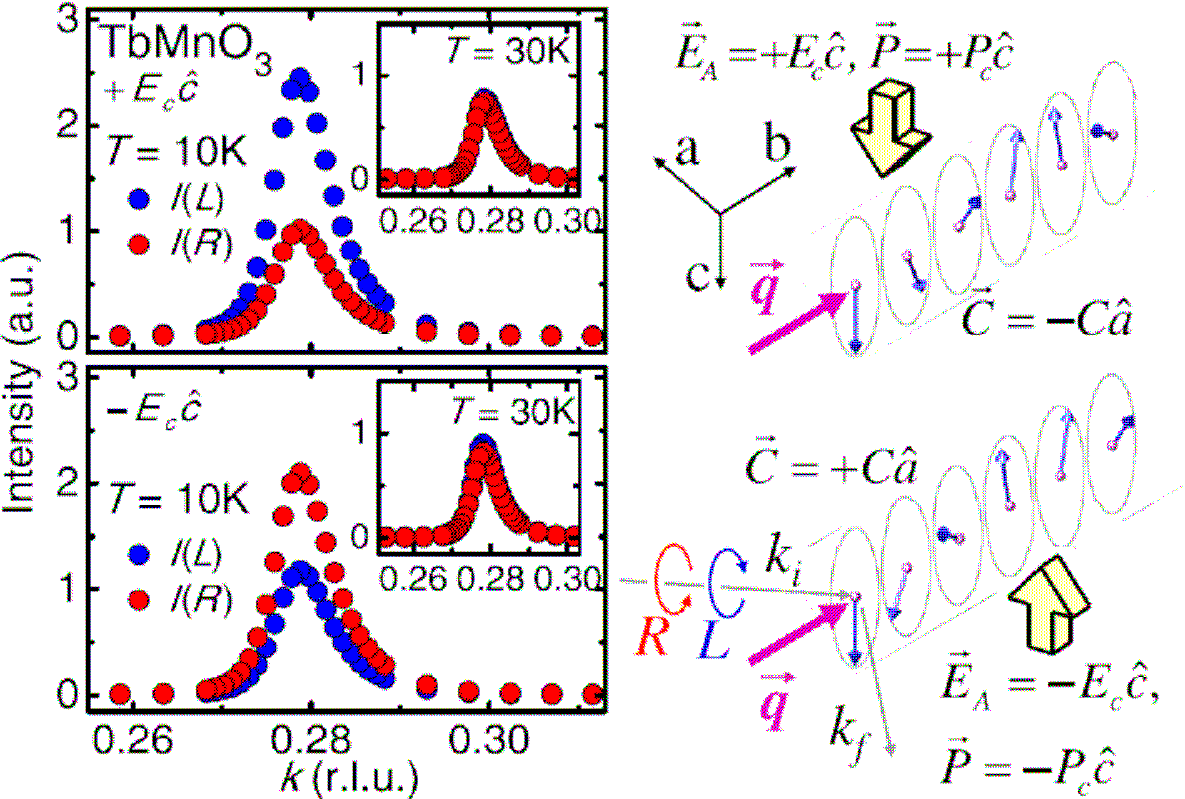}
\caption{\label{Jang2011-1} RSXS at the ($0~\tau~0$)  diffraction peak of  TbMnO$_3$. The circular dichroism  which is observed in the ferroelectric phase, is due to handedness of the magnetic cycloid depicted on the right panel. The  cycloids can be switched by an external electric field. (Reprinted with permission from \cite{Jang2011}. Copyright © 2011, American Physical Society.)}
\end{figure}

Further details  of the magnetic structure were revealed by the application of circular polarised x-rays at the Mn $L_2$ resonance (see figure\,\ref{Jang2011-1}). The magnetic (0~$\tau$~0) reflection exhibits pronounced circular dichroism, thus representing an F-type magnetic cycloid that is coupled to the previously observed A-type cycloid\,\cite{Jang2011}. This clear dependence of the scattered intensity on the helicity of the x-rays is not to be confused with x-ray circular dichroism in absorption, as it is due to the magnetic structure factor\,\cite{Mulders2010} and not an effect of the resonant form factor.    
The occurrence of an F-type magnetic cycloid is explained by antisymmetric exchange, the so-called Dzyaloshinskii-Moriya interaction that favours canting of the magnetic moments towards the $c$ axis\,\cite{Jang2011}. 

Figure\,\ref{Jang2011-1} also indicates another application of resonant soft x-ray diffraction to multiferroic materials. With the help of an external electric field, the electric polarisation of TbMnO$_3$ is switched from pointing along -c to pointing along c. According to Eq.\,\ref{eqnpolarisation}, this corresponds to magnetic cycloids of opposite chirality as depicted on the right panel of figure\,\ref{Jang2011-1}. Therefore, the asymmetry in the scattering of circular polarised light changes sign, as can be inferred from the changing response to left ($L$) and right ($R$) circularly polarised light. Hence, besides general refinements of the magnetic structures, RSXS can provide a useful contrast mechanism to identify and study microscopic details of the magnetic structure upon the application of external fields. 

\begin{figure}[t!]
\centering 
\includegraphics* [scale=1, trim= 0 0 0 0, angle= 0] {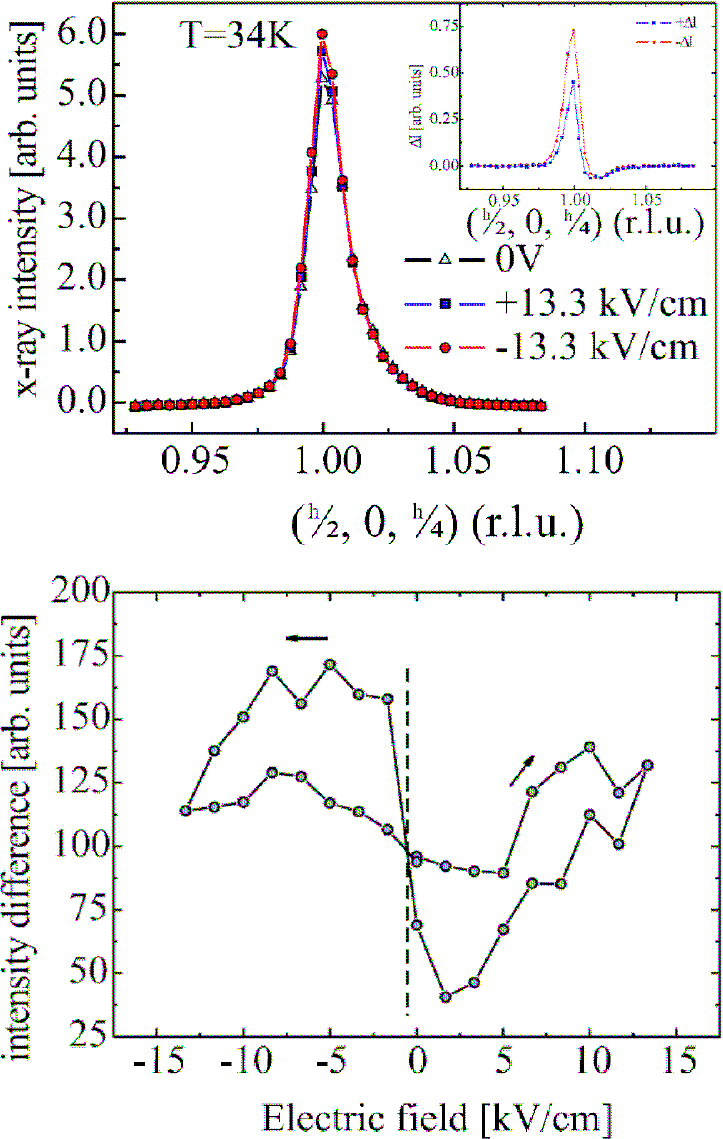}
\caption{\label{Bodenthin2008-1} (1/2 0 1/4) reflection of ErMn$_2$O$_5$ recorded at the Mn $L_3$ resonance in the ferroelectric phase for three different poling states of the sample (top). The inset shows difference curves with respect to the case of the  unpoled state. Hysteretic behaviour of the (1/2 0 1/4) peak intensity as a function of external electric  field. (Reprinted with permission from \cite{Bodenthin2008}. Copyright © 2008, American Physical Society.)}
\end{figure}

The first demonstration of such an application was provided by Bodenthin et al. for the case of ErMn$_2$O$_5$\,\cite{Bodenthin2008}. The contrast was obtained at the Mn $L_3$ resonance using the commensurate (1/2 0 1/4) reflection, which is closely connected to the ferroelectric phase in this material. As shown in figure\,\ref{Bodenthin2008-1} (upper panel), differences in the intensities of this peak are observed with an electric field applied compared to the situation without field. The differences are in fact small but significant, as shown in the inset, and the peak intensity reveals hysteretic behaviour, in this way demonstrating the manipulation of the magnetic structure by an external electric field\,\cite{Bodenthin2008}.  

\begin{figure}[t!]
\centering 
\includegraphics* [scale=0.4, trim= 0 0 0 0, angle= 0] {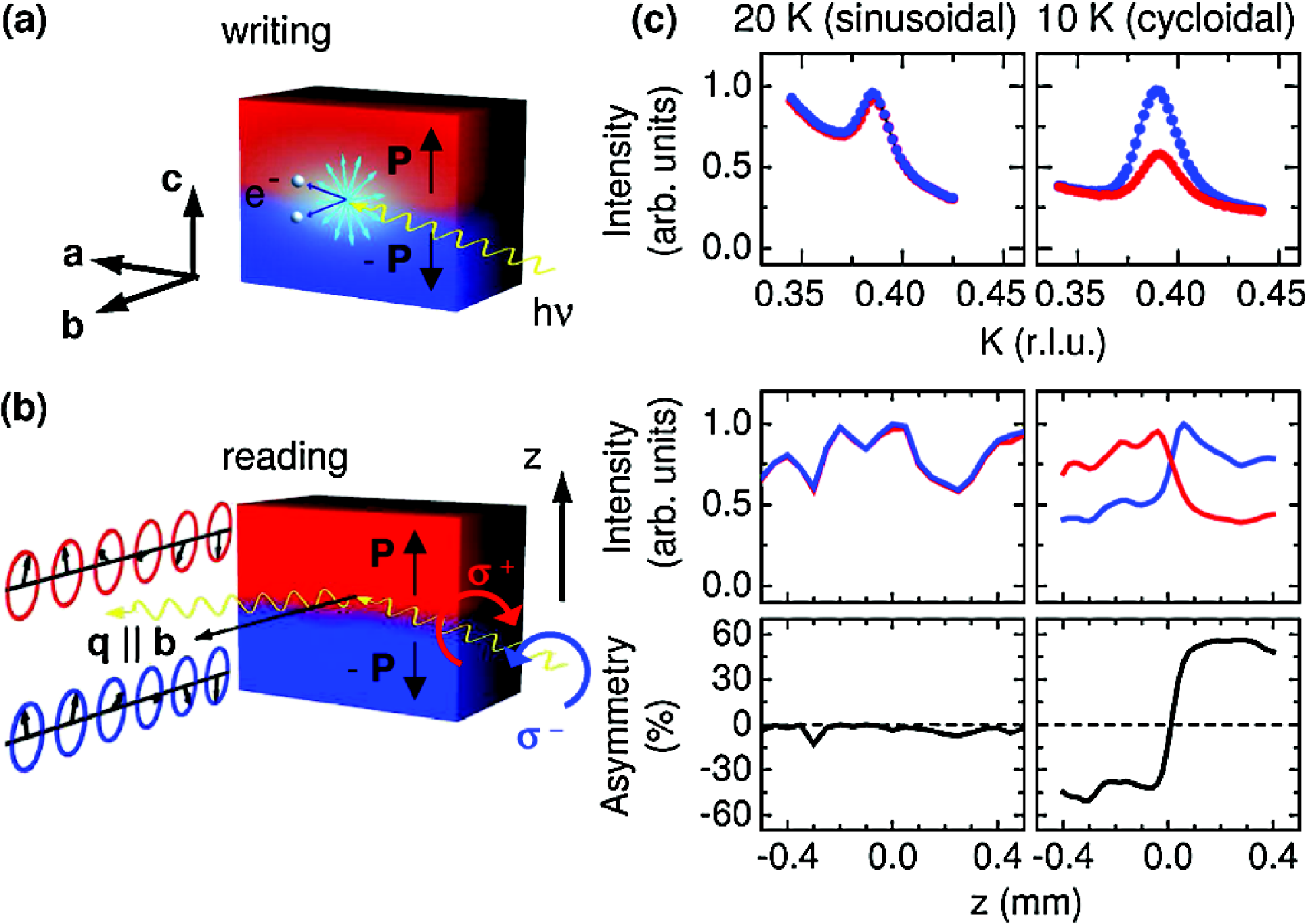}
\caption{\label{Schierle2010-1} Manipulating and imaging multiferroic domains in DyMnO$_3$ using photon energies at the Dy $M_5$ resonance. (a) The photoelectric effect results in local charging of the sample, thus inducing two regions of opposite polarisation. (c) The circular dichroic contrast due to the cycloidal magnetic structure changes sign as a function of the sample position $z$, however, only in the ferroelectric cycloidal phase. (Reprinted  from \cite{Schierle2010}. Copyright © 2010, American Physical Society.)}
\end{figure}

The studies reported so far were mainly concerned with the magnetic order of the Mn spins.  However, also the ordering of the $4f$ moments is expected to play a role for the ferroelectric properties of the $RE$MnO$_3$ compounds. This may not be so important for TbMnO$_3$, but the situation is definitely different in case of DyMnO$_3$. Here, the ferroelectric polarisation is much larger than observed for TbMnO$_3$\,\cite{Goto2004}, and its temperature dependence is closely linked to $4f$ ordering\,\cite{Prokhnenko2007}. By tuning the photon energy to the $RE$ $M_{4,5}$ resonances, one takes advantage of the element-selectivity of the method that allows to study the $4f$ magnetic ordering separately. Using circularly polarised x-rays, it was shown that the $4f$ moments in DyMnO$_3$ themselves form a magnetic cycloid with the proper chirality  to support ferroelectric polarisation along the $c$ axis according to Eq.\,\ref{eqnpolarisation}, the same direction, as also promoted by the Mn cycloid in the material\,\cite{Schierle2010}. 
This is demonstrated in figure\,\ref{Schierle2010-1}\,(c) where the upper 
panels display the incommensurate ($0 \tau 0$) diffraction peak, recorded at the Dy $M_5$ resonance using left and right circularly polarised light. Obviously, a large intensity difference is observed in the ferroelectric phase at 10~K, which is absent in the paraelectric phase at 20~K. This latter phase is characterized by a collinear sinusoidal magnetic modulation that exhibits no handedness. Figure\,\ref{Schierle2010-1} also suggests an interesting approach to manipulate ferroelectric domains: as the photoelectric effect induces local charging of the sample surface in the highly insulating DyMnO$_3$, a radial electric field is induced that creates two regions of opposite electric polarisation along the crystallographic $c$ axis (cf. figure\,\ref{Schierle2010-1}\,(a)). They are characterized by magnetic cycloids of opposite handedness. When the sample is scanned along $z$, i.e.,  parallel to the $c$ axis, the asymmetry of the magnetic diffraction signal, defined as $A=(I_{\sigma^+} - I_{\sigma^-})/(I_{\sigma^+} + I_{\sigma^-})$, changes its sign as seen in figure\,\ref{Schierle2010-1}\,(c).

\begin{figure}[t!]
\centering 
\includegraphics* [scale=0.4, trim= 0 0 0 0, angle= 0] {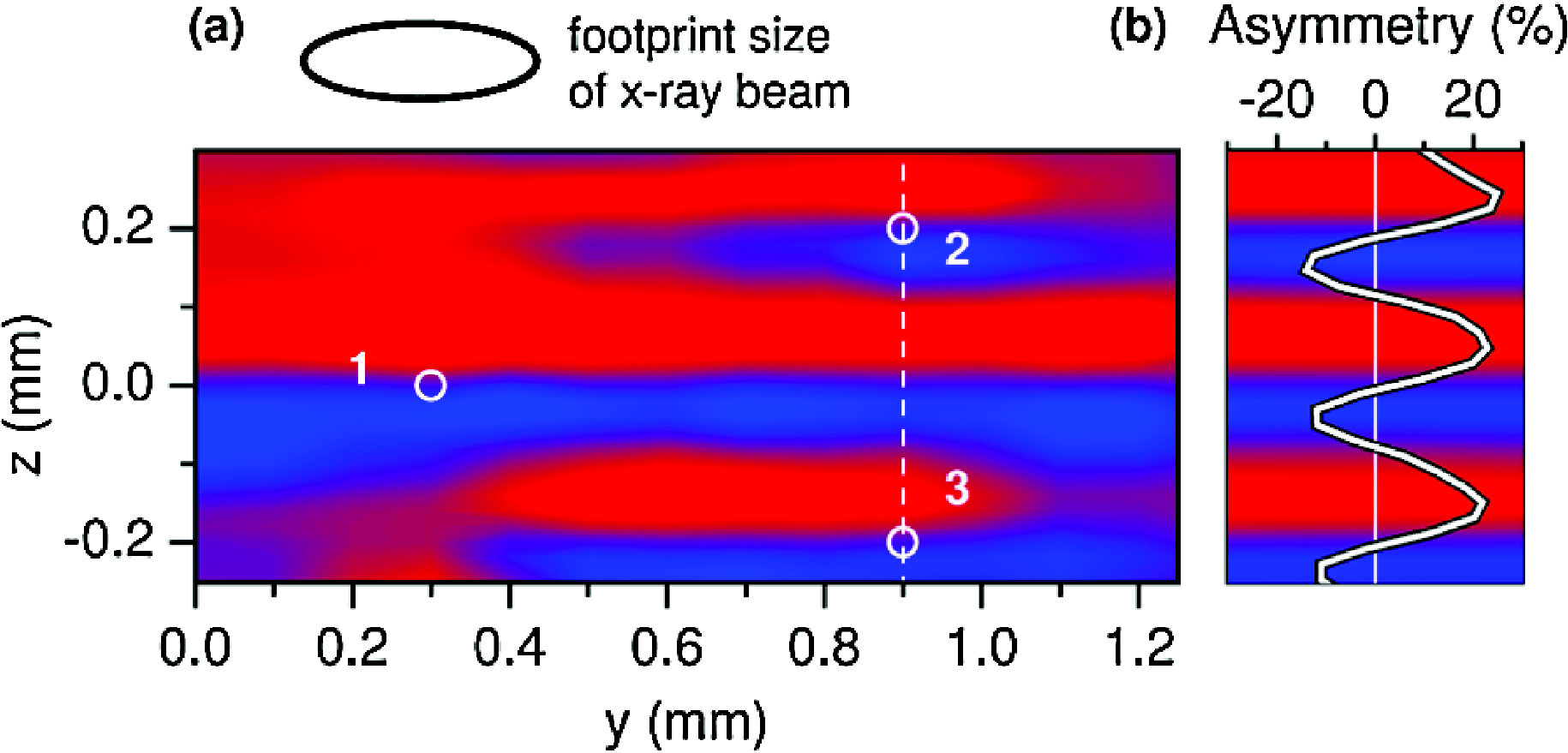}
\caption{\label{Schierle2010-2} Image of a multiferroic domain pattern imprinted on the surface of DyMnO$_3$ by the x-ray beam. The contrast was obtained from the circular dichroic asymmetry shown in figure\,\ref{Schierle2010-1}.(Reprinted  from \cite{Schierle2010}. Copyright © 2010, American Physical Society.)}	
\end{figure}

The contrast obtained by the circular dichroic asymmetry in scattering could eventually be used to image multiferroic domains in the material as shown in figure\,\ref{Schierle2010-2}. The domain pattern was imprinted by the x-ray beam either while cooling through the paraelectric/ferroelectric phase transition (position 1) or during heating close to the transition temperature in order of facilitate switching of the electric polarisation (positions 2 and 3). Using the circular dichroic contrast of the incommensurate ($0 \tau 0$) diffraction peak, the resulting domain pattern was imaged by scanning the sample\,\cite{Schierle2010}.  

While it was argued that the electric field generated by the local charging via the photoelectric effect\,\cite{Schierle2010} is inducing the ferroelectric domains, later studies of YMn$_2$O$_5$ and ErMn$_2$O$_5$ provided evidence that small currents flowing through the samples upon the application of the electric fields may also play a role\,\cite{Souza2011} . 

\begin{figure}[t!]
\centering 
\includegraphics* [scale=0.8, trim= 0 0 0 0, angle= 0] {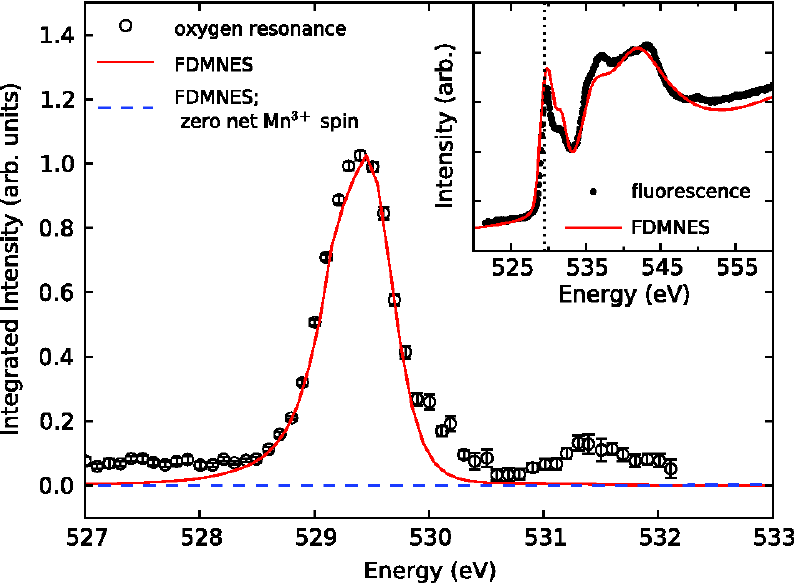}
\caption{\label{Beale2010-1} Resonant behaviour of the antiferromagnetic (1/2 0 1/4) reflection of TbMn$_2$O$_5$ across the O\,$K$ resonance, revealing spin polarisation of O $2p$ states. Inset: absorption edge near the O\,$K$ edge measured in fluorescence yield. The solid red lines  in both panels and and the blue dashed line display model calculations with and without spin on the Mn$^{3+}$ ions, respectively. (Reprinted with permission from\,\cite{Beale2010b}. Copyright © 2010, American Physical Society.)}	
\end{figure}

 After essentially considering the $3d$ and $4f$ magnetic ordering in the multiferroic $RE$ manganites, recent studies have eventually also elucidated the role of oxygen in multiferroics. As discussed before, an incommensurate magnetic cycloid was the driving force for ferroelectricity in $RE$MnO$_3$ compounds. For TbMn$_2$O$_3$, on the other hand, the magnetic ordering connected to the ferroelectric phase is commensurate\,\cite{Chapon2004,Koo2007}, and almost collinear. Nevertheless, strong magnetoelectric effects are observed\,\cite{Hur2004}. A possible mechanism involving charge transfer between Mn$^{3+}$ and O was suggested to account for the large electric polarisation of the material\,\cite{Moskvin2008}. Resonant soft x-ray diffraction at the O\,$K$ edge of TbMn$_2$O$_5$ provided clear evidence for a magnetic polarisation of the oxygen along with the commensurate antiferromagnetic ordering in the ferroelectric phase\,\cite{Beale2010b}. Figure\,\ref{Beale2010-1} displays the intensity change of the antiferromagnetic (1/2 0 1/4) reflection when the photon energy is scanned across the O $K$ edge, revealing resonant behaviour at energies that correspond to unoccupied oxygen states characterized by hybridisation with Mn 3$d$ states. Model calculations of this resonance behaviour with and without Mn spin supported the interpretation that in fact a magnetic polarisation at the O site is observed that is antiferromagnetically correlated\,\cite{Beale2010b}. These results support the above mentioned theoretical models\,\cite{Moskvin2008} that the O magnetic  polarisation is key to the magnetoelectric coupling mechanism.

\begin{figure}[t!]
\centering 
\includegraphics* [scale=0.8, trim= 0 0 0 0, angle= 0] {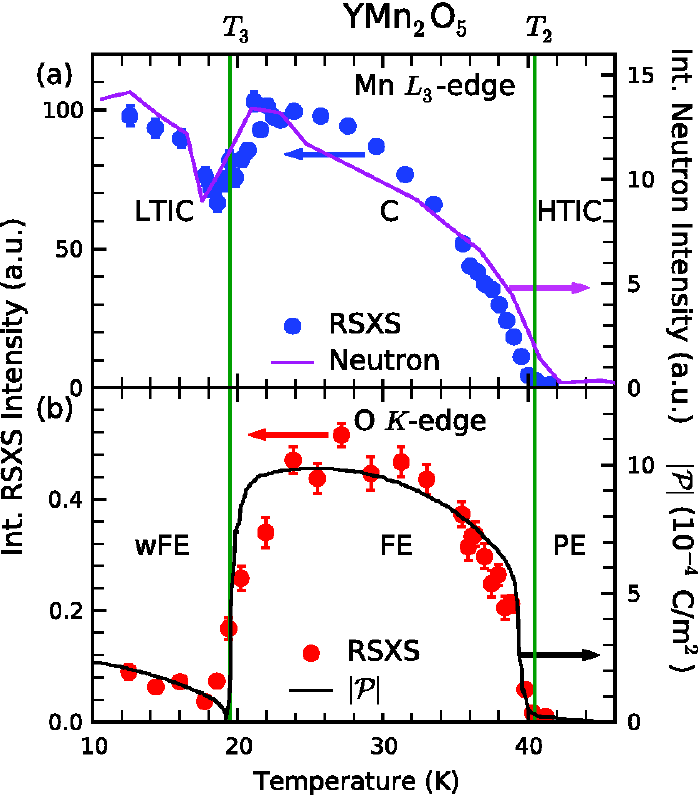}
\caption{\label{Partzsch2011-1} Temperature dependences of the antiferromagnetic (1/2 0 1/4) reflection of YMn$_2$O$_5$, recorded at the Mn $L_3$ (top) and the O $K$ resonance\,\cite{Partzsch2011}. The former tracks the magnetic order parameter as obtained from neutron scattering, while the latter tracks the ferroelectric polarisation.  (Reprinted  from\,\cite{Partzsch2011}.  Copyright © 2011, American Physical Society.)}	
\end{figure}

A similar result was obtained by Partzsch et al.\,\cite{Partzsch2011}, who compared the temperature dependence of the (1/2 0 1/4) reflection in the isostructural compound YMn$_2$O$_5$ both at the Mn $L_3$ resonance and the O $K$  edge (see figure\,\ref{Partzsch2011-1}). While the behaviour at the Mn $L_3$ resonance closely resembles the magnetic order parameter as measured by neutron diffraction (top), the temperature dependence at the O\,$K$ edge  is rather different and follows the spontaneous polarisation in the material, i.e.,  tracking the ferroelectric order parameter (bottom). This provides evidence that the covalency of the Mn and O atoms plays a central role for the ferroelectric polarisation in YMn$_2$O$_5$. Density-functional calculations indeed show that the spin order drives a redistribution of the valence band electrons resulting in a purely electronic contribution to the  ferroelectric polarisation\,\cite{Partzsch2011}. 
A further result in this context is to be  mentioned here: a study of TbMn$_2$O$_5$ revealed the existence of an internal field defined by $\mathbf{S}_{q}\times \mathbf{S}_{-q}$.
A quantity that is related to that expression and that was measured by resonant diffraction follows the same temperature dependence as the ferroelectric polarisation\,\cite{Okamoto2007}. 

\begin{figure}[t!]
\centering 
\includegraphics* [scale=0.8, trim= 0 0 0 0, angle= 0] {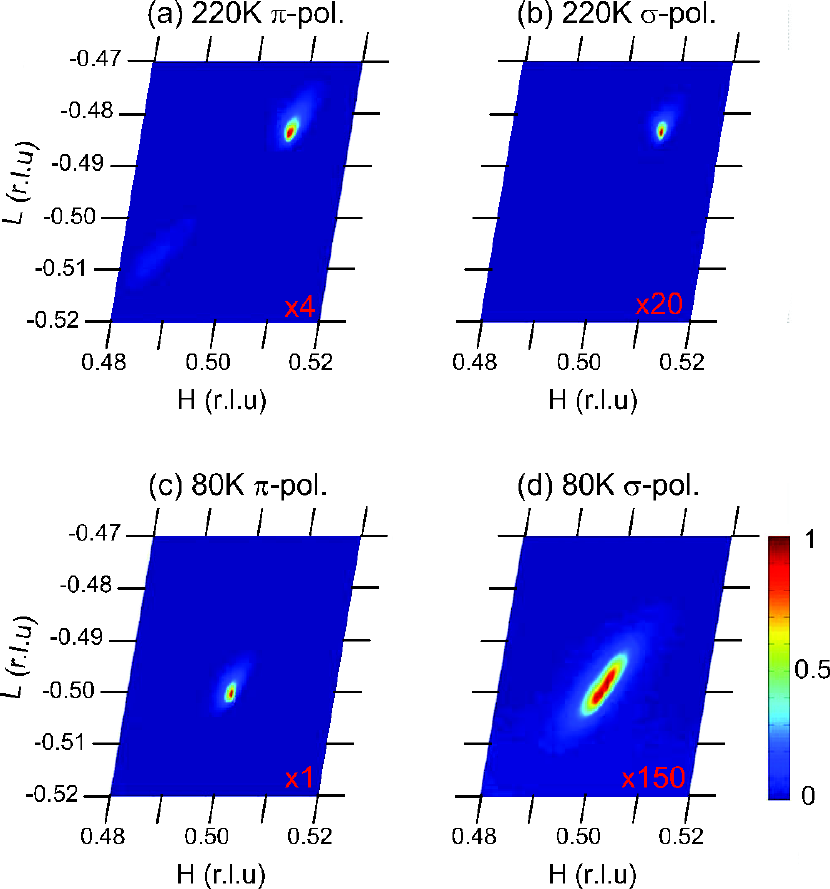}
\caption{\label{Wu2010-1} Two-dimensional distributions of magnetic
scattering. (a) and (b) Intensity distributions in the $a^*c^*$ plane
from resonant soft x-ray magnetic scattering of CuO with $\pi$ and $\sigma$
polarisations at 220 K. (c) and (d) Distributions at 80\,K. Note the substantial broadening in (d). It occurs along the propagation direction of the magnetic spiral. (Reprinted with permission from \cite{Wu2010}. Copyright © 2009, American Physical Society.)}	
\end{figure}

The majority of multiferroic materials studied by RSXS are manganites. The recent discovery that cupric oxide CuO is a multiferroic even at comparably high temperatures around 230\,K\,\cite{Kimura2008}, has triggered magnetic RSXS studies of this compound. RSXS has been used to identify ferroelectric nanoregions by diffuse scattering\,\cite{Wu2010} (see figure\,\ref{Wu2010-1}). An anomalous memory effect for the direction of the electric polarisation in the commensurate-incommensurate magnetic transition has been detected that coincides with the ferroelectric transition. From the RSXS results, incorporated with simulations of diffuse scattering, it was proposed  that a preserved spin
handedness in the multiferroic nanoregions is responsible for this memory effect in the magnetically induced
ferroelectric properties of CuO. Copper oxide was also recently studied  by Scagnoli et al.\,\cite{Scagnoli2011}. Rather than the ferroelectric properties, this investigation was focused on the low-temperature phase of CuO. Polarisation analysis as well as particular dependencies of diffracted intensities on incident circular polarisation revealed  the existence of so-called orbital currents in the material.

\section {Current developments}

The increasing number of studies using RSXS is closely connected with the high intensity and the particular properties of the x-ray radiation available at modern synchrotron radiation sources. The control of the incident x-ray polarisation and its use was already discussed throughout this article. Another characteristic is the high transverse coherence of the radiation that can be provided by undulator sources. Together with the longitudinal coherence given by the high energy resolution, synchrotrons can provide fully coherent x-ray beams for correlation spectroscopy as well as imaging. And finally, synchrotron radiation has a well-defined time structure that can be used for time-resolved experiments. Recent development have pushed the resolution into the femtosecond region, which allows to access the time scales involved in the ordering phenomena discussed throughout this article. Both, high coherence and femtosecond time resolution is eventually provided by free-electron lasers, which will permit both spatially and time-resolved studies.

As the scope of this account is to provide a more general overview of RSXS, it cannot discuss these latter developments in depth. Coherent scattering using soft x-rays, in particular, is already well-developed, and a complete description of this field is beyond the scope of this  article. Time-resolved experiments using RSXS, on the other hand, are scarce, even though studies related to the physics of complex order in correlated materials have been carried out with femtosecond lasers already at length. But also here, the instrumental development is on the way, and many exciting result can be expected in the near future. While a complete description of these fields cannot be accomplished in this article, still, the last section is devoted to a short outlook to what can be achieved using the high sensitivity to magnetic and electronic ordering in the soft x-ray regime, starting with a classical method of x-ray diffraction that only recently was extended to the soft x-ray regime.  

\subsection{Powder diffraction}

Resonant magnetic x-ray diffraction is a method essentially applied to single-crystalline material as discussed throughout this article. And in fact, magnetic x-ray diffraction from powders is very difficult because of the weak signal intensity compared to charge scattering and fluorescence. For the determination of magnetic structures, magnetic neutron powder diffraction is the method of choice, because the available momentum transfer allows to measure a large number of reflections required for structural analysis and refinement. Nevertheless, resonant x-ray diffraction may be useful in cases, where the synthesis of new materials does not yield single crystals, even if the full potential of the method like azimuth dependent studies cannot be used. 

\begin{figure}[t!]
\centering 
\includegraphics* [scale=0.8, trim= 0 0 0 0, angle= 0] {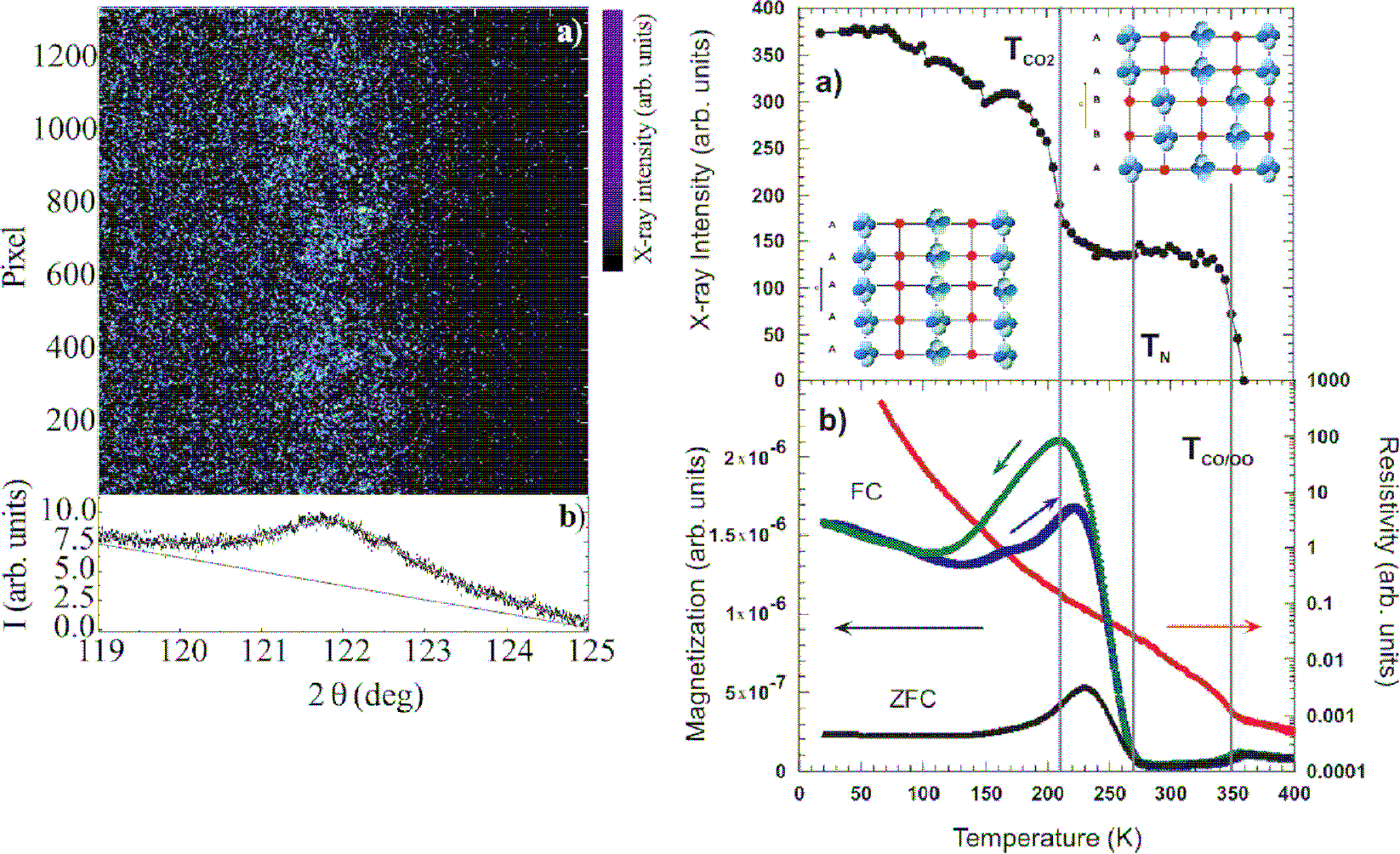}
\caption{\label{Garcia-Fernandez2008-1} Orbital powder diffraction from SmBaMn$_2$O$_6$. Left panel: section of the Debye-Scherrer ring of the (1/4 1/4 0) reflection recorded at the Mn $L_3$  resonance with an x-ray CCD camera together with the integrated intensity as a function of 2$\Theta$. Right panel: (a) temperature dependence of the intensity of the (1/4 1/4 0) reflection. The vertical lines indicate various phase transitions and the insets show the different orbital stackings.  (b) Temperature dependence of the zero-field-cooled
(ZFC) and field-cooled (FC) magnetization during cooling and heating compared to the temperature dependence of the resistivity. (Reprinted with permission from\,\cite{Garcia-Fernandez2008b}. Copyright © 2008, American Physical Society.)}	
\end{figure}

Even in the hard x-ray regime, resonant magnetic powder diffraction studies  are scarce. A recent investigation of GdNi$_2$Ge$_2$ at the comparably weak Gd $L_3$ edge used polarisation analysis and fluorescence background suppression to prepare the magnetic signal~\cite{Kim2005}. Already a decade earlier, resonant magnetic diffraction was demonstrated for the case of UO$_2$~\cite{Collins1995}, exploiting the huge enhancement at the U $M_4$ resonance, a situation comparable to soft x-ray resonances discussed in this article. However, these two studies  essentially demonstrated the feasibility of resonant magnetic x-ray powder diffraction, rather than providing new insights into the microscopic properties of the materials. The situation has changed with recent RSXS  studies~\cite{Staub2007,Garcia-Fernandez2008b,Garcia-Fernandez2009,Staub2009b,Bodenthin2011} that were carried out at the Swiss Light Source.

Figure~\ref{Garcia-Fernandez2008-1} (left panel) shows a two-dimensional image of the diffraction pattern obtained from a powder sample of SmBaMn$_2$O$_6$. The data were obtained at the Mn $L_3$ resonance using an in-vacuum x-ray CCD camera and represent a section of the Debye-Scherrer ring corresponding to orbital order in the material. Vertical integration yields a clearly discernible (1/4~1/4~0) diffraction peak. Diffraction from a powder, of course, provides an average over all azimuthal directions. Therefore, an important input for the study of the type of order by measuring azimuthal dependences is not available. But the energy dependence of the diffraction peak allows to further characterize the orbital order in this layered manganite material as $x^2-z^2/y^2-z^2$ type, compared to $3x^2-r^2/3y^2-r^2$ present in single-layer manganites~\cite{Garcia-Fernandez2008}. Further, the temperature dependence of the diffraction peak was studied (see figure~\ref{Garcia-Fernandez2008-1}, right panel), which provides a measure of the orbital order parameter. Below the ordering temperature of $T_{CO}~\approx 355$\,K, a second orbital phase transition around 210\,K is observed that is also reflected in resistivity and magnetization data. The intensity increase of the orbital reflection is quantitatively reproduced by a structure factor calculation that provides a scenario involving a change of the orbital stacking at the second transition~\cite{Garcia-Fernandez2008}.

\begin{figure}[t!]
\centering 
\includegraphics* [scale=0.8, trim= 0 0 0 0, angle= 0] {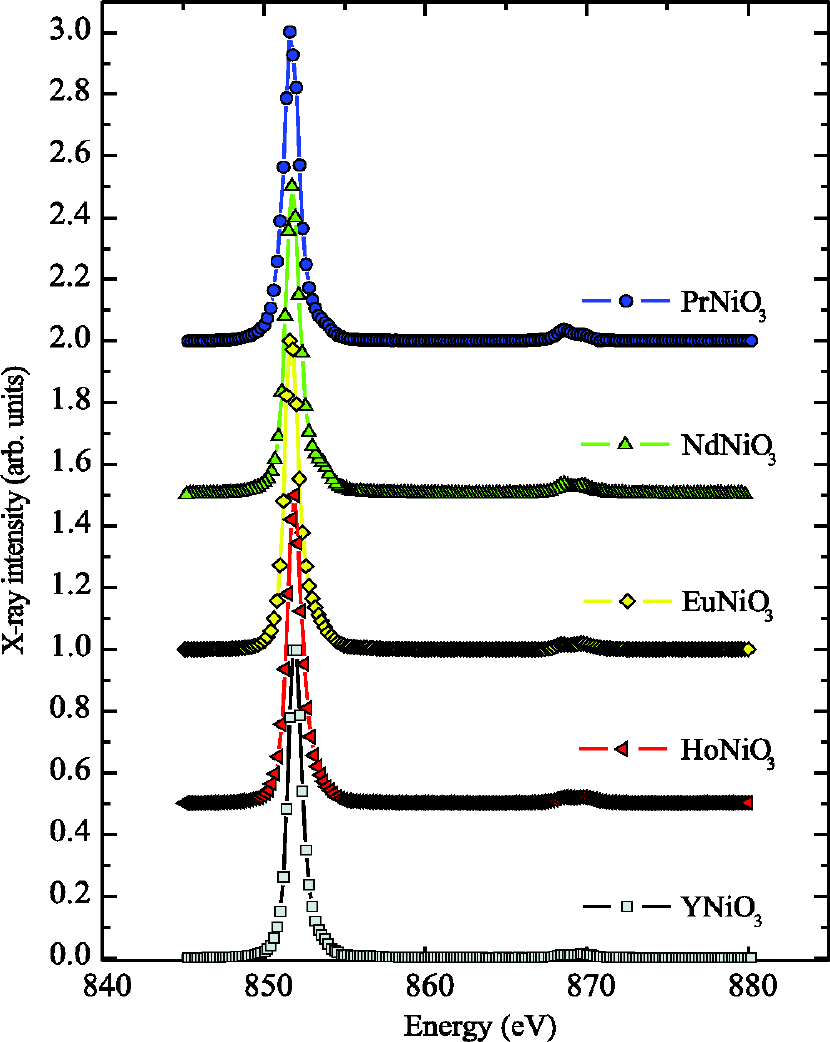}
\caption{\label{Bodenthin2011-1} Magnetic powder diffraction from a series of $RE$NiO$_3$ compounds. The energy dependencies of the (1/2 0 1/2) magnetic diffraction peak across the Ni $L_{2,3}$ resonances reveal a striking similarity indicating that changes in the charge disproportionation between inequivalent Ni sites due to charge order are very small across the  series.  (Reprinted with permission from\,\cite{Bodenthin2011}. Copyright © 20011, Institute of Physics.)}	
\end{figure}

Another example of resonant powder diffraction was given by a study of rare-earth nickelates $RE$NiO$_3$. The interest in these perovskite nickelates stems from metal-insulator transitions with transition temperatures that can be tuned by the ionic radius of the rare-earth ion and that are accompanied by magnetic and charge order. Figure~\ref{Bodenthin2011-1} displays the intensity of the (1/2 0 1/2) magnetic diffraction peak across the Ni $L_{2,3}$ resonance for a series of $RE$NiO$_3$ compounds, with the ionic radius of the $RE$ decreasing from Pr to Y. Again, data were obtained with an x-ray CCD camera and subsequent integration across the Debye-Scherrer ring. The energy dependencies shown in Fig.~\ref{Bodenthin2011-1} reveal a striking similarity, with no energy shifts and little variation of the spectral shape. This shows that the electronic structure of the materials essentially stays the same across the series. More specifically, model calculations of the energy dependencies revealed that the change in the charge disproportionation between inequivalent Ni sites connected with charge order in the material cannot be larger than 0.05 electron charges despite the substantial change in the metal-insulator transition temperature~\cite{Bodenthin2011}.

While the application of resonant soft x-ray diffraction to powder materials is still rare, these examples show the usefulness of the method beyond a mere demonstration of feasibility.

\subsection{Coherent scattering}

Conventional x-ray diffraction investigates predominantly periodic structures. The diffraction patterns are transformed into real-space atomic maps. The determination of non-periodic nanoscale structures by x-ray scattering is much more difficult since the inversion of the intensity profile suffers from the intrinsic loss of phase information. In standard experiments the diffuse scattering intensity around a Bragg peak yields some information on  the disorder of a system, as also discussed  in section\,\ref{sec:CaseStudies}  in connection with magnetic roughness. By limiting the diffraction volume to the coherence volume, extra information about the structure is available. In this case the measured intensity profile has strong fringes and which represent a specific microscopic configuration of the sample, which are called "speckles".  Speckle patterns can be recorded as a function of time, in this way probing the time-dependence of correlations in a material. In holography, on the other hand, a reference beam is used to interfere with the light scattered from the sample. This allows retrieval of the scattering phase and hence reconstruction of the scattering object by a simple Fourier transform. 

Scattering of coherent light is a technique that is well established in the optical regime, as optical lasers providing highly coherent light are available for a long time. Recent years have seen a transfer of the method to the x-ray range\,\cite{Veen2004}, which has become possible with the advent of undulator radiation sources at third-generation synchrotrons that provide x-ray beams with sufficient degree of coherence. A high degree of longitudinal coherence of x-rays is anyway provided by monochromatization, but only undulators with their high brilliance provide sufficient transverse  coherence as well, resulting in a highly coherent x-ray beam. The full power of coherent x-ray scattering will be exploited after the installation of x-ray free electron lasers. 
In the meantime, both, correlation spectroscopy as well as holographic imaging were successfully carried out using soft x-rays, exploiting the magnetic sensitivity at resonance.

\begin{figure}[t!]
\centering 
\includegraphics* [width=7cm,angle=90] {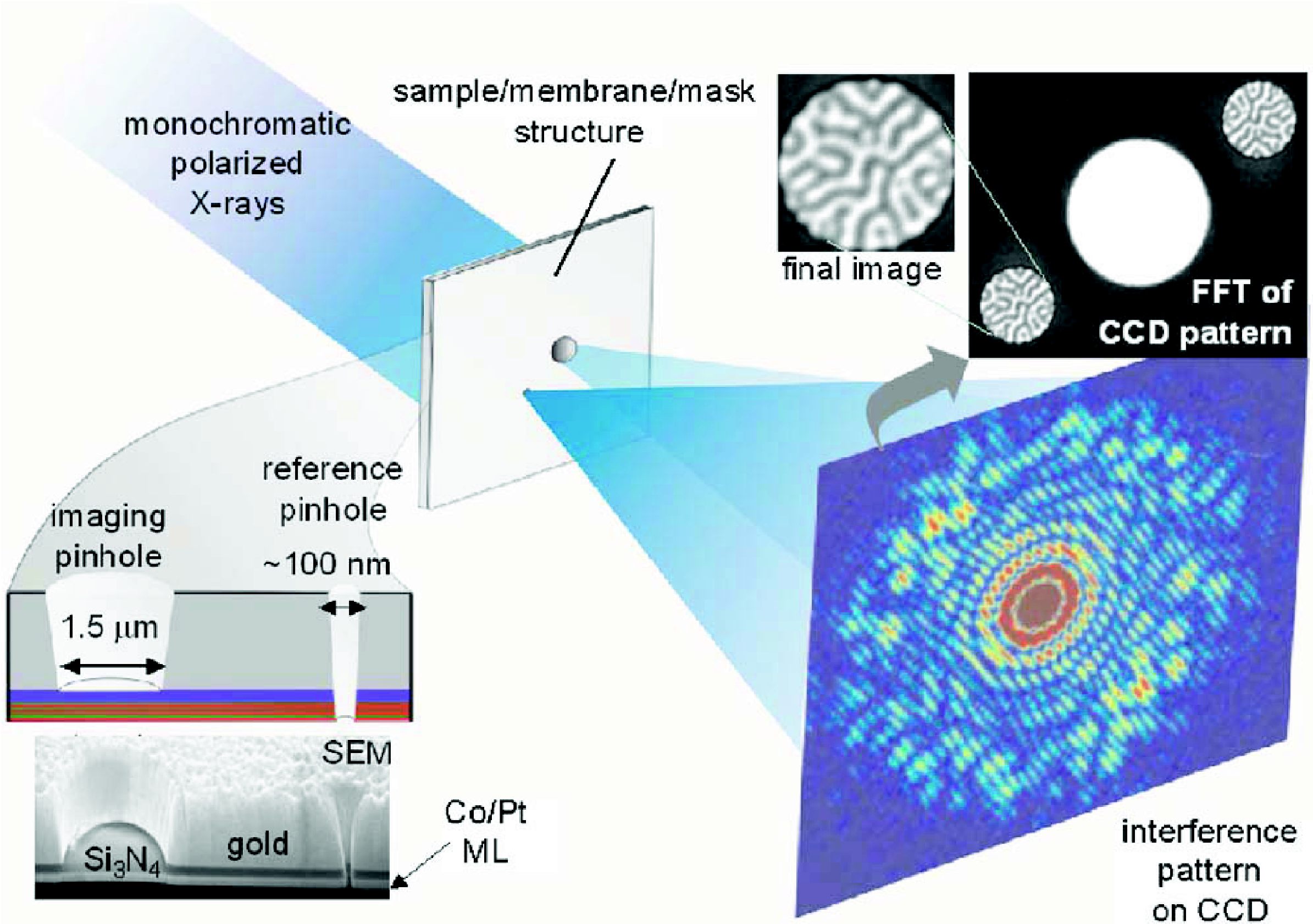}
\caption{\label{fig:Hellwig2006} Illustration of the spectroholography technique for magnetic domains in a Co/Pt  multilayer. The inset at the bottom left displays an illustration and a scanning electron micrograph of the cross section of the sample/membrane/mask structure. In the upper right corner the final image of the magnetic domains derived by a Fourier transform of the speckle pattern  is depicted. The magnetic contrast is provided by the dichroism at the Co $L_3$  resonance. (Reprinted with permission from \cite{Hellwig2006}. Copyright © 2006, American Physical Society.)}
\end{figure}

A first experiment that transferred the method of coherent scattering into the soft x-ray regime was carried out by Price et al.~\cite{Price1999}, who studied fluctuations in smectic liquid crystals. These experiments, however, did not exploit electronic resonances in the materials. The power of coherent scattering in RSXS for the imaging of magnetic domains has been demonstrated in a landmark experiment by Eisebitt et al.\,\cite{Eisebitt2004,Hellwig2006} in which the meandering maze-like  magnetic domains in Co/Pt multilayers with alternating up and down  magnetization perpendicular to the layer surface were investigated. The magnetic contrast has been achieved by exploiting the magnetic dichroism in resonance at the Co $L_{2,3}$ edges using circularly polarised photons.  A schematic picture of the experimental set-up is shown in figure\,\ref{fig:Hellwig2006}. The coherent part of the undulator radiation is selected by an aperture of micrometer diameter and hits a specially designed sample structure: Co/Pt multilayers are deposited on a silicon nitride membrane that is transparent at soft x-ray energies. The imaging pinhole  is of micrometer diameter. The sample also comprises the nanometer-sized reference pinhole that defines a reference beam. The x-rays scattered by the magnetic structure  interfere with the reference beam and form a holographic interference pattern, i.e., the speckle pattern, which encodes the image of the magnetic nanostructure.   The final  image derived by a Fourier transform and  shown in the inset in the upper right corner of figure\,\ref{fig:Hellwig2006} demonstrates the detection of the labyrinth stripe domain pattern which is characteristic for  magnetic systems with perpendicular anisotropy. Magnetic field dependent studies revealed a transformation of the labyrinth domain into isolated domains and, finally near saturation, into isolated bubble domains with increasing field.
Following this first demonstration of lens-less imaging with soft x-rays, further
experiments improved the efficiency of the method~\cite{Marchesini2008} and demonstrated that the resolution of the method can be substantially enhanced beyond the fabrication limit of
the reference structure~\cite{Zhu2010}. It was also shown that lens-less imaging can even be accomplished without reference beam by oversampling methods~\cite{Tripathi2011}.
These exploring experiments demonstrated that with photon sources with shorter pulses and higher brightness, such as free electron lasers, there will be a pathway to study dynamics of magnetic structures in real time to foster progress in rapid  magnetic reading and writing processes.

The availability of free electron lasers in the soft x-ray region with a spectacular increase of some nine order of magnitude in peak brilliance in combination with femtosecond time resolution and high coherence\,\cite{Treusch2010} allow now ground breaking, completely new experiments on the dynamics of magnetic nanoscaled structures.

\begin{figure}[t!]
\centering 
\includegraphics* [scale=0.73, trim= 0 0 0 0, angle= 0] {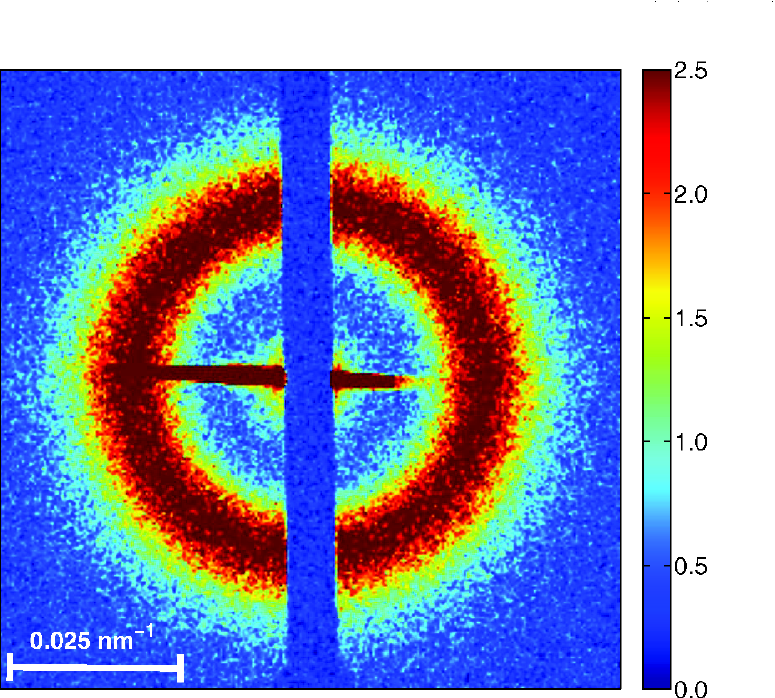}
\caption{\label{Gutt2010-1} Resonant magnetic small-angle scattering
pattern of a Co/Pt multilayer recorded with a single 30 fs free electron laser pulse. The photon wavelength was in resonance with the Co $M_{2,3}$ edge (20.8 nm corresponding to the energy  of 59 eV) (Reprinted with permission from \cite{Gutt2010}. Copyright © 2010, American Physical Society.)}
\end{figure}

 As one of the first examples,  we present the investigation of meandering maze-like domains in Co/Pt multilayers with alternating up and down  magnetization perpendicular to the layer surface (see above). Speckle patterns recorded at the free electron laser FLASH with a time-resolution of 30 fs are depicted in figure\,\ref{Gutt2010-1}\,\cite{Gutt2010}. Magnetic contrast for the domains was obtained by using photons at the Co\,$M_{2,3}$ edge and a geometry  which maximizes the second term in Eq.\,\ref{eqn:Hannon}.
Information on the mean value of the size of the domains could be obtained from a single 30 fs free electron laser pulse. In this work it was also demonstrated that the speckle signal for subsequent pulses provided information on the magnetization dynamics as a function of time. These pioneering results point to the possibility of ultrafast magnetization studies down to the 100 fs timescale.

All these experiments are carried out in transmission geometry, with sample and reference beam scattered from the same substrate, which inherently provides a rigid coupling between sample and reference. In these cases, samples require a special design with particularly thin films in the nanometer range that remain transparent at resonance energies in the soft x-ray regime. In reflection geometry, the scope of coherent scattering can be substantially broadened towards application to surfaces of bulk materials as discussed throughout this paper. While this is more demanding, a successful demonstration at 500 eV photon energy using a test structure was achieved~\cite{Roy2011}. 

\begin{figure}[t!]
\centering 
\includegraphics* [scale=1, trim= 0 0 0 0, angle= 0] {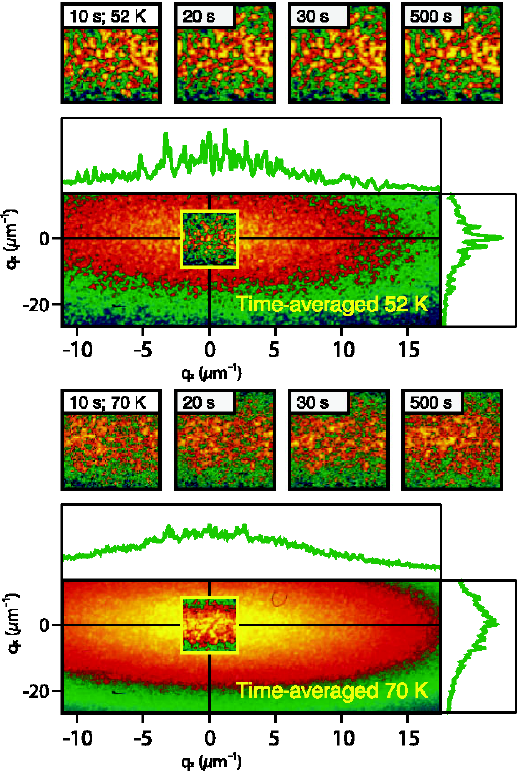}
\caption{\label{Konings2011-1} Speckle patterns recorded from the magnetic Bragg peak of a thin Ho film well below (52~K, top) and close to the ordering temperature (70~K, bottom). The change from a static speckle pattern to a time-dependent pattern can be observed. (Reprinted  from \cite{Konings2011}. Copyright © 2011, American Physical Society.)}	
\end{figure}

Another example of coherent scattering carried out in reflection geometry is given by correlation spectroscopy across the phase transition of helical magnetic ordering in a Ho thin film\,\cite{Konings2011} as shown in figure\,\ref{Konings2011-1}. The helical magnetic structure of Ho leads to a pronounced magnetic diffraction peak that can be probed by x-rays at the Ho $M_5$ resonance\,\cite{Weschke2004} (cf. \ref{sec:long-periodAFM}). At 52\,K, well below the ordering temperature of the film ($T_N \approx 76$\,K), the speckle pattern connected with this diffraction peak is static, as shown on the top panel. It shows that magnetic fluctuations at this temperature are practically absent. Upon heating to 70\,K close to the  ordering temperature, the speckle patterns change with time, indicating the development of magnetic fluctuations at the phase transition. A close inspection of the speckle patterns as well as the time-averaged intensity distribution on the bottom of figure\,\ref{Konings2011-1} shows that some parts still remain fixed. Hence, fluctuations at the phase transition do not affect the whole sample, revealing a non-ergodic behaviour of the system.  An analogous experiment on the orbital order of the half-doped manganite Pr$_{0.5}$Ca$_{0.5}$MnO$_3$  revealed similar behaviour near the phase transition, with both pinned and slowly fluctuating orbital domains~\cite{Turner2008}, 

The examples of coherent scattering described here open a broad field of applications to the materials and their various electronic ordering phenomena discussed throughout this paper.

\subsection{Time-resolved scattering}

The development of lasers with pulse duration in the femtosecond range opened a new field to study ultrafast processes. Magnetic and other complex ordering phenomena can be studied using photons in the optical range, as changes of the symmetry in a system upon ordering are often reflected in the optical properties of the material, particularly involving second-order processes that are excited using intense laser light. Thus, optical wavelengths provide contrast to study the ultrafast dynamics connected with the decay of ferromagnetic or more complex ordering phenomena in transition metal oxides\,\cite{Kimel2004}. However, as soft x-ray resonances provide highest sensitivity to spin, orbital, and charge order in combination with element selectivity, the development of free electron lasers in this energy range represents a large step in this field. In addition to free electron lasers, also slicing facilities at third-generation synchrotron sources provide the ultrashort photon pulses required for time-resolved x-ray diffraction studies with femtosecond time resolution. A setup installed at the Swiss Light Source provides ultrashort x-ray pulses for hard x-ray diffraction. Using structural superlattice reflections associated with charge and orbital order, it was possible to study the dynamics in La$_{0.42}$Ca$_{0.58}$MnO$_3$\,\cite{Beaud2009}. 

\begin{figure}[t!]
\centering 
\includegraphics* [scale= 0.57, trim= 0 0 0 0, angle= 0] {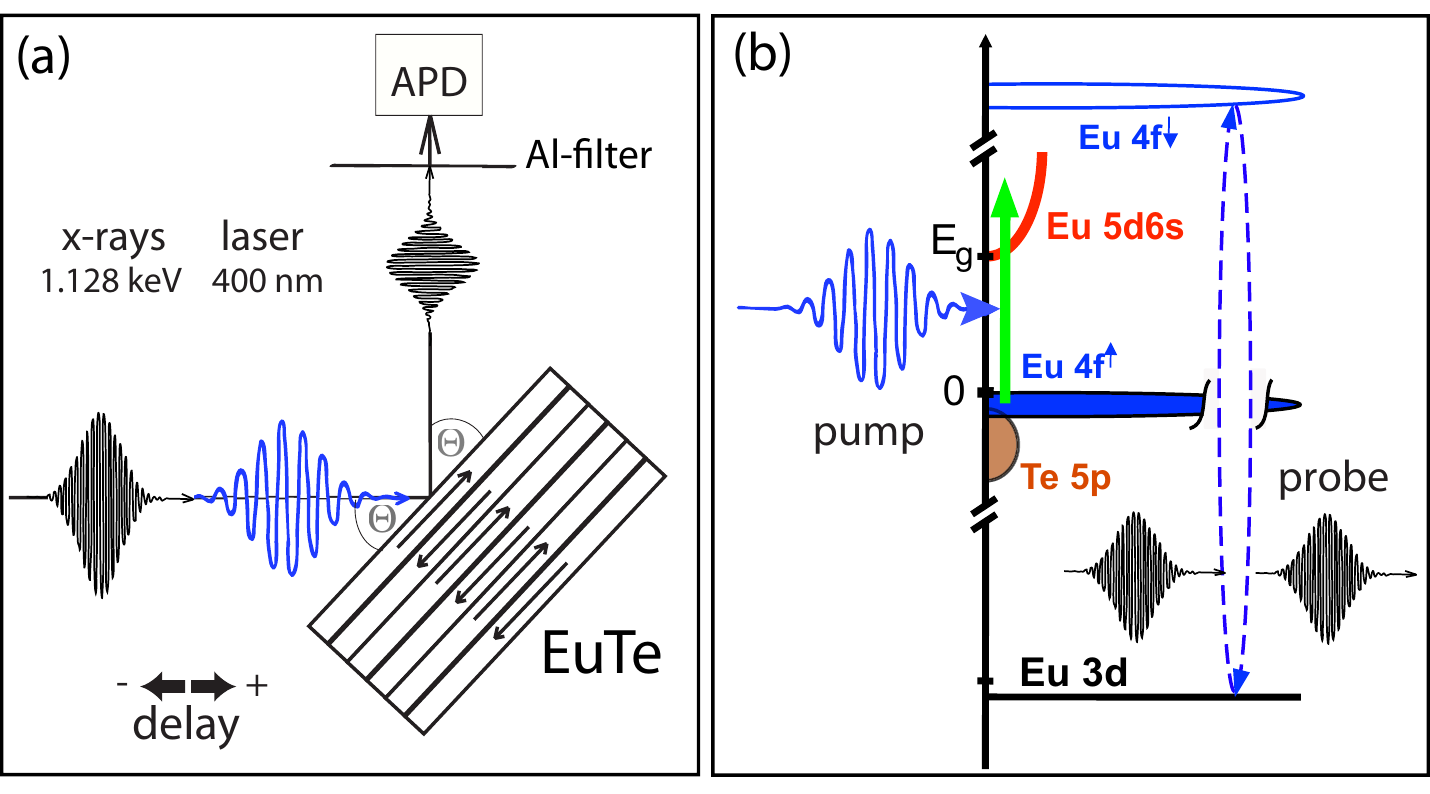}
\caption{\label{Holldack2010-1} (a) Pump-probe setup for time-resolved magnetic x-ray diffraction from antiferromagnetic EuTe. (b) Schematic energy diagram illustrating the 400~nm pump excitation (green arrow) and the resonant 1128~eV x-ray probe (dashed blue arrows). (Reprinted  from \cite{Holldack2010}. Copyright © 2010, American Physical Society.)}
\end{figure}

A slicing beamline covering the soft x-ray region and providing linear as well as circularly polarised light is installed at BESSY II at the Helmholtz-Zentrum Berlin\,\cite{Zholents1996,Khan2006}. Here, studies of demagnetization dynamics could be extended to the soft x-ray region, with the particular advantage of element selectivity and the possibility to access spin and orbital moment separately. First experiments were concerned with x-ray absorption only\,\cite{Stamm2007,Boeglin2010}, but recently, also RSXS experiments with fs time resolution were carried out successfully. For that purpose, EuTe films were used that were already discussed in section\,\ref{magneticdepth}, as they exhibit magnetic Bragg peaks of unprecedented quality at the Eu-M$_5$ resonance\,\cite{Schierle2008}. Time resolved experiments were carried out using a pump-probe scheme as illustrated in figure\,\ref{Holldack2010-1}. The system was excited by a laser using a wavelength of 400 nm that is sufficiently small for an excitation across the band gap of the material and corresponds approximately to a $4f\rightarrow5d$ excitation. The photon energy of the probe pulse was tuned to the Eu $M_5$ resonance and the scattering geometry was chosen to correspond to the ($\frac{1}{2} \frac{1}{2} \frac{1}{2}$) AFM Bragg peak that occurs almost at a scattering angle of 
$90^\circ$. 

\begin{figure}[t!]
\centering 
\includegraphics* [scale=0.73, trim= 0 0 0 0, angle= 0] {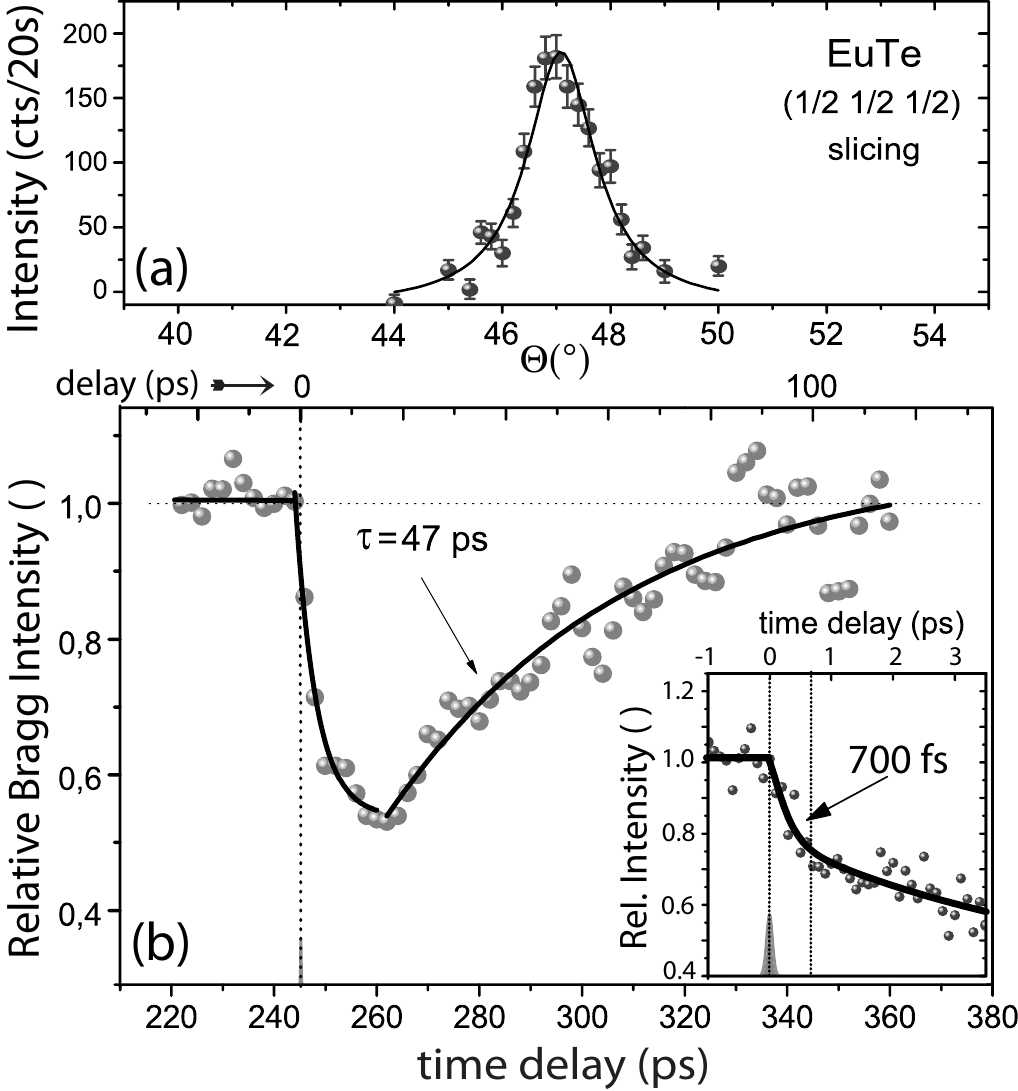}
\caption{\label{Holldack2010-2} (a) Rocking scan through the magnetic ($\frac{1}{2} \frac{1}{2} \frac{1}{2}$) diffraction peak of EuTe recorded with  100~fs probe pulses without laser pumping and Lorentzian fit (solid line). (b) Normalized peak intensity as a function of pump-probe delay. Inset: details of the initial decay. Solid lines represent the results of exponential fits. Remarkably, a fast decay with a time constant of 700 fs is observed. (Reprinted  from \cite{Holldack2010}. Copyright © 2010, American Physical Society.)}
\end{figure}

Figure\,\ref{Holldack2010-2} displays the intensity of the AFM diffraction peak as a function of the time delay between the optical pump and the x-ray probe pulse. Interestingly, a very fast decay of the AFM order is observed, which is faster than ($700\pm 200$)~fs. This is quite remarkable for a semiconductor, as the fast demagnetization mechanisms involving conduction electrons known from metals are not available. 
Figure\,\ref{Holldack2010-2} also displays a rocking scan through the AFM diffraction peak recorded with a 100 fs probe pulse without laser pumping. This demonstrates that in this type of experiments the capabilities of diffraction beyond recording the mere peak intensity can be exploited to study the temporal evolution of spatial correlations in the ordered phase.

\begin{figure}[t!]
\centering 
\includegraphics* [scale=0.4, trim= 0 0 0 0, angle= 0] {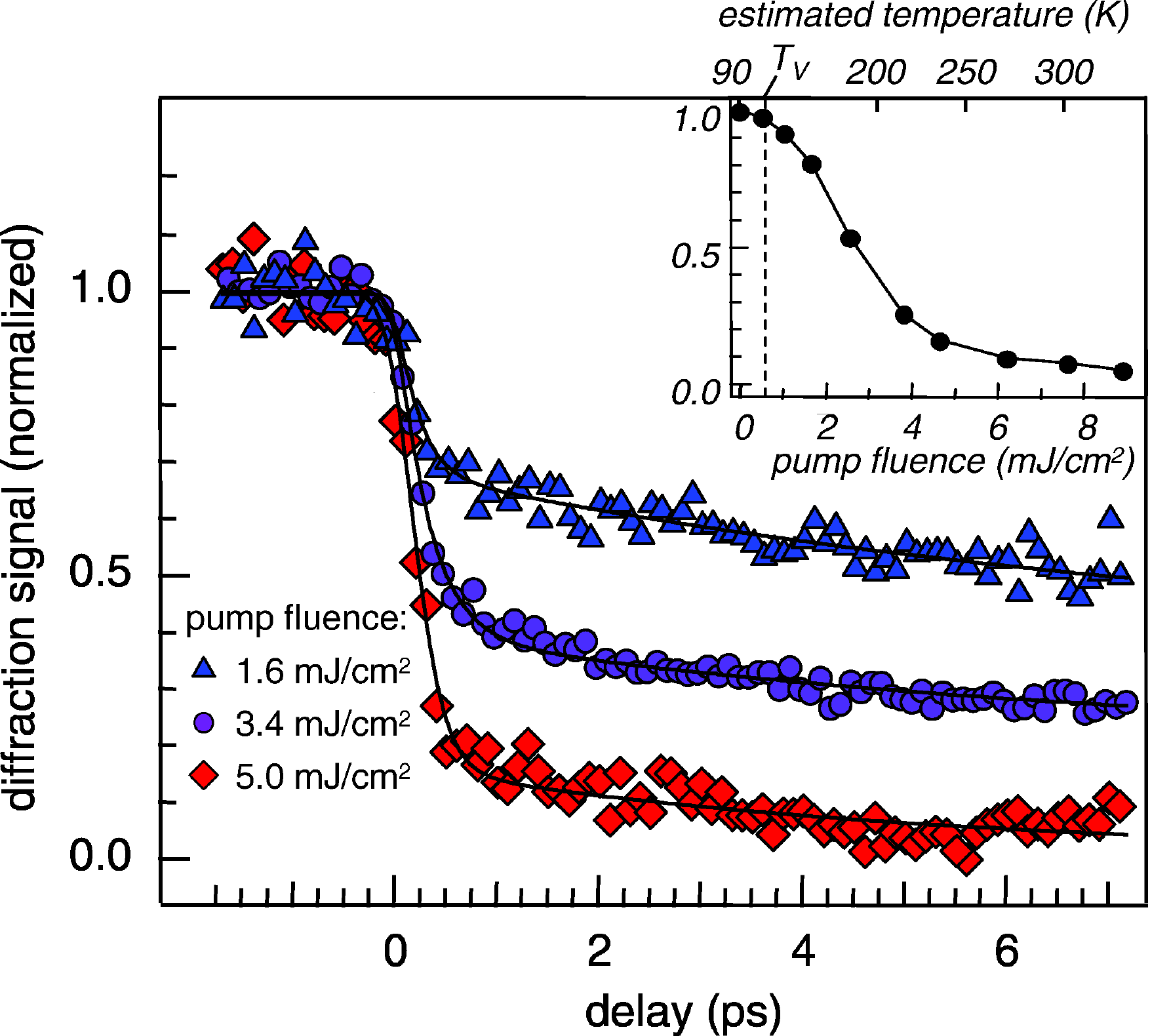}
\caption{\label{Pontius2011-1} 
Time-dependent O\,$K$ edge (0,0,1/2) RSXS signal of Fe$_3$O$_4$
vs. laser pump and x-ray probe time-delay for three different fs-laser
pulse fluences. The lines represent double exponential fits. The
inset shows the fluence dependence of the (0,0,1/2) RSXS signal measured
at 200 ps time delay.(Reprinted with permission from \cite{Pontius2011}. Copyright © 2011, American Physical Society.)}
\end{figure}

While experimental facilities at the Free-Electron Laser in Hamburg (FLASH) at Desy  (Deutsches Elektronen-Synchrotron) and the Linac Coherent Light Source (LCLS) at the SLAC National Accelerator Laboratory have paved the way to ultrafast studies, time-resolved scattering experiments for the study of correlated materials using soft x-rays are presently still scarce. An experiment recently carried out at FLASH is concerned with the decay of orbital order in the prototypical correlated material magnetite Fe$_3$O$_4$\,\cite{Pontius2011}(cf. section\,\ref{Magnetite}). In figure\,\ref{Pontius2011-1} the time-dependent intensity of the (0 0 1/2) reflection, measured with photons at the O\,$K$ edge, is shown as a function of time after a femtosecond  laser pump pulse, for several pump fluences. Since the intensity of the peak in resonance is directly related to the unoccupied density of states, this intensity provides direct information about the electronic order and hence of the melting of charge/orbital order induced with the infrared pulse. An ultrafast melting of the charge/orbital order has been found in the rapid decrease of the intensity of the superstructure peak. The persistent intensity of the peak after 200 ps even for pump fluences that correspond to sample heating above the Verwey transition (inset of figure\,\ref{Pontius2011-1}) was interpreted in terms of  a transient phase characterized by the existence of partial charge/orbital order which has not been observed in equilibrium. In this way the time-dependent RSXS experiment have demonstrated that important information on the dynamics of phase transitions in correlated systems can be obtained.

\begin{figure}[t!]
\centering 
\includegraphics* [scale=0.2, trim= 0 0 0 0, angle= 0] {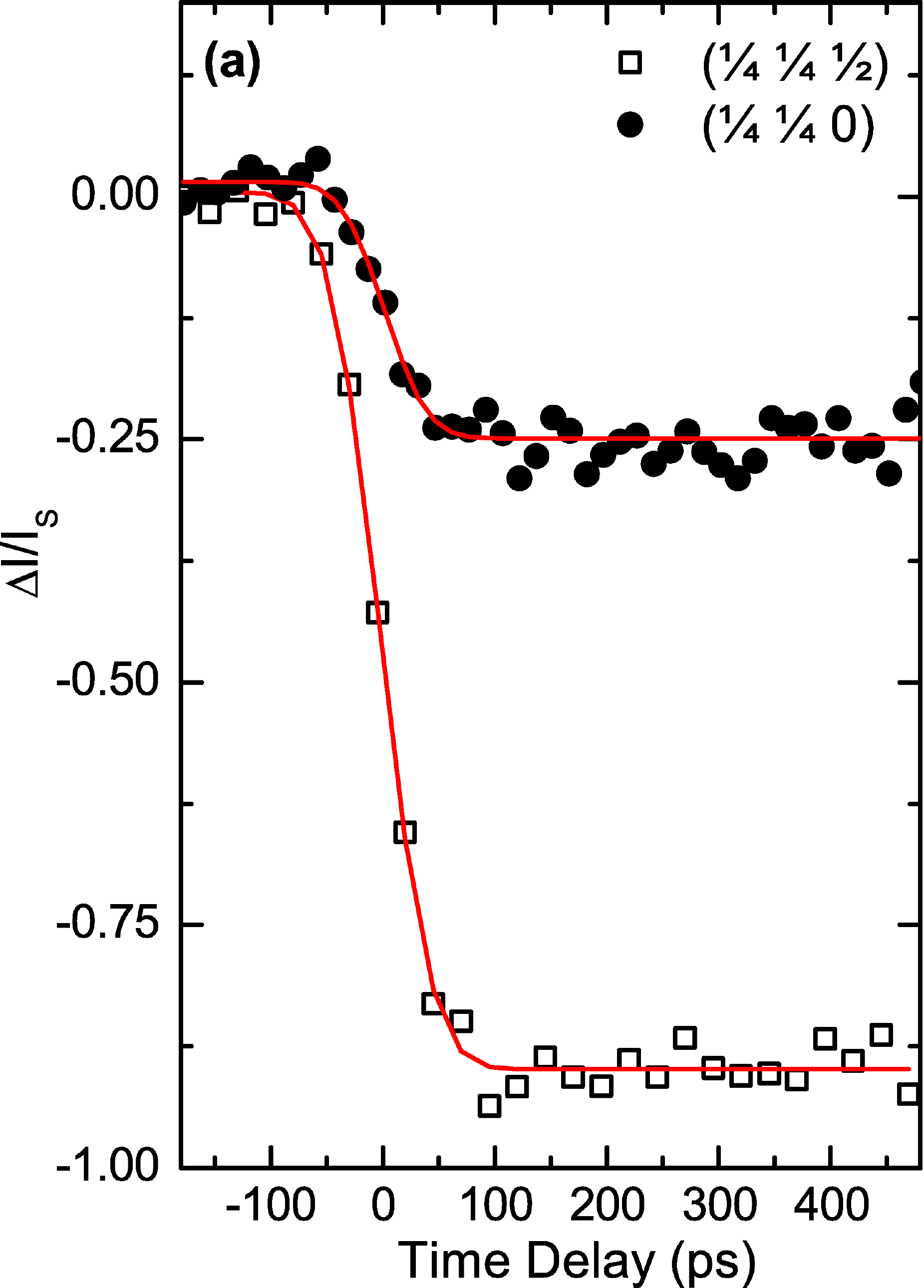}
\caption{\label{Ehrke-1} Time-dependence of the ($\frac{1}{4} \frac{1}{4} \frac{1}{2}$) antiferromagnetic and the ($\frac{1}{4} \frac{1}{4} 0$) orbital diffraction peak of  La$_{0.5}$Sr$_{1.5}$MnO$_4$ measured at the Mn\,$L_3$ edge after excitation with a femtosecond laser pulse. (Reprinted with permission from \cite{Ehrke2011}. Copyright © 2011, American Physical Society.)}
\end{figure}

A third time-resolved experiment, albeit only with picosecond  resolution, was reported recently from Diamond Light Source\,\cite{Ehrke2011}. 
Time resolution was achieved here by gating the synchrotron radiation pulses synchronized with the laser pulses. The material studied was La$_{0.5}$Sr$_{1.5}$MnO$_4$, which is characterized by spin and orbital order (cf. section\,\ref{sec:Manganites}). These two types of order are represented by the ($\frac{1}{4} \frac{1}{4} \frac{1}{2}$) AFM Bragg peak, and the ($\frac{1}{4} \frac{1}{4} 0$) orbital diffraction peak, respectively. Figure\,\ref{Ehrke-1} displays the evolution upon laser excitation of the two diffraction peaks, revealing that spin order can be quickly destroyed, while the orbital order persists. The results were interpreted in terms of an intermediate transient phase  in which the spin order is completely removed  by the photoexcitation while the orbital order is only weakly perturbed. In this way the time-dependent RSXS experiments have demonstrated that it is possible to separate the spin dynamics determined by the electronic structure from Jahn-Teller contributions which are more related to the lattice dynamics.

\section{Summary and Outlook}

During the last two decades RSXS has established itself as a powerful
tool in modern solid state physics to investigate magnetic,
orbital and charge ordering phenomena associated with
electronic degrees of freedom. Since the technique is coupled to specific core excitations it is element specific. The techniques combines diffraction methods with x-ray absorption spectroscopy and therefore delivers structural nanoscale  information on the modulation of the valence band electron density, on the bond orientation, i.e., orbital ordering, and on the spin density. In particular RSXS has provided important structural and spectroscopic information on 3$d$ transition metal compounds and 4$f$ systems which comprise interesting solids such as high-$T_c$ superconductors, charge density compounds, giant and colossal magneto-resistance systems which may be important for spintronics.
Since RSXS has a finite probing depth, the method is not only suited for the investigation of bulk properties: RSXS
provides also important structural information on small crystals, ultra thin films, on multilayer systems and interfaces, and on all kind of nanostructures.

The future of the technique will strongly depend on the instrumental development of the photon sources and the diffractometers. Variable polarisation  is already realized at various beamlines at synchrotron radiation facilities. Fast switching of the polarisation will help to detect small polarisation effects. In several RSXS diffractometers the application of suitable polarisation analyzers will allow to extend investigations of magnetic and orbital ordering in a large class of interesting materials. Arrays of channel plate detectors or charge coupled devices (CCD) will enable to perform faster diffraction experiments. Sample environment will play an important role in future RSXS experiments: the possibility to apply high magnetic fields and low temperatures to the sample will offer interesting studies on new ground states in correlated systems. Furthermore, the development of intense coherent soft x-ray sources by the development of fourth generation synchrotron radiation facilities (energy recovery linacs and free electron lasers) will allow to reveal, element specific  and with nanometer spacial resolution, complicated non-periodic lattice, magnetic, and charge structures.  Here important issues will be phenomena related to fluctuation near phase transitions, chemical reactions, defects and precipitations in materials etc. Finally, with the advent of free-electron laser photon sources in the soft x-ray region, interesting time-dependent pump-probe studies on the structural evolution of photo-excited solids will be feasible on a large scale. Possibly movies of chemical reactions, studies on matter under extreme conditions, watching magnetic spin flips in real time, and unraveling the functional dynamics of biological materials will be feasible.
The future of RSXS will be not only related to bulk studies of the charge, orbital, and spin order in transition metal compounds. Rather the focus will be also on similar studies in thin films, all kind of artificial nanostructures and in particular on the structure and the electronic properties of interfaces. Therefore those RSXS experiments will strongly depend on the development of preparation stations for artificial nanostructures, e.g. by laser deposition or molecular beam epitaxy.

One focus of future RSXS experiments will still be related to complex bulk properties of correlated materials as a function of composition, temperature, and magnetic field. There, mainly charge, spin, and orbital ordering as well as Jahn-Teller distortions will be studied in correlated transition metal and rare earth compounds. New bulk experiments could be considered in the fields of vortex lattices of superconductors since the charge inside and outside of vortexes is different. Since similar to the O $K$ edges in TM oxides, the C $K$ edges in soft matter are strongly dependent on the chemical bonding of C atoms, RSXS will develop to a great tool for investigations of nanoscale structures and spacial distributions in polymer blends, multiblock copolymers, polymer solutions, lipid membranes, colloids, micelles, and polymeric biological composite particles. Furthermore, using coherent light sources, disordered systems will be studied applying  speckle methods. Finally RSXS will be extended to the time domain via setting up pump-probe experiments to study the dynamics of phase transitions.

The other focus of future RSXS will be certainly related in the analysis of structural and electronic properties of artificial structures on the  nanometer scale which have great promise for novel functionality. In particular, RSXS will help to resolve the problem to characterize the electronic structure of oxide hetero-epitaxial devices  in which, as already discussed above, at the interfaces between two band insulators metallicity and even superconductivity has been detected. At present it is not decided whether this observation is due to the polar nature of the layers or due to O vacancies. Future RSXS experiments will certainly help to resolve open questions in this field.

Thus one can be confident that   RSXS will have a great impact  on a vast variety of scientific fields ranging from solid state physics across materials science and chemistry to biology and medicine.

\section*{Acknowledgments}
The authors thank J. Hill, U. Staub, and S. Wilkins for helpful suggestions.

\section*{References}


\end{document}